\documentclass[preprint,11pt]{JHEP3} 

\JHEPspecialurl{http://jhep.sissa.it/JOURNAL/JHEP3.tar.gz}

\usepackage{epsfig,multicol}

\def\beq{\begin{equation}}
\def\eeq{\end{equation}}
\def\bea{\begin{eqnarray}}
\def\eea{\end{eqnarray}}

\def\eq#1{{Eq.~(\ref{#1})}}
\def\fig#1{{Fig.~\ref{#1}}}
\newcommand{\bas}{\bar{\alpha}_S}
\newcommand{\as}{\alpha_S}

\newcommand{\Lb}{\left(}
\newcommand{\Rb}{\right)}

\parskip=\itemsep               

\newcommand{\D}{\partial}
\newcommand{\h}{\frac{1}{2}}

\newcommand{\x}{\vec{x}}
\newcommand{\y}{\vec{y}}

\newcommand{\de}{\delta}

\def\ap#1#2#3{     {\it Ann. Phys. (NY) }{\bf #1} (19#2) #3}
\def\arnps#1#2#3{  {\it Ann. Rev. Nucl. Part. Sci. }{\bf #1} (19#2) #3}
\def\npb#1#2#3{    {\it Nucl. Phys. }{\bf B#1} (19#2) #3}
\def\plb#1#2#3{    {\it Phys. Lett. }{\bf B#1} (19#2) #3}
\def\prd#1#2#3{    {\it Phys. Rev. }{\bf D#1} (19#2) #3}
\def\prep#1#2#3{   {\it Phys. Rep. }{\bf #1} (19#2) #3}
\def\prl#1#2#3{    {\it Phys. Rev. Lett. }{\bf #1} (19#2) #3}
\def\ptp#1#2#3{    {\it Prog. Theor. Phys. }{\bf #1} (19#2) #3}
\def\rmp#1#2#3{    {\it Rev. Mod. Phys. }{\bf #1} (19#2) #3}
\def\zpc#1#2#3{    {\it Z. Phys. }{\bf C#1} (19#2) #3}
\def\mpla#1#2#3{   {\it Mod. Phys. Lett. }{\bf A#1} (19#2) #3}
\def\nc#1#2#3{     {\it Nuovo Cim. }{\bf #1} (19#2) #3}
\def\yf#1#2#3{     {\it Yad. Fiz. }{\bf #1} (19#2) #3}
\def\cpc#1#2#3{    {\it Comp. Phys. Commun. }{\bf #1} (19#2) #3}

\def\thefootnote{\fnsymbol{footnote}}

\title{\huge \bf  The BFKL Pomeron Calculus:  Probabilistic \\Interpretation and High Energy
Amplitude.}
\author{\Large M. Kozlov \thanks{Email: kozlov@post.tau.ac.il}\,\,,\,\,E. Levin\thanks{Email:
leving@post.tau.ac.il, levin@mail.desy.de;}\,\,  and\,\, A.
Prygarin\thanks{Email:
prygarin@post.tau.ac.il;} \\
Department of Particle Physics, School of Physics and Astronomy\\ Raymond and Beverly Sackler Faculty
of Exact Science\\  Tel Aviv University, Tel Aviv, 69978, Israel}

\abstract{In this paper we continue to pursue a goal of finding the effective theory for high energy
interaction
in QCD based on the colour dipole approach, for which the BFKL Pomeron Calculus gives a low energy
limit.   The two key problems, that we consider are the following:
the probabilistic interpretation of the BFKL Pomeron Calculus and the possible
scenario for the asymptotic behaviour of the scattering amplitude at high energy
in QCD. In this paper we show that the generating functional approach is equivalent to the BFKL Pomeron
Calculus and leads to a clear interpretation of this calculus as an alternative description of the
system of interacting colourless dipoles.

 We calculated the two BFKL Pomerons into one BFKL Pomeron vertex directly in the dipole approach
and  show that this merging  can be described as the decay of
two dipoles into four dipoles in the dipole approach. This result means the JIMWLK approach can give us all needed
vertices for the BFKL Pomeron calculus and, therefore, they together give the selfconsistent and complete theoretical
 approach to high energy scattering in QCD.

In this paper, we developed the
semi-classical
approach to the functional evolution equation for the generating functional. Using this method  we found
the solution in the
entire kinematic region starting from low energies  for the simplified case of
the toy model in which
we assume that interaction of dipoles does not depend on their sizes.  In the general case we find the
asymptotic solution
at ultra high energies as well as the first correction to this solution.

We considered the semi-classical approach to the BFKL Pomeron Calculus and  recovered the
Mueller-Shoshi band for the kinematic region where we can trust the linear BFKL equations.
The solution for the scattering amplitude deeply in the saturation region was found and
discussed.}


\preprint{ \bf TAUP - 2825-06\\
\today\\
{\tt hep-ph/0606260}}
\begin{document}

\def\thefootnote{\arabic{footnote}}
\section{Introduction}
\label{sec:Int}
The simplest approach that we can propose for  high energy interaction is
based \cite{GLR,MUQI} on the BFKL Pomeron \cite{BFKL} and reggeon-like diagram technique
for the BFKL Pomeron interactions \cite{BART,BRN,NP,BLV}. This technique , which is a
generalization of Gribov Reggeon Calculus \cite{GRC},  can be written
in the elegant form of the functional integral (see Ref. \cite{BRN} and the next
section); and it is a challenge to solve this theory in QCD  finding the high energy
asymptotic behaviour.  However, even this simple approach has not been solved during
three decades of attempts by the high energy community. This failure stimulates a search
for deeper understanding of physics which is behind the BFKL Pomeron Calculus.
On the other hand, it has been known for three decades that  Gribov Reggeon Calculus has intrinsic
difficulties \cite{KAN} that are related to the overlapping of Pomerons. Indeed, due to this overlapping
we have no hope that the  Gribov Reggeon Calculus could be correct in describing the ultra high energy
asymptotic behaviour of the amplitude. The way out of these difficulties we see in searching for  a new
approach which will  coincide
with the BFKL Pomeron Calculus at high, but not very high,  energies (our correspondence principle)
but it
will be different in the region of ultra high energies. In a  spirit of the  parton approach we
believe that
this effective theory should be based on the interaction of `wee' partons.  We consider, as an important
step in this direction, the observation that has been made at the end of the Reggeon era
\cite{GRPO,LEPO,BOPO},
  that
the Reggeon Calculus
can be reduced to a Markov process \cite{GARD} for the probability of finding a given
number of Pomerons at fixed rapidity $Y$.  Such an interpretation, if it would be
reasonable in QCD,  can be useful,
 since it allows us to use powerful methods of statistical
physics in our search for the solution.

The logic and scheme of our approach looks as follows.
The first step is the Leading Log (1/x) Approximation (LLA)  of perturbative QCD in which we sum all
contributions of the order of $ \left(\as\,\ln(1/x) \right)^n$. In the LLA we consider such high
energies that
\beq \label{LLA}
\as\,\ln(1/x)\,\,\approx\,\,1\,;\,\,\,\,\,\,\mbox{while}\,\,\,\,\,\,\,\as\,\,\ll\,\,1
\eeq
It is well known that the LLA approach generates the BFKL Pomeron (see Ref. \cite{BFKL} and the next
section)  which leads to the power-like
increase of the scattering amplitude ($ A \,\propto \,\frac{1}{x^{\omega(n=0,\nu=0)}}$ with \\
 $
\omega(n=0,\nu=0)\,\,\propto\,\as$).

The second step is the BFKL Pomeron Calculus in which we sum all contributions of the order of
\beq \label{PC}
\left( \frac{\bas^2}{N^2_c}\,\frac{1}{x^{\omega(n=0,\nu=0)}} \right)^n
\,\,\,\,\,\,\mbox{therefore}\,\,\,\,\,\,\,
\frac{\bas^2}{N^2_c}\,\frac{1}{x^{\omega(n=0,\nu=0)}}\,\,\approx\,\,1
\eeq
where $\bas\,\,=\,\,\as N_c/\pi$.

The structure of this approach as well as its  parameter has been understood before QCD \cite{VEN}
and was confirmed in QCD (see Refs. \cite{GLR,MUQI,BART,BRN,NP,BLV,IM,KOLE}). This calculus extends the
region of energies from $\ln (1/x) \approx\,1/\bas$ of LLA to $\ln (1/x)
\approx\,(1/\bas)\ln(N^2_c/\bas)$. The BFKL Pomeron Calculus describes correctly the scattering
process in the region of energy:
\beq \label{REPC}
\frac{1}{\bas}\,\,\ln\left(\frac{N^2_c}{\bas}\right)\,\,\ll\,\,ln
\frac{1}{x}\,\,\ll\,\,\,\frac{1}{\bas^2}
\eeq
For higher energies the corrections of the order of $\left(\bas^2 \,\ln (1/x)\right)^n$ should be taken
into account making all calculations very complicated.

Our credo is that we will be able to  describe the high energy processes outside of the region
of \eq{REPC}, if we
could  find an effective theory which  describes the BFKL Pomeron calculus in
the kinematic region given by \eq{REPC}, but based on the microscopic degrees of freedom and  not on the
BFKL Pomeron.  In so doing, we hope that we can avoid all intrinsic difficulties of the  BFKL Pomeron
calculus and build an approximation that will be in an agreement with all general theorems like the
Froissart bounds and so on. Solving this theory,  we can create a basis for moving forward considering
all corrections to this theory due to higher orders in $\bas$ contributions, running QCD coupling and
so on.

The goal of this paper is to consider two main problems: the probabilistic
interpretation of the  BFKL Pomeron Calculus based on the idea that colour dipoles
are the correct degrees of freedom at high energy QCD \cite{MUCD}; and the possible
solution for scattering amplitude at high energy.  We believe that colourless dipoles
and their interaction will lead to a future theory at high energies which will have the BFKL Pomeron
Calculus as the low energy limit (see \eq{REPC})  and which will allow us to avoid all difficulties of
dealing with BFKL
Pomerons at ultra high energies.

Colourless dipoles play two different roles in our approach. First, they are
partons (`wee' partons) for the BFKL Pomeron. This role is not related to the large
$N_c$ approximation and, in principle, we can always speak about probability to
find a definite number of dipoles instead of defining the probability to find a
number of the BFKL Pomerons. The second role of the colour dipoles is that at high
energies we can interpret the vertices of Pomeron merging and splitting in terms
of probability for two dipoles  annihilate in one dipole and of probability for
decay of one dipole into two ones. It was shown in Ref. \cite{MUCD} that $P
\rightarrow 2P$ splitting can be described as the process of the dipole decay into
two dipoles. However, the relation between the Pomeron merging ($2 P \rightarrow
P$) and the process of annihilation of two dipoles into one dipole is not so
obvious
and it will be discussed here.

 This paper is a next step in our programme of searching the simplest but
correct approach to high energy scattering in QCD in which we continue the line of thinking presented in
Refs. \cite{L1,L2,L3,L4}.
The outline of the paper looks as follows.

 In the next section we will discuss the
BFKL Pomeron Calculus in the elegant form of the functional integral,  suggested by
M. Braun about five years ago \cite{BRN}. We  will show that the  intensive
recent work on this subject \cite{IT,MSW,L3} confirms the BFKL Pomeron Calculus in
spite of the fact that these attempts were based on slightly different but not more
general  assumptions.

In the third section the general approach based on the generating functional will
be discussed. The set of  equations for the amplitude of $n$-dipole interaction
with the target will be obtained and the interrelations between vertices of the
Pomeron interactions and the microscopic dipole processes will be considered.
It will be shown that the generating functional approach is equivalent to the BFKL Pomeron Calculus in
the kinematic region of \eq{REPC} and leads to a clear interpretation of the BFKL Pomeron Calculus as an
alternative description of the system of interacting colorless dipoles.

The  Fourth section is devoted to the probabilistic interpretation of the BFKL Pomeron Calculus and, in
particular, the process of merging of two Pomerons in one. We calculate the vertex of this merging
directly in the dipole approach and show that it is related to the process of two dipole decay into four
dipoles. We discuss the toy model which simplify the QCD interaction
and allows us to see the main properties of high energy amplitude. In particular,
we are going to discuss the solution of the equations for the asymptotic behaviour
of the scattering amplitude at high energies. We find the solution for the generating functional in the
entire phase space of both variables: rapidity $Y$ and the variable $u$ which is a conjugate variable
to the number of dipoles.

In the fifth section we discuss the solution to the generating functional at high energies. We confirm
the analysis of Ref. \cite{L4} and  find  the asymptotic solution suggested there. However, we  study in more details
the mechanism of  approaching the ultra high energy limit for the scattering amplitude.

In conclusion we are going to compare our approach with other approaches on the
market.

\section{The BFKL Pomeron Calculus}
\label{sec:Pomcal}
\subsection{The general structure of the BFKL Pomeron calculus.}
We start with a general structure of the BFKL Pomeron calculus in QCD. The BFKL Pomeron exchange can
be written in the form (see \fig{bfklcal}-1)

\beq \label{BFKLC1}
A( \mbox{\fig{bfklcal}-1})\,\,=\,\,\,V_u \bigotimes G_P(r_1,r_2;b|Y_1 - Y_2)\bigotimes
V_d\,\,\propto\,\,\frac{\bas^2}{N^2_c}\,\exp\left(\omega(n=0,\nu=0)\,(Y_1 - Y_2)\right)
\eeq
with $\omega(n=0,\nu=0)\,\propto \,\bas$ and $\bigotimes$ denotes the all needed integrations.
$Y_1 - Y_2 = \ln(1/x)$.
To understand the main parameters of the BFKL Pomeron calculus, it is enough to compare the
contribution of the first `fan' diagram of \fig{bfklcal}-2 with the one BFKL Pomeron exchange.
This diagram has the following contribution
\bea \label{BFKLC2}
A( \mbox{\fig{bfklcal}-2})&=&\int^{Y_1}_{Y_2}\,d Y' \, V_u\,\bigotimes\, G_P(r_1,r';b|Y_1
-
Y')\,\bigotimes \,\Gamma(1 \to 2)
 G^2_P(\{r'\},r_2;b|Y' - Y_2)\bigotimes
V^2_d\ \\
&\propto& \frac{V_u\,V^2_d\,\Gamma(1 \to
2)}{\omega(n=0,\nu=0)}\,\exp\left(2\,\omega(n=0,\nu=0)\,
Y\right)\,\,\propto\,\left(\frac{\bas^2}{N^2_c}\right)^2\,\,\exp\left(2\,\omega(n=0,\nu=0)\,Y\right)\nonumber
\eea
where $Y = Y_1 - Y_2$;  $R_1$ and $r_2$ are the sizes of the projectile and target dipoles while
$\{r'\}$ denotes all
dipole variables in Pomeron splitting and/or merging.

One can see that the ratio of this two diagrams is proportional to $\frac{\bas^2}{N^2_c}\,\exp \left(
\omega(n=0,\nu=0)(Y_1 - Y_2)\right)$ which is the parameter given by \eq{PC}. When this ration is
about 1 we need to calculate all diagrams with the Pomeron exchange and their interactions (see
\fig{bfklcal}-a - \fig{bfklcal}-f ). All vertices, that are shown in \fig{bfklcal},   has been
calculated in Refs. \cite{BART,BRN} and they have the following order in $\bas$ \footnote{In
\eq{VRTS} we use the normalizations of these vertices which are  originated from calculation of the
Feynman diagrams. In  the dipole approach we use a different normalization
(see below section 3 and 4) but all conclusions do not depend on the normalization.}:
$$
\omega(n=0,\nu=o)\,\,\,\propto\,\, \,\bas\,\,; \,\,\,\,\,\,\,\,\Gamma( 1 \to 2)\,\,\,\propto\,\,
\,\frac{\bas^2}{N_c}\,\,; \,\,\,\,\,\,\,\,\,\Gamma( 2 \to
1)\,\,\,\propto\,\,\,\frac{\bas^2}{N_c}\,\,;$$
\beq  \label{VRTS}
\Gamma( 2 \to 2)\,\,\,\propto\,\,\,\frac{\bas}{N^2_c}\,\,;\Gamma( 2 \to
3)\,\,\,\propto\,\,\,\frac{\bas^2}{N^2_c}\,\,; V_u\,\,\propto\,\,\,\frac{\bas}{N_c}\,\,;
V_d\,\,\propto\,\,\,\frac{\bas}{N_c}\,;
\eeq
Using \eq{VRTS} we can easily estimate the contributions of all diagrams in \fig{bfklcal}. Namely,
\bea
A(\mbox{\fig{bfklcal}-3})&\,\,\,\propto\,\,\,& V_u\,V_d\,\,
\frac{\Gamma(1 \to 2)\,\Gamma(2 - 1)}{\omega^2(n=0,\nu=0)}\exp\left( 2 \omega(n=0,\nu=0) (Y_1 - Y_2)
\right)\nonumber \\
\,\,\,&\propto&\,\,\,\left(
\frac{\bas^2}{N^2_c}\,\exp \left( \omega(n=0,\nu=0) (Y_1 - Y_2)
\right)\right)^2\,\,=\,\,L^2\left(Y\right)\,\,;\label{diapc3}\\
A(\mbox{\fig{bfklcal}-4})&\,\,\,\propto\,\,\,&  V_u\,V^2_d\,\,\frac{\Gamma(1 \to 2)}{\omega(n=0,\nu=0)}
\,\Gamma(2 \to 2)\,(Y_1 - Y_2)\,\exp\left( 2 \omega(n=0,\nu=0) (Y_1 - Y_2)
\right) \nonumber\\
&\propto&\,\,\,\frac{\bas}{N^2_c}\,\left(Y_1 - Y_2 \right)\,\left(
\frac{\bas^2}{N^2_c}\,\exp \left( \omega(n=0,\nu=0) (Y_1 - Y_2)
\right) \right)^2\,\,=\,\,\frac{\bas}{N^2_c}\,Y\,L^2\left(Y\right)\,\,;\label{diapc4}\\
A(\mbox{\fig{bfklcal}-5})&\,\,\,\propto\,\,\,& V_u\,V^3_d\,\,\frac{\Gamma(1 \to
2)\,\,\Gamma(2 \to 3)}{2\,\omega^2(n=0,\nu=0)}\,\,\exp\left( 3 \omega(n=0,\nu=0) (Y_1 - Y_2)
\right))  \nonumber \\
 &\propto&\,\,\,\frac{1}{N^2_c} \,\left(
\frac{\bas^2}{N^2_c}\,\exp \left( \omega(n=0,\nu=0) (Y_1 - Y_2)
\right)\right)^3\,\,=\,\,\frac{1}{N^2_c}\,\,L^3\left(Y\right);\label{diapc5}\\
A(\mbox{\fig{bfklcal}-6)}&\,\,\,\propto\,\,\,& V_u\,V_d\,\,\frac{\Gamma(1 \to 2)\,\Gamma(2 \to
1)}{\omega^2(n=0,\nu=0)}\,\Gamma(2 \to 2)\,(Y_1 - Y_2)\,\exp\left( 2 \omega(n=0,\nu=0) (Y_1 - Y_2)
\right))\nonumber \\
&\propto&\,\,\,\frac{\bas}{N^2_c}(Y_1 - Y_2)
\,\,\,\left(
\frac{\bas^2}{N^2_c}\,\exp \left( \omega(n=0,\nu=0) (Y_1 - Y_2)
\right)\right)^2\,\,=\,\,\frac{\bas}{N^2_c}\,Y\,\,L^2\left(Y\right);\label{diapc6}
\eea
with $L\left(Y\right)\,\,=\,\,\left(\bas^2/N^2_c\right)\,\,\exp \left( \omega(n=0,\nu=0)\,Y \right)$.

As we have mentioned in the introduction the BFKL calculus sums all diagrams at so high energy that
 parameter $L(Y) $ is of the order of 1 (see \eq{PC}). In this kinematic region we need to take into
account the diagrams of \fig{bfklcal} -1 ,  \fig{bfklcal}-2  and   \fig{bfklcal}-3 (see
\eq{BFKLC1},\eq{BFKLC2} and \eq{diapc3}). Indeed, diagrams of  \fig{bfklcal}-4 and  \fig{bfklcal}-6
( see  \eq{diapc4} and  \eq{diapc6})
are small since $(\bas^2/N^2_c)\,Y \,\,\ll \,\,1/N^2_c\,\,\ll\,\,1$ in the kinematic region of
\eq{REPC}, while the diagrams of  \fig{bfklcal}-5 (see \eq{diapc5}) is small at $L\left(Y\right)
\approx 1$.

The first conclusion that we can derive from this analysis that in the kinematic region where
$L\left(Y\right)\,\,\approx\,\,1$ we need to take into account all diagrams with $\Gamma(1 \to 2) $
and $\Gamma( 2 \to 1)$ vertices while the diagrams with $\Gamma( 2 \to 2)$ and $\Gamma(2 \to 3)$
vertices give small, negligible contributions.

However, if $ L\left(Y\right)\,\,\propto\,\,N_c$ one can see from \eq{BFKLC1} - \eq{diapc6} that all
diagrams give so essential contributions that we have to take them  into account. For what follows
it is interesting to  notice that the vertex $\Gamma(2 \to 2) \,\propto\,\,\bas^4/N^2_c$ can be
neglected even at such large values of $ L\left(Y\right)$.

Finally, we can conclude that the first step of our approach can be summing of  the diagrams
with $\Gamma(1 \to 2)$ and $\Gamma(2 \to 1)$ vertices. Namely, this approach we discuss below in
details.

 \FIGURE[ht]{\begin{tabular}{l}
 \centerline{\epsfig{file=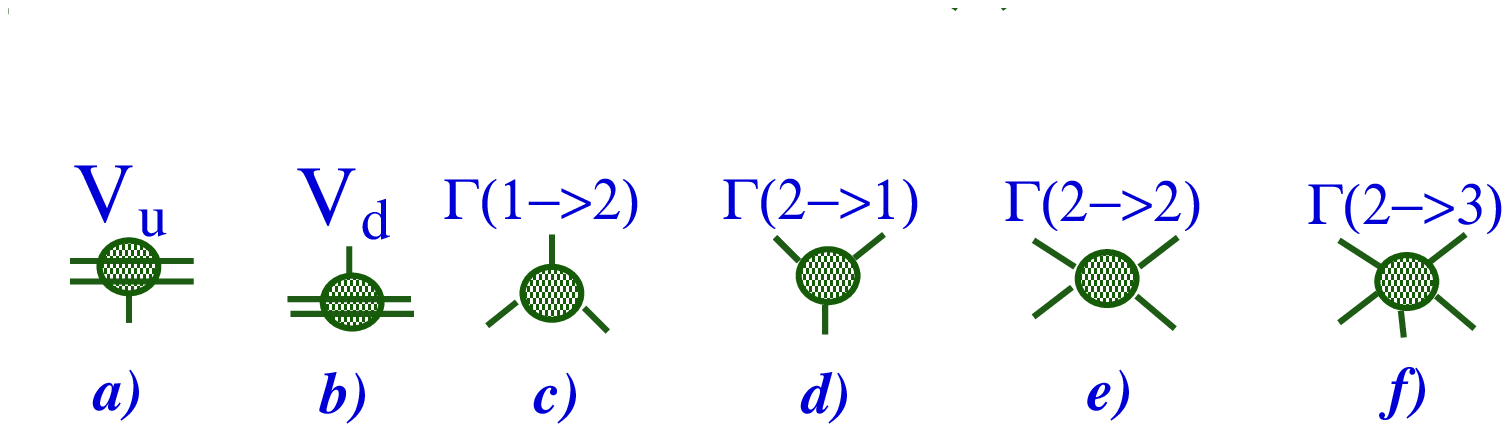,width=160mm}}\\
 \centerline{\epsfig{file=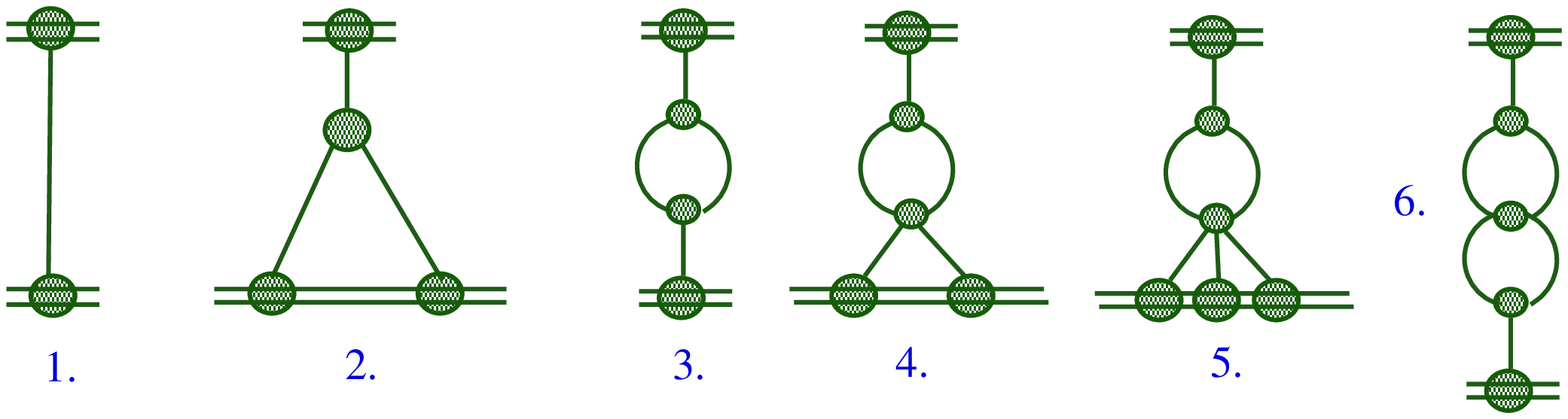,width=160mm}}
 \end{tabular}
 \caption{The BFKL Pomeron interactions and the examples of the
diagrams of the BFKL Pomeron  calculus in QCD. The solid line
describes the Pomeron exchange while the double line stands for the
 dipole.}
 \label{bfklcal}
 }

However, we would like to stress that we need to make an additional assumption inherent for the BFKL
Pomeron calculus: the multi-gluon states in $t$-channel of the scattering amplitude lead to smaller
contribution at high energies than the exchange of the appropriate number of the BFKL Pomerons (see
more in Ref.\cite{L4}). The real argument, supporting this  assumption, stems from the numerous
attempts to find the intercept of these states, larger than the one for the multi-Pomeron exchanges
\cite{KKM} that failed.

\subsection{The path integral formulation of the BFKL calculus}
The main ingredient of the  BFKL Pomeron calculus is the Green function  of the BFKL
Pomeron describing the propagation of a pair of gluons from rapidity $Y'$ and
points
$x'_1$ and $x'_2$ to rapidity $Y$ and points $x_1$ and $x_2$ \footnote{Coordinates
$x_i$ here are two dimensional vectors and, strictly speaking, should be denoted by
$\vec{x}_i$ or $\bf{x}_i$. However, we will use notation $x_i$ hoping that it will
not cause difficulties in understanding.} Since the Pomeron does not carry colour
in $t$-channel we can treat initial and final coordinates as coordinates of quark
and antiquark in a colourless dipole. This Green function is well known\cite{LI}
and has a form:
\begin{equation} \label{BFKLGF}
G(x_1,x_2;Y | x'_1,x'_2;Y')\,\,=\,\,\Theta(Y - Y')\times
\end{equation}
$$
\times\,\sum^{\infty}_{n=-\infty} \,\int\,d \nu\,\,d^2 x_0\,e^{\omega(n,\nu) (Y - Y')}\,\lambda(n,\nu)\,
E(x_1,x_2;x_0|\nu)\,E^*(x'_1,x'_2;x_0|\nu)
$$
where vertices $E$ are given by
\begin{equation} \label{BFKLE}
E(x_1,x_2;x_0|\nu)\,=\,
\left(\frac{x_{12}}{x_{10}\,x_{20}}
\right)^h\,\left(\frac{x^*_{12}}{x^*_{10}\,x^*_{20}} \right)^{\tilde{h}}
\end{equation}
where $x_{ik} = x_i - x_k $, $x_i = x_{i,x} + i x_{i,y}$ \footnote{$x_{i,x}$ and
$x_{i,y}$  are components of the two dimensional vector $x_i$ on $x$-axis and $y$-
axis} ,$ x^*_i = x_{i,x} + i x_{i,y} $ ; $h = (1 - n)/2 + i\nu$ and $\tilde{h} = 1
- h^*$. The energy levels $\omega(n,\nu)$  are the BFKL eigen values
\begin{equation} \label{BFKLOM}
\omega(n,\nu)\,=\,\bar{\alpha}_S \left( \psi(1) - Re{\, \psi\left(\frac{|n| + 1}{2} + i
\nu\right)}
\right)
\end{equation}
where $\psi(z) = d \ln \Gamma(z)/d z$ and $\Gamma(z)$ is the Euler gamma function.
 Finally
\begin{equation} \label{BFKLLA}
\lambda(n,\nu)\,=\frac{1}{[ ( n + 1)^2 + 4 \nu^2] [(n - 1)^2 + 4 \nu^2]}
\end{equation}

 \FIGURE[ht]{
\centerline{\epsfig{file=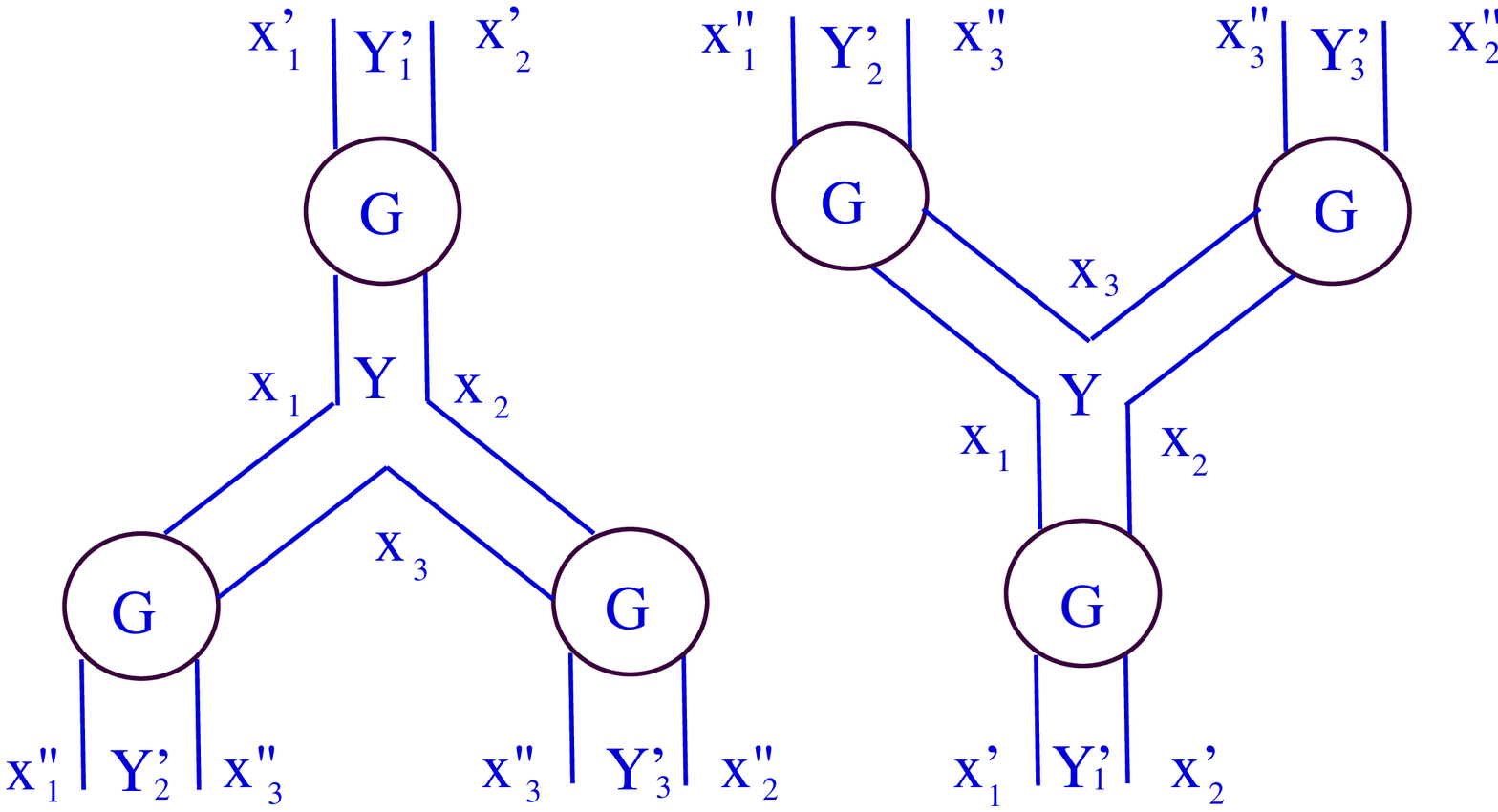,width=120mm}} \caption{The
graphic form of the triple Pomeron vertex in the coordinate
representation.} \label{triple-p} }

The interaction between Pomerons is depicted in Fig. \ref{triple-p}
and  described by the triple Pomeron vertex which can be written in
the coordinate representation \cite{BRN} for the following process:
two gluons with coordinates $x'_1$ and $x'_2$ at rapidity $Y'_1$
decay into two gluons pairs with coordinates $x^{\prime \prime}_1$
and $x^{\prime \prime}_3$ at rapidity $Y'_2$ and $x^{\prime
\prime}_2$ and $x^{\prime \prime}_3$ at rapidity $Y'_3$ due to the
Pomeron splitting at rapidity $Y$. It looks as
\begin{equation} \label{BFKL3P}
2\frac{\pi\,\bar{\alpha}^2_S}{N_c}\,\int\,\frac{d^2 x_1\,d^2\,x_2
\,d^2
x_3}{x^2_{12}\,x^2_{23}\,x^2_{13}}\,\left(G(x'_1,x'_2;Y'_1|x_1,x_2;
Y) \,\,L^{\!\!\!\!\!\!\!\leftarrow}_{1,2} \right) \cdot
\end{equation}
$$
G(x_1,x_3;Y|x^{\prime\prime}_1,
x^{\prime\prime}_3;Y'_2)\,G(x_3,x_2;Y|x^{\prime\prime}_3,x^{\prime\prime}_2;Y'_3)
$$
where
\begin{equation} \label{BFKLL}
L^{\!\!\!\!\!\!\!\leftarrow}_{1,2}\,\,=\,\,r^4_{12}\,p^2_{1}\,p^2_{2}\,\,
\,\mbox{with}\,\,p^2\,=\,-\,\nabla^2
\end{equation}
and arrow shows the direction of action of the operator $L$.
For the inverse process of merging of two Pomerons into one we have
$$
2\frac{\pi\,\bar{\alpha}^2_S}{N_c}\,\int\,\frac{d^2 x_1\,d^2\,x_2
\,d^2 x_3}{x^2_{12}\,x^2_{23}\,x^2_{13}}\,
G(x^{\prime\prime}_1,x^{\prime\prime}_3;Y'_1|x_1,x_3;Y)\,G(x^{\prime\prime}_3,x^{\prime\prime}_2;Y'_2|x_3,x_2;Y)
\,\cdot
$$
\begin{equation} \label{BFKLP3}
\left( L^{\!\!\!\!\!\rightarrow}_{1,2}\, G(x_1,x_2;
Y|x'_1,x'_2;Y'_1) \right)
\end{equation}

The theory with the interaction given by Eq.~(\ref{BFKL3P}) and  Eq.~(\ref{BFKLP3})
can be written through the functional integral \cite{BRN}
\begin{equation} \label{BFKLFI}
Z[\Phi, \Phi^+]\,\,=\,\,\int \,\,D \Phi\,D\Phi^+\,e^S \,\,\,\mbox{with}\,S \,=\,S_0
\,+\,S_I\,+\,S_E
\end{equation}
where $S_0$ describes free Pomerons, $S_I$ corresponds to their mutual interaction
while $S_E$ relates to the interaction with the external sources (target and
projectile). From Eq.~(\ref{BFKL3P}) and  Eq.~(\ref{BFKLP3}) it is clear that
\begin{equation} \label{S0}
S_0\,=\,\int\,d Y \,d Y'\,d^2 x_1\, d^2 x_2\,d^2 x'_1\, d^2 x'_2\,
\Phi^+(x_1,x_2;Y)\,
G^{-1}(x_1,x_2;Y|x'_1,x'_2;Y')\,\Phi(x'_1,x'_2;Y')
\end{equation}
\beq \label{SI} S_I\,=\,\frac{2\,\pi \bas^2}{N_c}\,\int \,d Y\,\int
\,\frac{d^2 x_1\,d^2 x_2\,d^2 x_3}{x^2_{12}\,x^2_{23}\,x^2_{13}}\,
\cdot \{ \left( L^{\!\!
\!\!\!\rightarrow}_{1,2}\Phi(x_1,x_2;Y)\,\right)\,\cdot\,
\Phi^+(x_1,x_3;Y)\,\Phi^+(x_3,x_2;Y)\,\,+\,\,h.c. \} \eeq For $S_E$
we have local interaction both in rapidity and in coordinates,
namely,
 \begin{equation} \label{SE}
S_E\,=\,-\,\int \,dY\,d^2 x_1\,d^2 x_2\, \{
\Phi(x_1,x_2;Y)\,\tau_{pr}(x_1,x_2;Y)\,\,+\,\,\Phi^+(x_1,x_2;Y)\,\tau_{tar}(x_1,x_2;Y)
\}
\end{equation}
where $\tau_{pr}$ ($\tau_{tar}$)  stands for the projectile and target, respectively.
The form of functions $\tau$  depend on the non-perturbative input in our problem
and for the case of nucleus target they are written in Ref. \cite{BRN}.

For the case of the projectile being a dipole that scatters off a  nucleus
the scattering amplitude has the form
\begin{equation} \label{T}
T(x_1,y_1;Y)\,\,\equiv
T^{(1)}(x_1,y_1;Y)\,\,=\,\,-\,\frac{4\,\pi^2\,\bar{\alpha}_S}{N_c}\,
\frac{\int\,\,D \Phi\,
D\Phi^+\,\Phi(x_1,y_1;Y)\,e^{S[\Phi,\Phi^+]}}{ \int\,\,D \Phi\, D
\Phi^+\,\,e^{S[\Phi,\Phi^+]}|_{S_E =0}}
\end{equation}
where extra $\alpha_S$ comes from our normalization and where we neglect term with $\tau_{pr}$
in \eq{SE}.

Generally, for the amplitude of interaction of $n$ dipoles at
rapidity $Y$ we can write the following expression
\footnote{Starting from this equation we use notations $x_i$ for the
coordinates of quark while $y_i$ denote the  coordinates of
antiquarks. For  rapidity we will use  $Y$.} \beq \label{TN}
T^{(n)}(x_1,y_1,\dots x_n,y_n;Y)\,\,=\,\,(-1)^n\Lb\,
\frac{4\,\pi^2\,\bar{\alpha}_S}{N_c}\,\Rb^n\, \frac{\int\,\,D \Phi\,
D\Phi^+\,\prod^n_{i=1}\Phi(x_i,y_i;Y)\,\,e^{S[\Phi,\Phi^+]}}{
\int\,\,D \Phi\, D \Phi^+\,\,e^{S[\Phi,\Phi^+]}|_{S_E =0}} \eeq

The extra factor $(-1)^n$ stems from the fact that in $S_E$ the source for a projectile as
well as for a target, has an extra minus sign.

It is useful to introduce the Green function of the BFKL Pomeron that includes the Pomeron
loops. This function has the form
\beq \label{GF}
G\Lb x_1,y_1;Y|x_2, y_2;Y' \Rb\,\,=\,\,\frac{ \int\,\,D \Phi\,
D\Phi^+\,\Phi^+(x_1,y_1;Y)\,\Phi(x_2,y_2;Y')\,\,e^{S[\Phi,\Phi^+]}}{ \int\,\,D
\Phi\, D \Phi^+\,\,e^{S[\Phi,\Phi^+]}|_{S_E =0}}
\eeq

For further presentation we need some properties of the BFKL Green function
\cite{LI}:

1.  Generally,

\begin{equation}
G^{-1}(x_1,x_2;Y| x'_1,x'_2;Y')\,\, \,=\,p^2_1\,p^2_2\,\left(
\frac{\partial}{\partial Y} + H \right) \,\,=\,\,\left(
\frac{\partial}{\partial Y} + H^+ \right)\,p^2_1\,p^2_2; \label{G1}
\end{equation}

\begin{equation} \label{H}
H f(x_1,x_2;Y) \,\,= \,\,\frac{\bar{\alpha}_S}{2
\pi}\,\int\,\frac{d^2 x_3\,x^2_{12}}{x^2_{23}\,x^2_{13}}\,\left(
f(x_1,x_2;Y)\,-\,f(x_1,x_3;Y)\,-\,f(x_3,x_2;Y) \right) ; \label{G2}
\end{equation}

2. The initial Green function ($G_0$) is
 equal to
\begin{equation} \label{G0}
G_0(x_1,x_2;Y| x'_1,x'_2;Y)\,\,=\,\, \pi^2\,\ln
\frac{x^2_{1,1'}\,x^2_{2,2'}}{x^2_{1,2'}\,x^2_{1',2}}  \,\ln
\frac{x^2_{1,1'}\,x^2_{2,2'}}{x^2_{1,2}\,x^2_{1',2'}}
\end{equation}
 This form of $G_0$ has been discussed in Ref.\cite{LI}. In  appendix A we demonstrate that
this expression for $G_0(x_1,x_2;Y| x'_1,x'_2;Y)$ stems from $\omega
= \omega(n=0,\nu)$ term in sum of \eq{BFKLGF}. Only this term is
essential at high energies since all other terms lead to decreasing
with energy contributions.

3. It should be stressed that
\bea \label{G01}
\nabla^2_1 \,\nabla^2_2 \,G_0(x_1,x_2;Y| x'_1,x'_2;Y)\,\,&=& \\
& & \,\,(2\,\pi)^4\,\,\left(\delta^{(2)}( x_1 - x'_1)\,\delta^{(2)}( x_2
- x'_2)\,+\,\delta^{(2)}( x_1 - x'_2)\,\delta^{(2)}( x_2 -
x'_1)\,\right)\nonumber
\eea

4. In sum of Eq.~(\ref{BFKLGF}) only the term with $n=0$ is
essential for high energy asymptotic behaviour since all
$\omega(n,\nu) $ with $n   \geq 1$  are negative and, therefore,
lead to contributions that decrease with energy. Taking into account
only the first term one can see that $G$ is the eigen function of
operator $L_{13}$, namely \beq \label{LG} L_{12}\,G(x_1,x_2;Y|
x'_1,x'_2;Y')\,\,=\,\,\frac{1}{\lambda(0,\nu)} \,G(x_1,x_2;Y|
x'_1,x'_2;Y')\,\,\approx\,\,G(x_1,x_2;Y| x'_1,x'_2;Y') ; \label{L13}
\eeq The last equation holds only approximately in the region where
$\nu \,\ll\,1$, but this is the most interesting region which is
responsible for high energy asymptotic behaviour of the scattering
amplitude.

\subsection{The chain of equations for the multi-dipole amplitudes.}

Using  Eq.~(\ref{BFKLFI}) and Eq.~(\ref{T})
 we can easily obtain the chain equation for multi-dipole amplitude $T^{(n)}$
noticing that every dipole interacts only with one Pomeron (see Eq.~(\ref{T})).

These equations follow from the fact that a change of variables does not alter the value of
functional integral of \eq{BFKLFI}. In particular, $Z[\Phi,\Phi^+] \,=\,Z[\Phi,\Phi'^+]$(see
\eq{BFKLFI} ) where $\Phi'^+ \,=\,\Phi^+ \,+\,\epsilon(x,y)$ with small function $\epsilon(x,y)$.
Therefore,
\beq \label{VZ1}
\int \,D\,\Phi\,D\,\Phi^+ \,e^{S[\Phi\,,\Phi^+]}\,\,=\,\,\int \,D\,\Phi\,D\,\Phi'^+
\,e^{S[\Phi\,,\Phi'^+]}
\eeq
Substituting $\Phi'^+ \,=\,\Phi^+ \,+\,\epsilon(x,y)$ and expanding this equation to first order
in $\epsilon$, we find

\beq \label{VZ2}
0\,=\,\int\,D\,\Phi\,D\,\Phi^+\,e^{S[\Phi\,,\Phi^+]}\,\times \eeq
$$
[\,\int\,d Y \,d Y'\,d^2 x_1\, d^2 x_2\,d^2 x'_1\, d^2 x'_2\,
\epsilon(x_1,x_2;Y)\,
G^{-1}(x_1,x_2;Y|x'_1,x'_2;Y')\,\Phi(x'_1,x'_2;Y')
$$
$$
+\frac{2\,\pi \bas^2}{N_c}\,\int \,d Y\,\int \,\frac{d^2 x_1\,d^2
x_2\,d^2 x_3}{x^2_{12}\,x^2_{23}\,x^2_{13}}\, \cdot \{
 \left( \epsilon(x_1,x_2;Y)L^{\!\! \!\!\!\leftarrow}_{1,2}\,\right)\,\cdot\,
\Phi(x_1,x_3;Y)\,\Phi(x_3,x_2;Y)   +$$
$$
  2\left( L^{\!\!
\!\!\!\rightarrow}_{1,2}\Phi(x_1,x_2;Y)\,\right)\,\cdot\,
\epsilon(x_1,x_3;Y) \,\Phi^+(x_3,x_2;Y)
  \} -\,\int \,dY\,d^2 x_1\,d^2 x_2\,
  \epsilon(x_1,x_2;Y)\,\tau_{tar}(x_1,x_2;Y)]
$$

We redefine the integration variables in the third term as follows
$$
2\left( L^{\!\!
\!\!\!\rightarrow}_{1,2}\Phi(x_1,x_2;Y)\,\right)\,\cdot\,
\epsilon(x_1,x_3;Y) \,\Phi^+(x_3,x_2;Y) \longrightarrow 2\left(
L^{\!\! \!\!\!\rightarrow}_{1,3}\Phi(x_1,x_3;Y)\,\right)\,\cdot\,
\epsilon(x_1,x_2;Y) \,\Phi^+(x_2,x_3;Y)
$$

Using the expression for the Hamiltonian Eq. \ref{G1} and the
Casimir operator Eq. \ref{BFKLL} we define a new variation parameter
$\epsilon(x_1,x_2;Y)p^2_{1}\,p^2_{2}$. In terms of this variation
parameter Eq. \ref{VZ2} reads as
 \beq \label{VZ2-a1}
0\,=\,\int\,D\,\Phi\,D\,\Phi^+\,e^{S[\Phi\,,\Phi^+]}\,\times \eeq
$$
[\,\int\,d Y  \,d^2 x_1\, d^2 x_2 \;\; \epsilon(x_1,x_2;Y)p^2_1p^2_2
\Lb\frac{\partial}{\partial Y}+ H \Rb\Phi(x_1,x_2;Y)
$$
$$
+\frac{2\,\pi \bas^2}{N_c}\,\int \,d Y\,\int \,\frac{d^2 x_1\,d^2
x_2\,d^2 x_3}{x^2_{12}\,x^2_{23}\,x^2_{13}}\, \cdot \{
 \left( \epsilon(x_1,x_2;Y)p^2_1 p^2_2\right)\,\cdot\,
\Phi(x_1,x_3;Y)\,\Phi(x_3,x_2;Y)   +$$
$$
  2\left( L^{\!\!
\!\!\!\rightarrow}_{1,3}\Phi(x_1,x_3;Y)\,\right)\,\cdot\,
\epsilon(x_1,x_2;Y) \, \frac{p^2_1p^2_2}{p^2_1p^2_2}
\,\Phi^+(x_3,x_2;Y)
  \} -\,\int \,dY\,d^2 x_1\,d^2 x_2\,
  \epsilon(x_1,x_2;Y)\frac{p^2_1p^2_2}{p^2_1p^2_2}\,\tau_{tar}(x_1,x_2;Y)]
$$

We denote the new variation parameter by
$\tilde{\epsilon}(x_1,x_2;Y) =p^2_1p^2_2\epsilon(x_1,x_2;Y)$ and use
the property of the initial Green function Eq. \ref{G01} to rewrite
$\frac{1}{p^2_1p^2_2}$ in terms of $G_0$ as follows
 \beq \label{I0}
 \frac{1}{p^2_1p^2_2}\tilde{\epsilon}(x_1,x_2;Y)\,\,=\,\,
 \frac{1}{(2\pi)^4}\int G_0(x_1,x_2;Y| x'_1,x'_2;Y)\tilde{\epsilon}(x'_1,x'_2;Y)
 d^2x'_1 d^2x'_2
\eeq
Thus, Eq. \ref{VZ2-a1} can be written as
 \beq \label{VZ2-a2}
0\,=\,\int\,D\,\Phi\,D\,\Phi^+\,e^{S[\Phi\,,\Phi^+]}\,\times \eeq
$$
[\,\int\,d Y  \,d^2 x_1\, d^2 x_2 \;\; \tilde{\epsilon}(x_1,x_2;Y)
\Lb\frac{\partial}{\partial Y}+ H \Rb\Phi(x_1,x_2;Y)
$$
$$
+\frac{2\,\pi \bas^2}{N_c}\,\int \,d Y\,\int \,\frac{d^2 x_1\,d^2
x_2\,d^2 x_3}{x^2_{12}\,x^2_{23}\,x^2_{13}}\, \cdot \{
  \tilde{\epsilon}(x_1,x_2;Y) \, \,
\Phi(x_1,x_3;Y)\,\Phi(x_3,x_2;Y)   +$$
$$
  2\left( L^{\!\!
\!\!\!\rightarrow}_{1,3}\Phi(x_1,x_3;Y)\,\right)\,\cdot\,
\left\{\frac{1}{(2\pi)^4}\int G_0(x_1,x_2;Y|
x'_1,x'_2;Y)\tilde{\epsilon}(x'_1,x'_2;Y)
 d^2x'_1 d^2x'_2 \right\}
\,\Phi^+(x_3,x_2;Y)
  \}
$$
$$
   -\,\int \,dY\,d^2 x_1\,d^2 x_2\,
  \left\{\frac{1}{(2\pi)^4}\int G_0(x_1,x_2;Y|
x'_1,x'_2;Y)\tilde{\epsilon}(x'_1,x'_2;Y)
 d^2x'_1 d^2x'_2 \right\}\,\tau_{tar}(x_1,x_2;Y)]
$$

 Noting that the r.h.s. of \eq{VZ2-a2} should vanish for any
possible variation $\tilde{\epsilon}(x_1,x_2;Y)$
 we obtain

 \beq \label{VZ2-a3}
0\,=\,\int\,D\,\Phi\,D\,\Phi^+\,e^{S[\Phi\,,\Phi^+]}\,\times [
\Lb\frac{\partial}{\partial Y}+ H \Rb\Phi(x_1,x_2;Y)  \eeq
 $$
+\frac{2\,\pi \bas^2}{N_c}\, \,\int \,\frac{x^2_{12}\,d^2
x_3}{x^2_{23}\,x^2_{13}}
 \Phi(x_1,x_3;Y)\,\Phi(x_3,x_2;Y)   +$$
$$
+\frac{2\,\pi \bas^2}{N_c}\,\frac{ 2}{(2\pi)^4} \,\int \,
\frac{\,d^2x'_1\,d^2x'_2\,d^2x_3}{x^2_{1'2'}x^2_{2'3}\,x^2_{1'3}}
 \left( x^4_{1' 3}\;p^2_{1'}p^2_3\Phi(x_1',x_3;Y)\,\right)
   G_0(x'_1,x'_2;Y|
x_1,x_2;Y) \,\Phi^+(x_3,x'_2;Y)
  \}
$$
$$
   -\frac{1}{(2\pi)^4} \,\int d^2x'_1 d^2x'_2
    G_0(x'_1,x'_2;Y|
x_1,x_2;Y)
  \,\tau_{tar}(x'_1,x'_2;Y)]
$$
where in the third and last terms we interchanged
$(x'_1,x'_2)\leftrightarrow (x_1,x_2)$. We notice that the third and
last terms are independent of rapidity and  can be absorbed  in  the initial
condition. It is obvious for the last term which represents the
target source. To show this for the third term we use the property
of the Casimir operator at high energies ($n=0\;,\;\nu=0$)
$$
 L^{\!\!
\!\!\!\rightarrow}_{1,3}\Phi(x_1,x_3;Y)\simeq \Phi(x_1,x_3;Y)
$$
and the definition of the Green function(see Eq. \ref{GF}). We see
that the third term results into the product of two initial Green
functions which are independent of rapidity.

Now we can use the definition of the amplitude defined in Eq.
\ref{T} and Eq. \ref{TN} to rewrite Eq. \ref{VZ2-a3} in a simple
form
\begin{equation} \label{T1-chain}
\frac{\partial T^{(1)}(x_1,x_2;Y)}{\partial Y}\,\,=\,\,
\frac{\bar{\alpha}_S}{2 \,\pi}\,\int\,d^2\,z\,K(x_1,x_2;z)\,
\end{equation}
$$
\left( T^{(1)}(x_1,z;Y)\,+\,T^{(1)}(z,x_2;Y)\, -\,
T^{(1)}(x_1,x_2;Y)\,-\,T^{(2)}(Y;x_1,z;z,x_2;Y)\right)
$$
where kernel $K(x,y|z)$ is defined as
\begin{equation} \label{K}
K(x,y|z)\,\,=\,\,\frac{(x - y)^2}{(x - z)^2\,(z - y)^2}
\end{equation}
and the Hamiltonian is given by Eq. \ref{G2}.

This equation has a very simple meaning that is clear from
\fig{pomeq}.
 \FIGURE[ht]{
\centerline{\epsfig{file=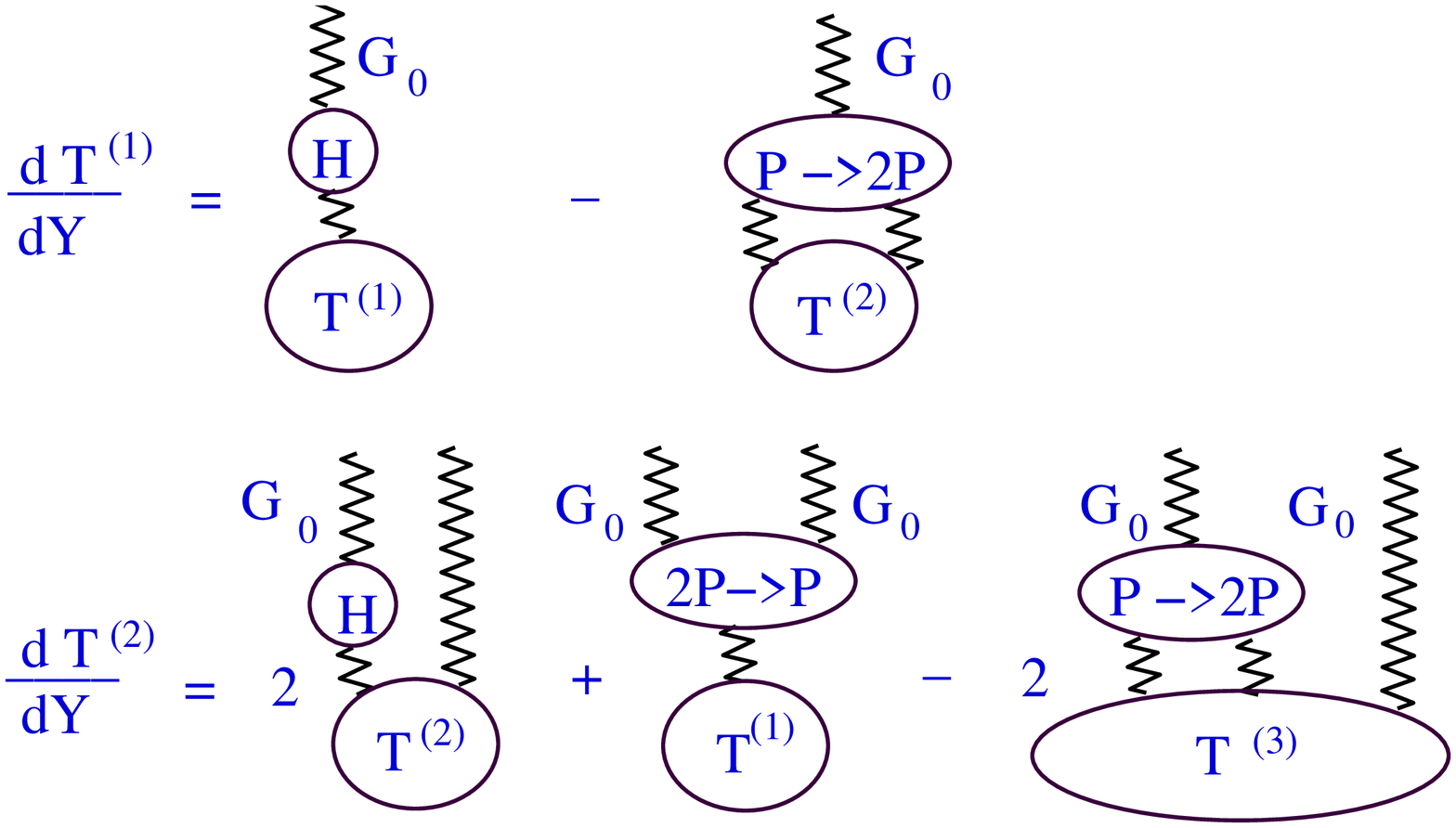,width=120mm}} \caption{The
graphic form of equations for the multi dipole amplitude.}
\label{pomeq} }

Starting from equation \beq \label{VZ4} \int \,D\,\Phi\,D\,\Phi^+
\,\Phi(Y; x_4,x_5)\,\, e^{S[\Phi\,,\Phi^+]}\,\,=\,\,\int
\,D\,\Phi\,D\,\Phi'^+\,\Phi(Y; x_4,x_5)\, \,e^{S[\Phi\,,\Phi'^+]}
\eeq we obtain the equation for the amplitude $T^{(2)}$, namely,
\begin{equation}  \label{T2-chain}
\frac{\partial T^{(2)}(x_1,x_2;x_3,x_4;Y)}{\partial Y} \,\,=\,\,
\frac{\bar{\alpha}_S}{2 \,\pi}\int\,d^2\,z\,K(x_1,x_2|z)\,
\end{equation}
$$
\left(  T^{(2)}(x_1,z;x_3,x_4;Y) \, +\,T^{(2)}(z,x_2;x_3,x_4;Y)
\,-\, \,T^{(2)}(x_1,x_2;x_3,x_4;Y)\, - \right.
$$
$$
\left. -\,T^{(3)}(x_1,z;z,x_2;x_3,x_4;Y) \right)\,+
$$
$$
+\,   2{\alpha}^2_S  \frac{\bar{\alpha}_S}{2 \,\pi} \int
\,d^2\,x'\,d^2\,x^{\prime \prime}\,\Gamma_{2 \to 1}(x_1,x_2;x_3,x_4|
x',x^{\prime \prime})\,\nabla^2_{x'}\nabla^2_{x^{\prime \prime}}
T^{(1)}(x',x^{\prime \prime};Y);
$$
 where function $\Gamma_{2\to 1}$ is equal to
\begin{equation} \label{V21}
\Gamma_{2 \to 1}(x_1,y_1;x_2,y_2|x,y)\,\,= \int\,d^2 z
K(x,y;z)\,\,G_0(x_1,y_1|x,z)\,G_0(x_1,y_1|z,y)
\end{equation}
Deriving Eq.~(\ref{T1-chain}) and Eq.~(\ref{T2-chain}) we use
Eq.~(\ref{G0}) and Eq.~(\ref{G01}) as well as normalization
condition (see Eq.~(\ref{T})) for the scattering amplitude. These
two equations are the same as in Ref. \cite{IST}. This
 shows that in the  papers \cite{IT,IST}, actually,  the same approach as the BFKL
Pomeron Calculus has been developed (much later, of course), in spite of the fact
that authors think that they are doing something more general.

Assuming that $T^{(2)}= T^{(1)}\,T^{(1)}$ we obtain the Balitsky-Kovchegov equation
\cite{B,K}. We can do this only if we can argue why the Pomeron splitting is more
important than the Pomeron merging. For example,  this assumption is reasonable for
scattering of the dipole with the nucleus target.   Generally speaking, the splitting
and merging have the same order in $\alpha_S$ ( see Eq.~(\ref{BFKL3P}) and
 Eq.~(\ref{BFKLP3}). In  Eq.~(\ref{T1-chain}) and  Eq.~(\ref{T2-chain}) these two processes  look
like having a  different order of magnitude in $\alpha_S$ but this fact does not
interrelate with any physics and reflects  only our normalization. However, we will
see that for a  probabilistic interpretation the correct normalization is very
important.

\section{Generating functional and probabilistic interpretation}
\label{sec:GenFPI}
\subsection{Statistical physics analogy: Langevin equation and directed percolation.}

The functional of \eq{BFKLFI} is reminiscent of the partition function of statistical mechanics.
Indeed, the partition function has a general form
\beq \label{SM1}
Z[H]\,\,=\,\,e^{ - \frac{1}{kT}\,F[H]}\,\,=\,\,\int d {\bf s}\,e^{ - \frac{1}{kT}\,\int\,d x \Lb {\cal
H}(s)\,- H \,s(x)\Rb}
\eeq
where $F[H]$ is the Helmholtz free energy.  As an example, \eq{SM1} is written for the system of spins
with the energy density ${\cal H}(s)$
in the external magnetic field $H$. The integration is over all possible configurations of spins in
the system.

Comparing \eq{SM1} and \eq{BFKLFI} one can see that \eq{BFKLFI} describes a statistical system with
$kT=1$ and  with $F[-\tau_{tr}] = - S$. The form of $S_E$ suggests that $H = - \tau_{tr}$ plays a role
of the external field.

\eq{BFKLFI} on the lattice looks as follows
\beq \label{SM2}
Z[\Phi,\Phi^+]\,\,=\,\,\int\,\prod_i
\,D\,\Phi_i(x_i,y_i;Y_i)\,D\,\Phi^+_i(x_i,y_i;Y_i)\,e^{S[\Phi_i,\Phi^+_i]}
\eeq

We can linearize the action using the Stratonovich transformation \cite{LELE}  but first
we simplify the expression for action $S$ introducing a new field  $\Phi^{+'}(x_1,x_2;Y)
\,=\,p^2_1\,p^2_2
\Phi^+(x_1,x_2;Y) $ and using \eq{G1} ,  \eq{L13} and \eq{I0}. In new notation the action has the form
\beq \label{S01}
S_0\,=\,\int\, dY\,dY'\,d^2 x d^2 y\,d^2 x'\,d^2 y'\,\Phi^{+'}(x,y;Y)\,\Lb \frac{\D}{\D
Y'}\,+\,H \,\Rb\,\Phi(x',y';Y')
\eeq
\beq \label{SI1}
S_I\,=\,\,\frac{2\,\pi \bas^2}{N_c}\,\,\Lb \int \,dY\,d^2 x \,d^2 y d^2 z\, K(x,y|z)
\,\Phi^{+'}(x,y;Y)\,\Phi(x,z;Y) \,\Phi(z,y;Y)\,\,+\,\,\right.
\eeq
$$
\left. \frac{1}{(2 \pi)^8}\,\,
\prod^3_{i=1}\,d^2\,x_i\,d^2\,y_i \,\frac{ \Gamma_{2
\to 1}(x_1,y_1;,x_2,y_2|x_3,y_3)}{(x_3 - y_3)^4}
\,\Phi^{+'}(x_1,y_1;Y)\,\Phi^{+'}(x_2,y_2;Y)\,\Phi(x_3,y_3;Y) \Rb
$$
while $S_E$ has the form
\beq \label{S0P}
S_E\,\,=\,\,-\,\,\frac{1}{(2 \pi)^4}\,\,\int\,dY\,\,d^2\,x\,d^2\,y\,d^2\,x'\, d^2\,y'\,\,
\Phi^{+'}(x,y;Y)\,G_0(x,y|x',y')\,\tau(x',y';Y)
\eeq

We demonstrate the idea of the statistical interpretation of our theory using the simplified version,
so called toy model, in which we assume that there is no dependence of the interaction on the
coordinate of dipoles (see Ref. \cite{MUCD} for details). The action of this model is
\beq \label{TMFI}
S\,\,=
\,\,\int\,dY\,\Lb \Phi^+(Y)\,\Lb \frac{d}{d Y}\,-\,\Delta \Rb \Phi(Y)\,+ \right.
\eeq
$$
\left.
\,G_{3P} \Lb
\Phi^+(Y)\,\Phi^2(Y) \,+\,\Phi(Y)\,(\Phi^+)^2(Y)\Rb - \Phi^+(Y)\,\tau_{tr}(Y) \delta(Y -Y_0) \Rb
$$
where $G_{3P}$ is the toy-model vertex that describes the one Pomeron to two Pomerons transition.

 Using the  Stratonovich transformation we
can linearize the action given by \eq{TMFI}, namely,
\beq \label{SM3}
\exp\Lb  G_{3P}\,\Delta Y \,\Phi_i\,(\Phi^+_i)^2\,\Rb\,\sim\,
\,\int\,\,d \eta_i\,\exp\Lb \,- \h \eta^2_i\,-\,\Phi^+_i\,\sqrt{2\, G_{3P}\,\Phi_i\, \Delta 
Y}\,\eta_i
\Rb
\eeq
Substituting \eq{SM3} into \eq{SM2}  and integrating explicitly over $\Phi^+_i$ we obtain
\beq \label{SM4}
Z\,\,\propto\,\int\,\prod_i
\,D\,\Phi_i \,D\,\eta_i\,e^{- \h \sum_i \int\,d Y \,\eta^2_i }\,\delta \Lb
d\,\Phi_i\-\,\,\Delta \Phi_i \,
+
\,\, G_{3P} \Phi^2_i -\,\,\sqrt{2\,G_{3P}\,\Phi_i\,d Y}\,\eta_i \Rb
\eeq
The $\delta$-function means that the only $\Phi$ which satisfy to the following (Langevin)
equation
\beq \label{LEQ}
d\,\Phi(Y)\,\,= \,\,\Delta\,\,\Phi(Y)\,\,-
\,\,G_{3P}\, \Phi^2 \,d\,Y\,\,+
\,\,\sqrt{2\,G_{3P}\,\Phi}\,d W(Y)
\eeq
(where $W(Y)$ is a Wiener process \cite{GARD}) contribute to the functional integral.
\eq{LEQ} is actually the  Langevin equation for directed percolation (see Ref. \cite{HH} for
details) which can be re-written in the form
\beq \label{EQRT}
\frac{\partial \,\,\Phi(Y)}{\partial Y}\,\,\,=\,\,\Delta\,\Phi(Y)\,\,\,-\,\,\,G_{3P}\,\,\Phi^2(Y)
\,\,+\,\,\zeta(Y)
\eeq
where $\zeta$ is a density ($\Phi$) dependent Gaussian noise field \cite{HH,GARD,WE,BIW,IT,MSW},
which is defined as
\beq \label{NOISE}
<|\zeta(Y) |> \,\,=\,\,0\,;\,\,\,\,\,\,\,\,<|\zeta(Y)\,\,\zeta(Y') 
|>\,\,=\,\,2\,G_{3P}\,\Phi(Y)\,\delta(Y - Y')\,\,.
\eeq

The  general functional of \eq{SM2}, we believe,  can be reduced to \eq{EQRT}. However, we found a way how to
do this only for \eq{SM2} only with addinional sipsimplificationse assume  (i) that \eq{L13} has the
form
\beq \label{SI10}
L_{12}\,\,G\left(x_1,x_2;Y| x'_1,x'_2;Y'\right)\,\,=\,\,G\left(x_1,x_2;Y| x'_1,x'_2;Y'\right)
\eeq
or, in other words, we consider $\Lambda(n,\nu)\,\,=\,\,\Lambda(n=0,\nu)\,\,=\,\,\Lambda(n=0,\nu=0)$ (see 
\eq{BFKLLA})  , which
gives the leading contribution for the BFKL calculus; and (ii)  that
\beq \label{SI2}  
\frac{x_1 \,+\,x_2}{2}\,\,\gg\,\,x_1 \,-\,x_2\,\,;
\eeq
Indeed, we expect that the typical size of the dipoles will be of the order of $1/Q_s(x)$ ($Q_s(x) $ is the 
satrusaturatione) while the typical impact parameter of the scattering dipole should be much larger (at least
of the order of the size of the hadron).

Using these two assumptions we can rewrite the functional integral in the form of  \eq{BFKLFI}
 but with action $S_I$ that can be written as follows
\beq \label{SIS}
S_I\,\,=\,\,\frac{2\,\pi\,\bas^2}{N_c}\,\int\,d\,Y\,\,
\int
\,\frac{d^2 x_1\,d^2 x_2\,d^2 x_3}{x^2_{12}\,x^2_{23}\,x^2_{13}}\,
\cdot\,\,  \{ \, \Phi(x_1,x_2;Y)\,\cdot\,
\Phi^+(x_1,x_3;Y)\,\Phi^+(x_3,x_2;Y)\,\,+\,\,h.c. \}
\eeq
and \eq{G1} reads as
\beq \label{G1S}
G^{-1}(x_1,x_2;Y| x'_1,x'_2;Y')\,\, \,=\,p^2_1\,p^2_2\,\left(
\frac{\partial}{\partial Y} + H \right) \,\,=\,\,\frac{1}{x^4_{12}}\left(
\frac{\partial}{\partial Y} + H \right)
\eeq
Collecting \eq{SIS}, \eq{G1S} and \eq{G2} one can see that the action has a simple form
\bea 
S_0\,+\,S_I\,\,&=&\,\,\int\,\frac{d^2 \,x_{1}\,d^2\,x_2}{x^4_{12}}\,\,\Phi^+\left(x_1,x_2; 
Y\right)\,\,\frac{\partial \Phi\left(x_1,x_2;Y\right)}{\partial\,Y}\,\,+\,\,\nonumber \\
\, &+&\,\, \int\,d\,Y\,\,
\int
\,\frac{d^2 x_1\,d^2 x_2\,d^2 x_3}{x^2_{12}\,x^2_{23}\,x^2_{13}} \,\,\left( \frac{\bas}{2 
\pi}\,\,\Phi^+\left(x_1,x_2;Y\right)\,\,\left\{\,\Phi\left(x_1,x_2;Y\right)\,\,-\,\,\Phi\left(x_1,x_3;Y\right)
\,\,-\,\,\Phi\left(x_3,x_2;Y\right) \right\}\,\,\right. \nonumber \\
 &+& \left. \,\,\frac{2\,\pi\,\bas^2}{N_c}\cdot\,\,
\{ \, 
\Phi(x_1,x_2;Y)\,\,\cdot\,
\Phi^+(x_1,x_3;Y)\,\Phi^+(x_3,x_2;Y)\,\,+\,\,h.c. \}\,\right)\,.\label{S0IS}
\eea
\eq{S0IS} can be reduce to more elegant form using \eq{SI2} and going toomentum representation, namely,
\beq \label{MR}
\Phi(x_1,x_2;Y)\,\,=\,\,x^2_{12}\,\int\,d^2\,k\,e^{i\,\vec{k} \cdot \vec{x}_{12}}\,\,\Phi\left(k,b;Y\right)
\eeq
where $ b\,=\,( x_1 + x_2)/2 \,\,\rightarrow\,\,(x_1 + x_3)/2\,\,=\,\,(x_2 + x_3)/2$ (see \eq{SI2}).

In terms of fields $\Phi\left(k,b;Y\right)$ and $\Phi^+\left(k,b;Y\right)$ the action looks as follows
\beq  \label{S0ISMR}
S\,\,=\,\,S_0\,+\,S_I\,\,=
\,\,\int\,d^2 \,b\,d^2\,k\,\,\Phi^+\left(b, k;
Y\right)\,\,
\eeq
$$
\left( \frac{\partial \Phi\left(b,k;Y\right)}{\partial\,Y}\,\,-\,\,
 \frac{\bas}{2\pi}\,\int\,d^2\,k'\,K\left(k,k'\right)\, 
\Phi\left(b,k';Y\right)\,\,+\,\,\frac{2\,\pi\,\bas^2}{N_c}\cdot \left\{\, \Phi^+b,k;Y)\,\Phi^+(b,k;Y)\,\,\,
+\,\,\,h.c. 
\right\}\,\right)
$$
where $K\left(k,k'\right)$ is the BFKL kernel in the momentum representation, namely, 
\beq \label{BFKLKER}
\int\,d^2\,k'\,\,K\left(k,k'\right)\,\,\Phi\left(b,k';Y\right)\,\,=\,\,
 \int\,d^2\,k'\,\,\frac{\Phi\left(b,k';Y\right)}{(k - 
k')^2}\,\,-\,\,k^2\,\int\,d^2\,k'\,\frac{\Phi\left(b,k;Y\right)}{(k - k')^2\,(k'^2 \,+\,(k - k')^2)}
\eeq

The action $S$ of \eq{S0ISMR} looks very similar to the action of the toy model (see \eq{TMFI}) and can be
easily reduced to the Langevin equation for directed percolation
\beq \label{DPG}
\frac{\partial \,\,\Phi\left(b,k; Y\right)}{\partial Y}\,\,\,
=\,\,\frac{\bas}{2\pi}\,\int\,d^2\,k'\,K\left(k,k'\right)\,
\Phi\left(b,k';Y\right)\,\,\,-\,\,\,\frac{2\,\pi\,\bas^2}{N_c}\cdot \, \Phi( b,k;Y)\,\Phi^(b,k;Y)
\,\,+\,\,\zeta(b,k; Y)
\eeq
with
\beq \label{DPNOISE}
<|\zeta(b,k; Y)|>\,\,=\,\,0\,;\,\,\,\,\,\,\,\,\,\,\,<|\zeta(b,k; Y)\,\zeta(b',k'; Y')
|>\,\,=\,\,\frac{4\,\pi\,\bas^2}{N_c}\cdot\Phi(b,k;Y)\,\delta^{(2)}(\vec{b} -\vec{ b}')\,\delta^{(2)}(\vec{k}
- 
\vec{k}')\,\delta\left(Y - Y'\right)
\eeq

Actually, we have shown (see Refs.{L3,L4}) and will discuss below that in QCD we expect a different form for 
the noise , namely,
\beq \label{DPNOISE1}
<|\zeta(b,k; Y)|>\,\,=\,\,0\,;
\eeq
$$
\,\,\,\,\,\,\,\,\,\,\,<|\zeta(b,k; Y)\,\zeta(b',k'; Y')
|>\,\,=\,\,\frac{4\,\pi\,\bas^2}{N_c}\cdot \Phi(b,k;Y)\,\left( 1 - \Phi(b,k;Y) \right)\delta^{(2)}(\vec{b}
-\vec{ b}')\,\delta^{(2)}(\vec{k} -
\vec{k}')\,\delta\left(Y - Y'\right)
$$
which belongs to a diffrent universality class than \eq{DPNOISE} (\cite{HH,DMS}) \footnote{ \eq{DPNOISE1} for
the noise was  also suggested  in Ref. \cite{MSX}} .

 The Langevin
equation is the one of many ways to describe a diffusion process and the condidereable progress has been 
achieved in this approach. (see Refs.\cite{HH,WE,BIW,IT,MSW,SO}).

 However, we prefer
a
different way for description of the BFKL Pomeron interactions,  which will also  lead to diffusion
equation: the so called generating functional approach. The advantage of the generating functional approach is 
its tansparent relation to the partonic wave function of the fast hadron (dipole) and, because of this, this 
approach can be a source of the new Monte Carlo approach for high energy scattering. In this approach we see
in the most explicit way our main theoretical problem: the BFKL Pomeron calculus provides the amplitude that 
satisfies the $t$-channel unitarity while the $s$-channel unitarity is still a problem in the BFKL Pomeron
calculus. However, the probabilistic interpretation in the framework of the generating functional leads to the 
correctly normalized partonic wave function which takes into account the main properties of the $s$-channel 
unitarity as well.

\subsection{Generating functional: general approach}
In this subsection we discuss the main equations of the BFKL Pomeron Calculus in the
formalism of the generating functional,  which we consider as the most
appropriate technique for the probabilistic interpretation of this approach to
high energy scattering in QCD.

To begin with let us write down the definition of the generating functional
\cite{MUCD}
\begin{equation} \label{Z}
Z\left(Y\,-\,Y_0;\,[u] \right)\,\,\equiv\,\,
\end{equation}
$$
\equiv\,\,\sum_{n=1}\,\int\,\,
P_n\left(Y\,-\,Y_0;\,x_1, y_1; \dots ; x_i, y_i; \dots ;x_n, y_n
 \right) \,\,
\prod^{n}_{i=1}\,u(x_i, y_i) \,d^2\,x_i\,d^2\,y_i
$$
where $u(x_i, y_i) \equiv u_i $ is an arbitrary function of $x_i$ and $y_i$.
The coordinates $(x_i,y_i)$ describe the colorless pair of gluons or a dipole.
$P_n$ is a probability density to find $n$ dipoles with the size $x_i - y_i$,
and with impact
parameter $(x_i + y_i)/2$. Directly from the physical meaning of $P_n$ and
definition in Eq.~(\ref{Z}) it
follows
that the functional (Eq.~(\ref{Z})) obeys the condition
\begin{equation} \label{ZIN1}
Z\left(Y\,-\,Y_0;\,[u=1]\right)\,\,=\,\,1\,.
\end{equation}
The physical meaning of this equation  is that the sum over
all probabilities is equal to unity.

Introducing vertices for the dipole process: $1 \to 2$ ($V_{1 \to 2}( x,y \to
x_1,y_1 + x_2,y_2)$),  $2 \to 1$ ($V_{2 \to 1}( x_1,y_1 + x_2,y_2 \to x,y)$)
and $2 \to 3$ ($V_{2 \to 3}(x_1,y_1 + x_2, y_2 \to x'_1,y'_1 + x'_2, y'_2  +
x'_3,y'_3)$ we can write a typical birth-death equation in the form
\begin{equation}  \label{P}
\frac{\partial \,P_n(Y;\dots;x_i,y_i; \dots;x_n,y_n)}{\partial Y}\,\,\,=
\end{equation}
\begin{eqnarray}
 &=&\,\, \sum_{i} \  V_{1 \to
2}
\bigotimes \left(P_{n-1}(Y; \dots;x_i,y_i; \dots;x_n,y_n)\, -\,P_n (Y;\dots;x_i,y_i;
\dots;x_n,y_n)\right) \label{P1} \\
 &+ & \,\, \sum_{i > k} V_{2 \to 1}\,\bigotimes\,\left( P_{n+1}(Y; \dots;x_k,y_k; \dots ; x_i,y_i;\dots:x_n,y_n )
\,- \, P_n(Y;\dots;x_i,y_i;\dots;) \right)  \label{P2} \\
 &+ & \,\, \sum_{i > k}\, V_{2 \to 3}\bigotimes  \left( P_{n
-1}(Y;x_i,y_i;\dots ; x_k,y_k;\dots;x_n,y_n)\,-\,P_n(Y;x_i,y_i;\dots ; x_k,y_k;\dots;x_n,y_n)\right) \label{P3}
\end{eqnarray}
\eq{P} is the typical Markov's chain and the fact that we have the correct normalized partonic wave function
is written in \eq{P} by introducing for each microscopic (dipole) process two terms (see \eq{P1},\eq{P2} and 
\eq{P3}): the emission of dipoles (positive birth term ) and their  recombination (negative death term).

 Multiplying this equation  by the product $\prod^n_{i=1}\,u_i$
and integrating over all $x_i$ and $y_i$,  we obtain the
following linear equation for the generating functional:
\begin{equation}\label{ZEQ}
\frac{\partial \,Z\,\Lb Y-Y_0; [\,u\,]\Rb}{
\partial \,Y}\,\,= \,\,\chi\,[\,u\,]\,\,Z\,\Lb Y- Y_0; [\,u\,] \Rb
\end{equation}
with
\begin{eqnarray}
\chi[u]\,\,&=&\,\, \,\int\,d^4\,q d^4 q_1\,d^4 q_2 \,\,
  \left(  V_{1\,\rightarrow \,2}\left( q \to
q_1 + q_2 \right)\,  \left(  - u(q) \,+\,u(q_1) \,u(q_2)\,\right)
\,\frac{\delta}{\delta u(q)}\,- \right. \label{chi} \\
 & &\left. - V_{2\,\rightarrow \,1}\left(  q_1 + q_2 \to q  \right)\,
\left( u(q_1)  \,u(q_2) \,-\, u(q) \right) \,\,\frac{1}{2} \,\frac{\delta^2}{\delta
u(q_1)\,\delta
u(q_2)} \right)\,;
\label{VE21}
\end{eqnarray}

 Trying to make out presentation more transparent   we omitted  in
Eq.~(\ref{chi})
the
term that corresponds to   the $ 2 \to 3$ transition (see Ref. \cite{L3} for full
presentation). We also use notations $q_i$ for $(x_i,y_i)$ and $d^4 q_i $ for $d^2 x_i\,d^2\,y_i$
where $x_i$ and $y_i$ are positions of quark (antiquark) for the colourless dipole.

Eq.~(\ref{ZEQ})  is a typical diffusion equation or Fokker-Planck equation
\cite{GARD},   with the
diffusion
coefficient which depends on $u$. This is the master equation of our approach, and
the goal of this
paper is  to find the correspondence of this equation with the BFKL Pomeron Calculus
and   the asymptotic
solution to this equation. In spite of the fact that this is a
functional equation we intuitively feel that this equation could be useful since
we can develop a
direct method for its  solution  and, on the other hand, there exist many  studies
of such
an equation
in the literature ( see for example Ref. \cite{GARD}) as well as some physical
realizations in
statistical physics. The intimate relation between the Fokker-Planck equation, and
the high energy
asymptotic was first pointed out  by Weigert \cite{WE} in JIMWLK approach
\cite{JIMWLK},  and has
been discussed in
Refs.
\cite{BIW,IT,MSW}.

The scattering amplitude can be defined as a functional \cite{K,L2}
\begin{eqnarray}
 N\left(Y;[\gamma_i] \right)\,&= & \,- \sum^{\infty}_{n =1} \int
\gamma_n(x_1,y_1;\dots;x_n,y_n;Y_0) \prod^n_{i=1}\frac{\delta}{\delta
u_i}Z\left(Y,[u_i]\right)|_{u_i=1}\,d^2 x_i\,d^2 y_i \label{N}\\
 &=& \,-\,\sum^{\infty}_{n =1} (-1)^n \int
\gamma_n(x_1,y_1;\dots;x_n,y_n;Y_0)\,\,\rho(x_1,y_1;\dots;x_n,y_n;Y -Y_0) \nonumber
\end{eqnarray}

The physical meaning of functions $\gamma_n$ is the imaginary part of the amplitude
of interaction of $n$-dipoles with the target at low energies. All these functions
should be taken from the non-pertubative QCD input. However, in Refs.
\cite{L1,L2,L3}
it was shown that we can introduce the amplitude of interaction of $n$-dipoles
($\gamma_n(x_1,y_1;\dots;x_n,y_n;Y)$ at
high energies (large values of rapidity $Y$) and Eq. ~(\ref{Z}), Eq. ~(\ref{ZEQ})
and Eq. ~(\ref{N}) can be rewritten as a chain set of equation for
$\gamma_n(x_1,y_1;\dots;x_n,y_n;Y)$.  The equation has the form\footnote{This
equation is Eq.~(2.19) of  Ref. \cite{L3} but, hopefully, without  misprints ,
part of which has been noticed in Ref. \cite{IST}.}
\begin{eqnarray} \label{N5}
{}&{} & \frac{\partial\, \gamma_n\left( q_1\dots, q_n
\right)}{\partial\,Y}\,\,\,\,=  2\,\sum_{i=1}^n\,\int\,d^4q'\,d^4q \,
V_{1\,\rightarrow \,2} (q_i;\,q,\,q')\,
\gamma_n\left(  \dots q'\dots \right) \,\nonumber
\end{eqnarray}
\begin{eqnarray}
&-&\,   \sum_{i=1}^n \,\int d^4q_1'\,d^4q_2'\,
V_{1 \,\rightarrow\,2} (q_i;\,q_1',\,q_2')\,\gamma_n\left(  \dots, q_i\dots
\right)\,-\sum^{n - 1}_{i =1} \int d^4q\,d^4q' \\
& & V_{1\,\rightarrow \,2} (q_i;\,q,\,q')\,\gamma_{n+1}\left(  \dots q \dots
q'\right)
-\,\sum_{i > j}^n\,\int  d^4\,q\,
V_{2\,\rightarrow\,1}\left( q_i,\,q_j ;\,q \right) \nonumber \\
&\cdot&\gamma_{n-1}\left(
 q_i \dots  q_j \dots q\right)
\,+\, 2\,\sum_{i=1}^n \int d^4 q\, d^4 q' V_{2
\rightarrow 1} \left( q ,q_i;q'\right) \, \gamma_{n-1}\left( \dots q_i \dots
q\right)
\,+ \nonumber \\
&+&
\,\sum^n_{i,k,i > k}\,   \int d^4q\,
V_{2\rightarrow 1} \left( q_i ,q_k; q \right)\,\gamma_n\left(  \dots
q_i\dots q_k \dots \right)
\nonumber
\end{eqnarray}

Comparing this equation for $\gamma_1 \equiv T^{(1)}$ and $\gamma_2 \equiv T^{(2)}$
one can see that
\begin{eqnarray}
V_{1 \to 2} &=&\frac{\bar{\alpha}_S}{2\,\pi} \Gamma_{1 \to
2}\,\,=\,\,\frac{\bar{\alpha}_S}{2\,\pi}\,K\left(x,y;z\right)\,; \label{GA1}\\
V_{2 \to 1}\,&=&\, \,\frac{2\,\alpha^2_S}{\pi^2}
\frac{\bar{\alpha}_S}{2 \,\pi }\, \left(-\,\frac{\Gamma_{2 \to
1}(x_1,y_1 + x_2,y_2 \to x,y)}{(x - y)^4} \,+ \right.
\nonumber \\
&+& \left.
 \,
\,\int \,\frac{d^2x\,d^2y}{(x - y)^4}\,\Gamma_{2 \to 1}(x_1,y_1 +
x_2,y_2 \to x,y) \,\cdot \right. \nonumber
\\
 & \cdot& \left. \left(\delta^{(2)}(x_1 - x)\delta^{(2)}(y_1 - y)\,+\,\delta^{(2)}(x_2 -
x)\delta^{(2)}(y_2
- y)\right) \right)\label{GA2}
\end{eqnarray}
with $\Gamma_{2 \to 1}$ is given by Eq.~(\ref{V21}).
\subsection{A toy model: Pomeron interaction and probabilistic interpretation}
In this section we consider the simple toy model in which the probabilities to find
$n$-dipoles do not depend on the size of dipoles \cite{MUCD,L1,L3,L4}. In this
model the master equation (\ref{ZEQ}) has a simple form
\begin{equation} \label{TMZ}
\frac{\partial Z}{\partial Y}\,=\,- \Gamma(1 \to 2)\,u (1 - u)
\,\frac{\partial Z}{\partial u}+\,\Gamma(2 \to
1)\,u (1 - u)\,\frac{\partial^2 Z}{(\partial u)^2}
\end{equation}
and this equation generates: the Pomeron splitting $G_{P \to 2P}= - \Gamma(1 \to
2)$;  Pomerons merging $G_{2P \to P}=  \Gamma(2 \to 1)$  and also the two Pomeron
scattering $G_{2P \to 2P}\,\,=\,\,- \Gamma(2 \to 1)$.  It is easy to see that
neglecting $u^2 \partial^2 Z/(\partial u)^2$ term in Eq.~(\ref{TMZ}) we cannot
provide a correct sign for Pomerons merging $G_{2P \to P}$.

The description given by \eq{TMZ} is equivalent to the path integral of \eq{TMFI}. To see this we need
to notice that the general solution of \eq{TMZ} has a form
\beq \label{TMH1}
Z(Y;u)\,\,=\,\,e^{H(u)\,(Y - Y_0)}Z(Y_0;u)
\eeq
with operator $H$ defined as
\beq \label{TMH2}
H(u)\,\,=\,\,-\, \Gamma(1 \to 2)\,u (1 - u)
\,\frac{\partial }{\partial u}+\,\Gamma(2 \to
1)\,u (1 - u)\,\frac{\partial^2 }{(\partial u)^2}
\eeq
and
\beq \label{TMZ0}
Z(Y_0;u)\,\,=\,\,e^{\tau_{tr}( u -1 )}
\eeq

Introducing operators of creation ($a^+$) and annihilation ($a$)
\beq \label{TMH3}
\hat{a}\, =\,\frac{\D}{\D u}\,; \,\,\,\,\,\,\,\,\,\, \hat{a}^+ \,=\,u \,\,\,\mbox{that satisfy} \,\,[
\hat{a} ,\hat{a}^+] = 1
\,\,\, \mbox{at fixed}\,\, Y
\eeq
one can see that operator ${\cal H}$ has the form
\beq \label{TMH4}
{\cal H}\,\,=\,\,-\,\Gamma(1 \to 2)\,\hat{a}^+\,( 1 - \hat{a}^+)\,\hat{a} \,\,+\,\,\Gamma(2 \to
1)\,\hat{a}^+\,( 1 - \hat{a}^+)\,
\hat{a}^2
\eeq
and the initial state at $Y=Y_0$ is defined as
\beq \label{TMIS}
|Y_0>\,\,=\,\,e^{\tau_{tr}( \hat{a}^+ \,-\,1)} |0>
\eeq
with the vacuum defined as $\hat{a} |0> = 0$.

We need to discretize the development operator of \eq{TMH1} with ${\cal H}$ given by \eq{TMH4},
namely,
\beq \label{TMH5}
e^{{\cal H}\,(Y - Y_0)}=e^{{\cal H}\,\Delta Y}\, \dots\, e^{{\cal H}\,\Delta Y}
=\,\prod^N_{j=1}\,\Lb 1 \,+\,{\cal H}\,\Delta Y  \Rb
\eeq

and introduce the coherent states \cite{CION}
for a certain intermediate rapidity $Y_j=Y_0+j\Delta Y$
 as
\beq \label{TMH6}
|\phi_j>\,\,=\,\,e^{\phi_j\,\hat{a}^+ \,-\,\phi_j }\,|0>
\eeq
where $\phi_j$ are arbitrary complex numbers.
The initial state of \eq{TMIS} can be written as
\begin{center}
\begin{eqnarray}
|\phi_0(Y_0)>\,\equiv| \tau_{tr}>.
\end{eqnarray}
\end{center}

The unit operator in terms of the coherent states can be expressed as

\beq \label{TMH7}
1\,=\,\Lb \,\int\,\frac{d \phi_j d \phi_j^*}{\pi i}\,e^{ - \phi_j\,\phi_j^*\,+\,\phi_j\,+\,\phi_j^*} \Rb
|\phi_j>\,<
\phi_j|
\eeq

We want to calculate matrix element of some operator $A$ between states of initial $Y_0$ and
 final $Y$ rapidity $<Y|A|Y_0>$. This can be written as

\beq
<Y|A|Y_0>=<Y|A\left\{
\,\Lb \,\int\,\frac{d \phi_Y d \phi_Y^*}{\pi i}\,e^{ - \phi_Y\,\phi_Y^*\,+\,\phi_Y\,+\,\phi_Y^*} \Rb
|\phi_Y>\,<
\phi_Y|
\right\}|Y_0>
\eeq
here we denote $|Y>\equiv |\phi_Y>$.
Next we use the development operator given in \eq{TMH5} to find $<Y|Y_0>$. We split
the rapidity $Y-Y_0$ to $N$ intervals.And insert the development \eq{TMH5} and unit \eq{TMH7}
 operator  between the states of intermediate rapidity
\beq
<Y|   \prod^{N}_{j=1}\,\Lb 1 \,+\,{\cal H}\,\Delta Y  \Rb\, |Y_0>
\eeq
 We look at
\begin{eqnarray}
<\phi_{j+1}|\Lb 1 \,+\,{\cal H}\,\Delta Y
\Rb|\phi_j>&=&\exp\left\{-\phi^*_{j+1}\phi_{j+1}+\phi^*_{j+1}+\phi_{j+1}
-\phi^*_{j+1}-\phi_{j}+\phi^*_{j+1}\phi_j \right\}\, \nonumber \\
&\times&
\Lb 1 \,+ \,{\mathcal{H}( \phi^*_{j+1}, \phi_j
)}\,\Delta Y
 \Rb \nonumber \\
 & =&
\exp\left\{-\phi^*_{j+1}(\phi_{j+1}-\phi_{j})+\phi_{j+1}-\phi_{j}\right\}\Lb 1 \,+ \,{\mathcal{H}(
\phi^*_{j+1}, \phi_j )}\,
\Delta Y
  \Rb \nonumber \\
&= & \exp \left\{-\phi^*_{j+1}(\phi_{j+1}-\phi_{j})+\phi_{j+1}-\phi_{j}\right\}\,\exp\left(\mathcal{H}(
\phi^*_{j+1}, \phi_j )
\,\Delta Y\right)  \label{alex1}
\end{eqnarray}


Now we redefine an arbitrary function $\phi_j$ as
\beq
\Phi_j\,\,\,=\,\,\,-\phi_j,   \hspace{2cm} \Phi^+_{j}=\phi^*_{j}-1
\eeq
and rewrite \eq{alex1} in terms of $\Phi_j$ and $\Phi^+_j$
\begin{eqnarray}
\,\,\,\,\,\,\,\,\,\,\,\,& &
e^{-\phi^*_{j+1}(\phi_{j+1}-\phi_{j})+\phi_{j+1}-\phi_{j}}\,\,\exp\left\{\mathcal{H}( \phi^*_{j+1},
\phi_j
)\,\Delta Y\right\}\,\,= \nonumber \\
\,\,\,\,\,\,\,\,\,\,\,\, &=&
 exp\left\{\Phi^{+}_{j+1}(\Phi_{j+1}-\Phi_{j})+\mathcal{H}( \Phi^+_{j+1}+1, -\Phi_j
)\right\}\,\label{alex2} \\
& =& \exp\left\{  \left( \frac{\Phi^+_{j+1}(\Phi_{j+1}-\Phi_{j})}{\Delta Y}+\mathcal{H}( \Phi^+_{j+1}+1,
-\Phi_j
 )\,  \right )\Delta Y  \right\} \nonumber
\end{eqnarray}
 Summing over all rapidity intervals we have
\beq
<Y|A|Y_0>\,\,\,
\sim\,\,\,\prod_{j=0}^{N}\int d\Phi_j^+ d\Phi_j A(Y) e^{S}
\eeq
where $A(Y)$ is the expectation value of the operator $A$ at the final rapidity $Y$, and
\beq
S\,\,=\,\,\left( \frac{\Phi^+_{j+1}(\Phi_{j+1}-\Phi_{j})}{\Delta Y}+\mathcal{H}( \Phi^+_{j+1}+1, -\Phi_j
)\,  \right )\Delta Y
\eeq

In the continuous limit this becomes
\beq \label{FUIN}
<Y|A|Y_0>  \,\,\,=\,\,\,\frac{ \int \mathcal{D}\Phi^+ \mathcal{D}\Phi \,\,A(Y)\,\, e^{S}}{\int \mathcal{D}\Phi^+
\mathcal{D} \Phi\,\, e^{S}}
\eeq
with
\begin{eqnarray}
  S\,&=&\,\int \left( \Phi^+  \frac{d}{dY} \Phi +\mathcal{H} ( \Phi^+ +1, -\Phi ) \, \right)
dY \,\, =\, \,\int \left( \Phi^+  \frac{d}{dY} \Phi \,\,-\right. \label{TMHAC} \\
 &-& \left.
-\Gamma(1\rightarrow2) \Phi^+\Phi
+\Gamma(1\rightarrow2) \Phi^+(\Phi)^2
+\Gamma(2\rightarrow 1)( \Phi^+)^2\Phi
-\Gamma(2\rightarrow 1) (\Phi^+)^2\,(\Phi)^2
 \, \right) dY  \nonumber
\end{eqnarray}

This action is almost the action of \eq{TMFI} for $\Gamma(1 \to 2) = \Gamma(2 \to 1)$. In the
toy-model the difference between these two vertices is the normalization problem of function
$\Phi^+$ and $\Phi$. In our approach they are normalized in the way which allows us to treat them as
probabilities (see \eq{T} and \eq{TN}).  However, \eq{TMHAC} includes the new interaction: the
transition of two Pomerons to two Pomerons. The sign is such that this interaction provides the
stability of the potential energy. Indeed this term is responsible for the increase of the potential
energy at large values of both $\Phi^+$ and $\Phi$.

Comparing \eq{TMHAC} with \eq{SM1} one can see that we build the partition function and the
thermodynamic potential using the generating functional. It means that our \eq{ZEQ} is equivalent to
statistical description of the system of dipoles.

Eq.~(\ref{TMZ}) is the diffusion with the $u$ dependence in  diffusion coefficient.
In terms of the Langevin equation \eq{TMHAC} generates the noise term of the \eq{DPNOISE1} type.

 To out taste this equation is simpler than the Langevin equation of \eq{LEQ} and it will be easily
generalized for the case of QCD.     For $u<1$ the diffusion coefficient is positive and the equation
has a reasonable solution. If $u >1$, the sign of this coefficient changes and the equation gives a
solution which increases with $Y$ and $Z(Y)$  cannot be treated as the
generating function for the probabilities to find $n$ dipoles (Pomerons) (see Refs.
\cite{GRPO,BOPO,L4} for details). The same features we can see in the asymptotic
solution that is the solution to Eq.~(\ref{TMZ}) with the l.h.c. equal to zero.
It is easy to see that this solution has the form
\begin{equation}
Z(u;Y \to \infty) \,\,=\,\,\frac{1 \,-\,e^{\,\kappa\, u}}{1\,-\,e^{\,\,\kappa}}\,;\,\,\,\mbox{with}\,\,\,
\kappa\,=\,
\frac{\Gamma(1 \to 2)}{\Gamma(2 \to 1)}\,=\,\frac{2
\,N^2_c}{\bar{\alpha}^2_S}\,\gg\,1
\end{equation}
One can see that for negative $\kappa$ this solution leads to $Z>1$ for $u <1$.
This shows that we cannot give a probabilistic interpretation for such a solution.

\section{Probabilistic interpretation in QCD}
\label{sec:PIQCD}
\subsection{Several general remarks}
Eq. (\ref{GA1}) has a very simple physical meaning describing the Pomeron splitting
as the decay process of one dipole into two dipoles. It turns out that the vertices
for $2P \to 3P$ and $2P \to 2P$ can be easily understood as a dipole `swing'. What
we mean is that   with some probability two quarks of
a pair  of dipoles can exchange their antiquarks to form another pair  of dipoles
\cite{L3}.
Naturally, this process has $1/N_c^2$ suppression and it correctly reproduces the
splitting and rescattering of two Pomeron that has been explicitly calculated from
the diagrams \cite{BLV}.

Eq.~(\ref{GA2})\footnote{This equation is quite different from the equation which
is obtained in Ref. \cite{L3} (see also Ref. \cite{IST}). The main difference stems
from the  correct use of the BFKL Pomeron Calculus for determining this vertex
while in Refs. \cite{L3,IST} the Born diagram was used which does not and cannot
give a correct expression.}
 is more difficult to view as the vertex for the
transition of two  dipoles into one. Indeed, the integral over  coordinates of the
produced dipole  is positive, namely $\int V(2 \to 1) d^2 x d^2 y =  \int \Gamma(2
\to 1) d^2 x d^2 y$. However, Eq.~(\ref{GA2}) generally speaking leads to a
negative vertex in some regions of the phase space. Here, we want to point out that the key
problem is not in the probabilistic interpretation of the microscopic process of
two dipole to one dipole transition but the fact that a negative vertex $V(2 \to
1)$ means that in some kinematic region we have a negative diffusion coefficient
which results in a solution that increases at large values of rapidity $Y$.
To save such theory we need to introduce other Pomeron interactions like $2P \to
3P$ and/or $2P \to 2P$ transitions.

In Ref. \cite{L3,L4} the attempts were made to deal with such Pomeron interactions.
It turns out that at large values of $Y$ the $2 \to 1$ process contribute in a very
limited part of the kinematic region with very specific function $u(x,y)$. In this
particular region the vertex $V(2 \to 1)$ is positive. Therefore, we
could use the probabilistic interpretation but we need to study this process better
and deeper to obtain the final result.

In the toy model it has been shown \cite{GRPO,BOPO} that we can generate the $2P
\to 3P$ vertex without the process of two dipoles to one dipole transition. The
transition of two dipoles to two or more dipoles also leads to this vertex. The
similar ideas are developed in Ref. \cite{KOLU} in QCD. However, we need to pay a
price: the contribution to Eq.~(\ref{ZEQ}) will be negative to the correct sign
for $2P \to P$ interaction. In other word, we can add to Eq.~(\ref{ZEQ}) the
contribution
$$
\int\,\prod^2_{i=1}d^2 x_i\,d^2 y_i\,\prod^2_{i=1}d^2 x'_i\,d^2 y'_i\,
V_{2 \to 2}(x_1,y_1 + x_2,y_2 \to x'_1,y'_1 +
x'_2,y'_2)\,
$$
\begin{equation} \label{ZEQADD}
u(x'_1,y'_1)\,u(x'_2,y'_2)\,\frac{1}{2}
\frac{\partial}{\partial\,u(x_1,y_1)}\,\frac{\partial}{\partial\,u(x_2,y_2)}
\end{equation}
which will give the $2P \to P$ vertex in the form
\begin{equation} \label{V22}
\Gamma(2P \to P)(x_1,y_1 + x_2,y_2 \to x'_1,y'_1)\,=
\end{equation}
$$
 \,-\,\int d^2 x'_2\,d^2
y'_2\,\,V_{2 \to 2}(x_1,y_1 + x_2,y_2 \to
x'_1,y'_1 +
x'_2,y'_2)
$$
Therefore, we have to assume that $V_{2 \to 2} \,<\,0$. As we have discussed,
generally speaking, it means that our problem has no solution. However, in QCD
the situation is much better. Indeed, to generate  a correct  $2P \to P$ vertex
we need to introduce $V_{2 \to 2}\,\,\propto\,\bar{\alpha}^3_S$. The Feynman
diagrams for this transition is obvious (see Fig. ~(\ref{feymdi}-b). However, the
main contribution stems from the diagrams of Fig. ~(\ref{feymdi}-a) - type which
are of the order of $\bar{\alpha}_S/N^2_c$. Therefore, diagrams of
\fig{feymdi}-b - type are small corrections to the main contribution and could
be negative, in spite of the fact that the diagrams shown in  Fig.~(\ref{feymdi}-b)
actually gives a positive contribution.

\FIGURE[ht]{
\centerline{\epsfig{file=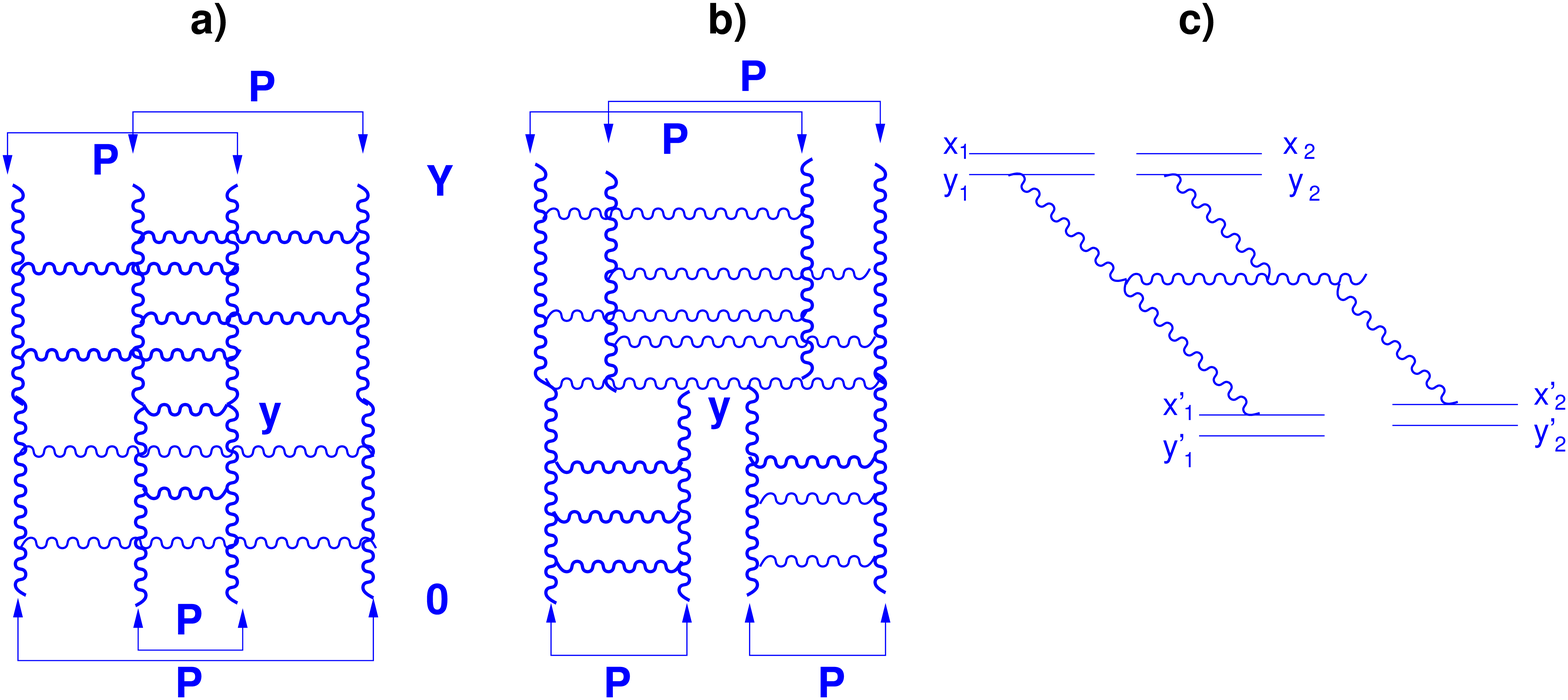,width=130mm}}
\caption{The diagrams for 2P to  2P process. \fig{feymdi}-a shows the
diagram which is of the order $\protect\bar{\alpha}_S/N^2_c$ while  \fig{feymdi}-a
leads to a contribution $\protect\propto \bar{\alpha}^3_S$. \fig{feymdi}-c shows the first diagram for
the 2
dipole to 2 dipole amplitude which square gives the probability for 2P to 2P
process.}
\label{feymdi}}

The situation with $V_{2 \to 1}$ is actually  more dramatic since even in this paper we have discussed four
different expression for this vertex:(i)\,\, \eq{GA2};\,\, (ii)\,\, \eq{T2-chain} leads to the vertex of the following
form (see Refs. \cite{IT,MSW,L3} where this form of the vertex has been discussed in details)
\beq \label{V22IT}
 \Gamma_{2 \to 1}(x_1,y_1;x_2,y_2|x,y)\,\,= \nabla^2_x\,\nabla^2_y\, \int\,d^2 z
K(x,y;z)\,\,G_0(x_1,y_1|x,z)\,G_0(x_1,y_1|z,y);
\eeq
(iii)\,\,\eq{S0IS} leads to
\beq \label{V22SM}
 \Gamma_{2 \to 1}(x,z; z,y|x,y)\,\,=\,\,K\left(x,y|z\right)
\eeq
and the expression for vertex $V_{2 \to 1}$ one can obtain substituting \eq{V22SM} into \eq{GA2};\,\,
(iv) \,\,\eq{S0IS} suggests the simple vertex
\beq \label{V22MR}
V_{2 \to 1}(b,k; b,k|b,k)\,\,=\,\,\left(\frac{4\,\pi^2\,\bas}{N_c}\right)^2\,\frac{\bas}{2\,\pi}.
\eeq
These expressions are equivalent in the framework of the BFKL Pomeron calculus but lead to quite different 
physics. For example, \eq{V22IT} does not generate $2 \to 2$ dipole term in the functional integral \cite{IST} 
and, therefore,  the noise term has a form of \eq{DPNOISE}, while all other expression correspond to the noise
term of \eq{DPNOISE1} which belongs to a different universality class.

As we have mentioned that in the kinematic region where $L(Y) \approx 1$ (see \eq{diapc6}) the $2P \to 2P$ 
transitionj gives only neglitransitionibution. However, it changes crucially the behaviour of the scattering
amplitude at ultra high energy.

Our strategy of the discussion looks as follows. First, we are going to discuss  
the two Pomeron to one Pomeron transition directly in terms of the dipole approach calculating Feyman 
diagrams.
The key question that we want to answer is what process of the dipole recombination is described by the 
diagram that contribute to the $2P \to P$  vertex.  Second, we wish to find what term in  $\chi[u]$
(see \eq{chi}) is responsible for this contribution. Third, we find the solution for the toy model that 
includes this term. Finally, we develop the generating functional approach for the approach that correspond to 
the action given by \eq{S0ISMR} which we think will be a practical way in searching the solution.

\subsection{Scattering of two dipoles in QCD}
We start to approach the problem of taking into account the two
Pomerons to two or more Pomerons transition with the clear
understanding of $2 P \to P$ merging in the dipole approach. In this
section we will obtain the same \eq{V21} for $\Gamma_{2 \to 1}$ but
directly from the dipole picture of interaction without using the
trick suggested in Ref. \cite{IT}.
 Namely, we will calculate
the contribution of the simple diagrams like that of \fig{v21pic}-a
but directly from the dipole picture of interaction \cite{MUCD}
shown in \fig{v21pic}-b.

 Starting to calculate the diagram
of \fig{v21pic} we would like to draw your attention to the fact
that these diagrams actually describe three different processes. In the
Born approximation all of them look quite similar but if you
imagine, that gluon with momentum $k$ and $l$ can produce more
gluons (see dotted lines in \fig{v21pic}, \fig{v211pic} and
\fig{v212pic}) ,  we can see a great  difference between three
diagrams. Indeed, the diagram of  \fig{v21pic} represents the
process where two bunches of additional gluons could be produced
from gluons with momenta $\vec{k}_\perp$ and $\vec{l}_\perp$. This
diagrams give a positive contribution and has clear probabilistic
interpretation.  \fig{v212pic} shows the process  where none of
additional gluons can be produced. The gluons, that were produced in
the process of the diagrams of \fig{v21pic}, contribute to the
structure of the BFKL Pomerons. The diagram of \fig{v211pic}
corresponds to a process where only one of the bunches of gluons is
produced while the second one is absorbed by the BFKL Pomeron. This
diagram belongs to  an interference type of the diagram which can
have a negative sign.

It is clear that these extra diagrams correspond to different AGK
cuts of the BFKL Pomeron diagrams. Fortunately,  they  differs by the
factor and sign in front of the same expression \cite{AGK}, namely,
\beq \label{AGK} \fig{v212pic}\,\,:\,\, \fig{v211pic}\,\,:\,\,
\fig{v21pic}\,\,=\,\,1 \,\,:\,\,-4\,\,:\,\,2 \eeq

Before approaching our problem we want to show how one can calculate
well known Lipatov vertex using light cone perturbation theory. The
calculations are performed in the infinite momentum frame, large
$N_c$, eikonal and Regge limits (\cite{GRIB1,LEPAGE}.

\FIGURE[ht]{
\centerline{\epsfig{file=  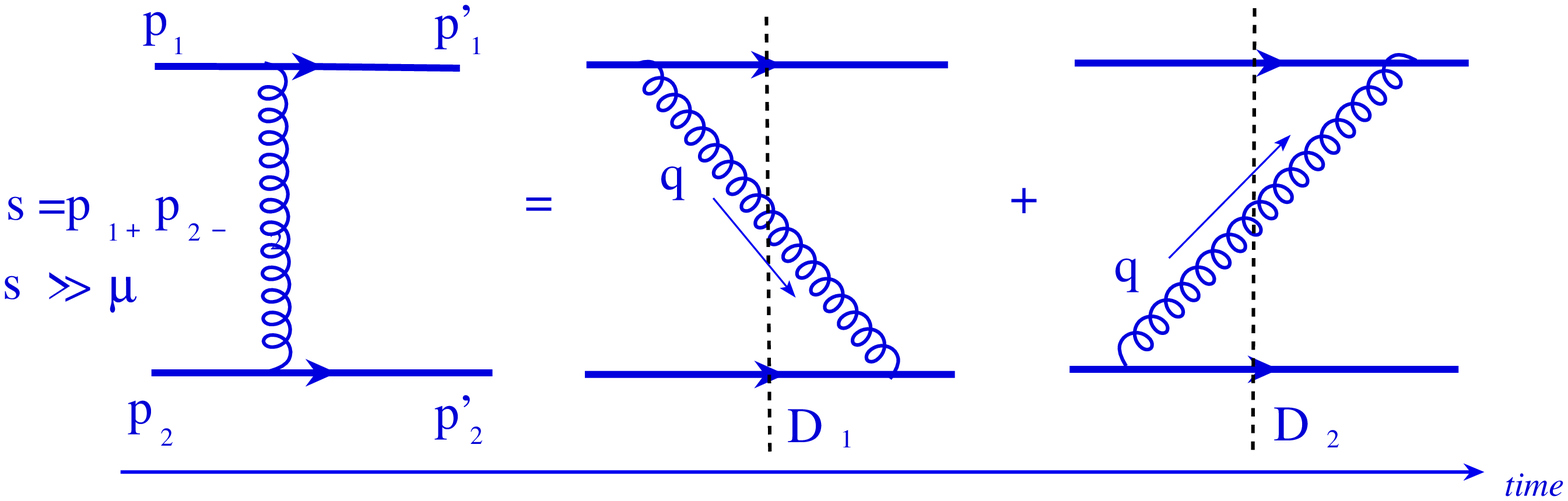,width=130mm}}
\caption{The diagram for the gluon exchange in the infinite momentum frame. The first diagram
survives while the second one gives a small,  negligible contribution at high energies.}
\label{glexch}}
We re-calculate the simplest diagram of one gluon  exchange in the  infinite momentum frame
technique to illustrate the simple space-time structure of the high energy scattering that
first have been formulated by Feynman and Gribov \cite{GRIB1,FEYN}.

It is well know that the one gluon exchange in Feynman diagram approach is equal to
\beq \label{GEX1}
A_G\,\,=\,\,\frac{\Gamma_{\mu} \,\Gamma_{\mu}}{q^2_{\perp}}
\eeq
In the infinite momentum frame we need to sum the contributions of two diagrams shown in \fig{glexch}
which are equal to
\bea
D_1 &=& \frac{\Gamma_{\mu} \,\Gamma_{\mu}}{2\,E_q}\,\frac{1}{E_{p_1}\, +\,
E_{p_2}\,-\,E_{p_2}\,-\,E_{p'_1}\,-\,E_q}\,\,=\,\,\frac{\Gamma_{\mu}
\,\Gamma_{\mu}}{2\,E_q}\,\frac{1}{-
\frac{p'^2_{1,\perp}}{2\,p'_{1,L}}\,\,-\,\,\frac{q^2_{\perp}}{q_{L}}}\,\,|_{s \ll
\mu}\rightarrow\,\,
\,\frac{\Gamma_{\mu} \,\Gamma_{\mu}}{q^2_{\perp}} \,;\,\label{GEX2}\\
D_1 &=& \frac{\Gamma_{\mu} \,\Gamma_{\mu}}{2\,E_q}\,\frac{1}{E_{p_1}\, +\,
E_{p_2}\,-\,E_{p'_2}\,-\,E_{p_1}\,-\,E_q}\,\,\approx\,\, \frac{\Gamma_{\mu}
\,\Gamma_{\mu}}{2\,E_q}\,\frac{1}{2E_q}\,\,\propto\,\,O(1/s)\,\ll 1; \label{GEX3}
\eea

The difference between these two diagrams is the following. The first one describes the process
with natural ordering in time. Indeed, the typical time for the relativistic particles is
$E/p^2_{\perp}$ where $E$ is the energy of the particle and $p_{\perp} $ is its transverse momentum.
Therefore, the first diagram describes the process of the emission of long living gluon with the
lifetime\,\,$\tau_q \,\propto\,E_q/q^2_{\perp}$ by the parent particle with its lifetime
$\tau_{p_1}\,\propto\,E_{p_1}/p^2_{1,\perp}\,\gg\,\tau_q$. In the second diagram, the particle with
short lifetime $\tau_{p_2}\,\propto\,E_{p_2}/p^2_{2,\perp}\,\,\ll\,\,\tau_q$ emits the gluon with
larger typical time. For emission of large number of gluons we have  a space-time picture shown in
\fig{sptime}. The square of these ladder-time diagram we will call the BFKL Pomeron. The general
feature of this diagrams that the cross section can be written as the product of two factors: the
square of the multi-gluon wave function ($\Psi$)  and the cross section of the interaction of the
slowest,`wee'
gluon with the target
\beq \label{PARTON}
\sigma\,\,=\,\,|\Psi |^2 \bigotimes \sigma(Gluon-target)
\eeq
where $\bigotimes$ stands for integration over all kinematic variables  of the wee gluon and $|\Psi
|^2$ is the wavefunction integrated over all kinematic variables
 of the gluons that are faster than
the wee one.

\FIGURE[ht]{
\centerline{\epsfig{file= 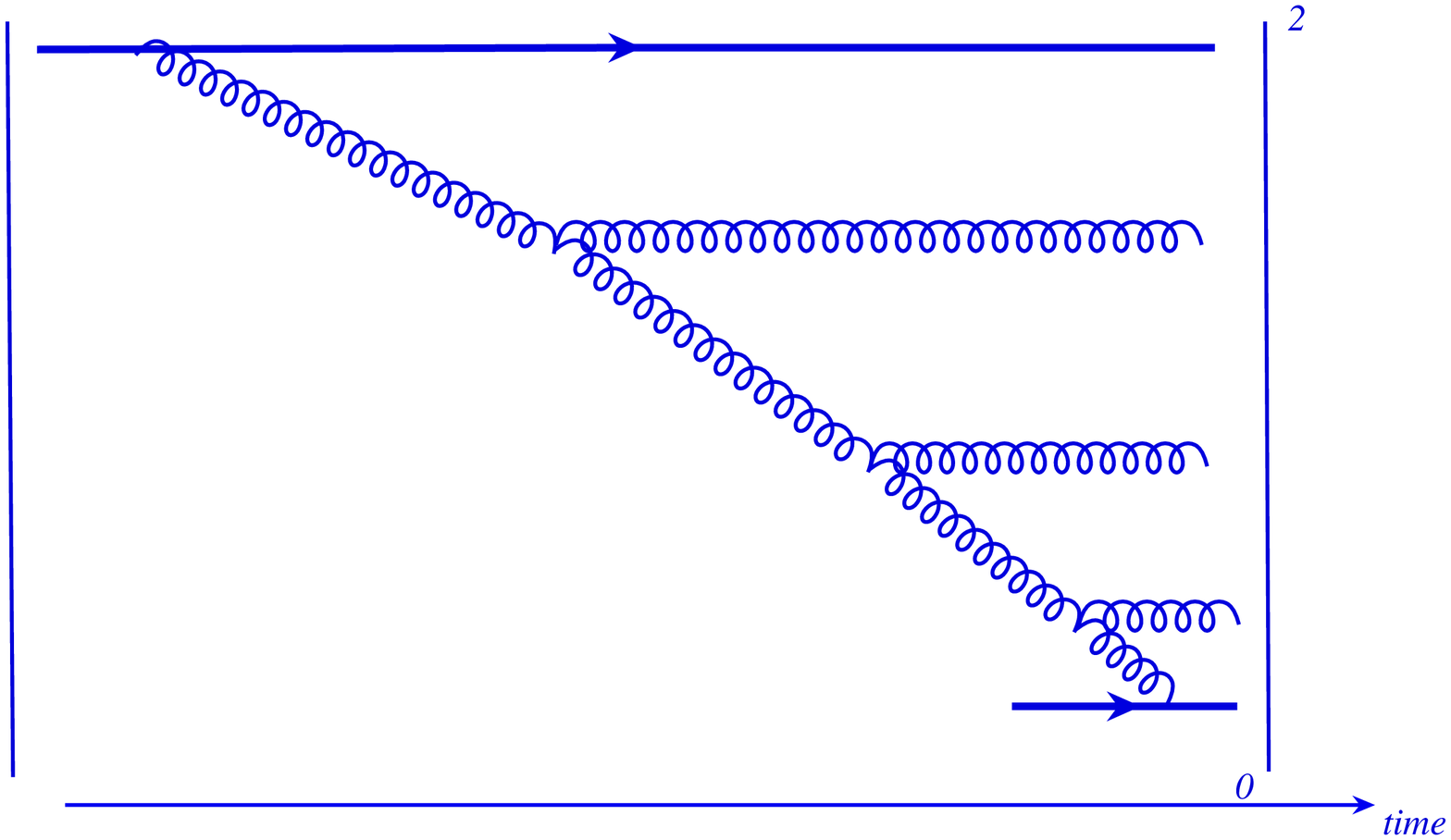,width=130mm}}
\caption{The space-time structure of the BFKL Pomeron in QCD.}
\label{sptime}}

\FIGURE[ht]{\centerline{\epsfig{file= 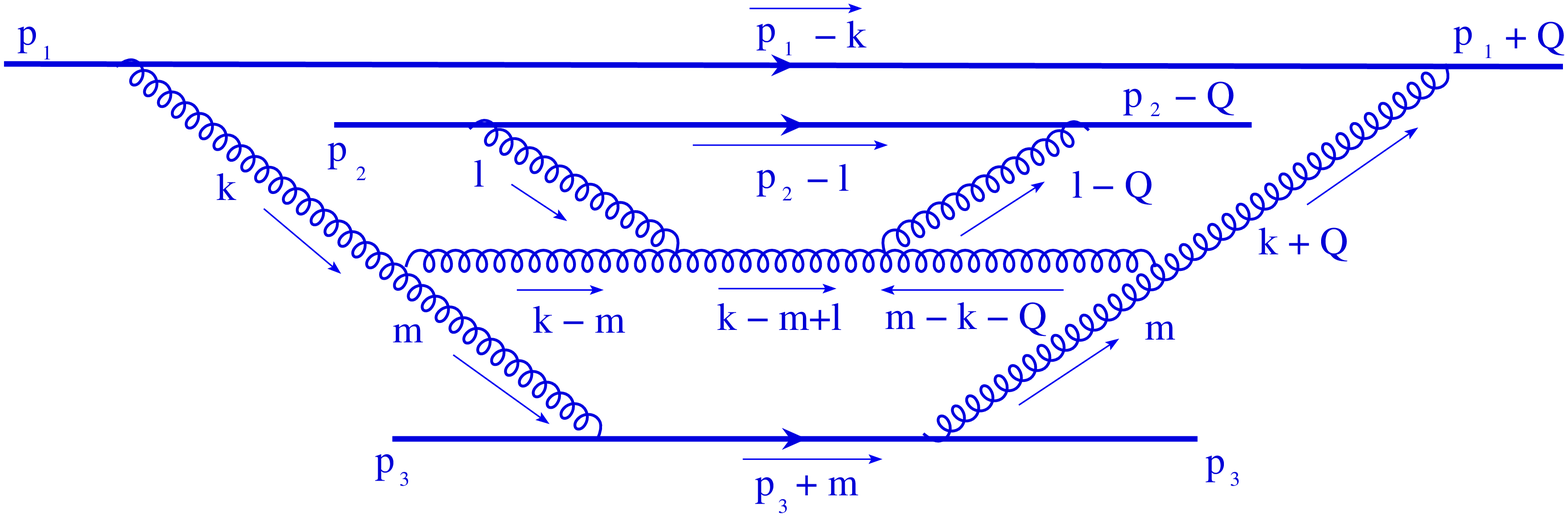,width=135mm}}
\caption{The total cross section of two quarks on one scattering}\label{diag_6}
}

\DOUBLEFIGURE[ht]{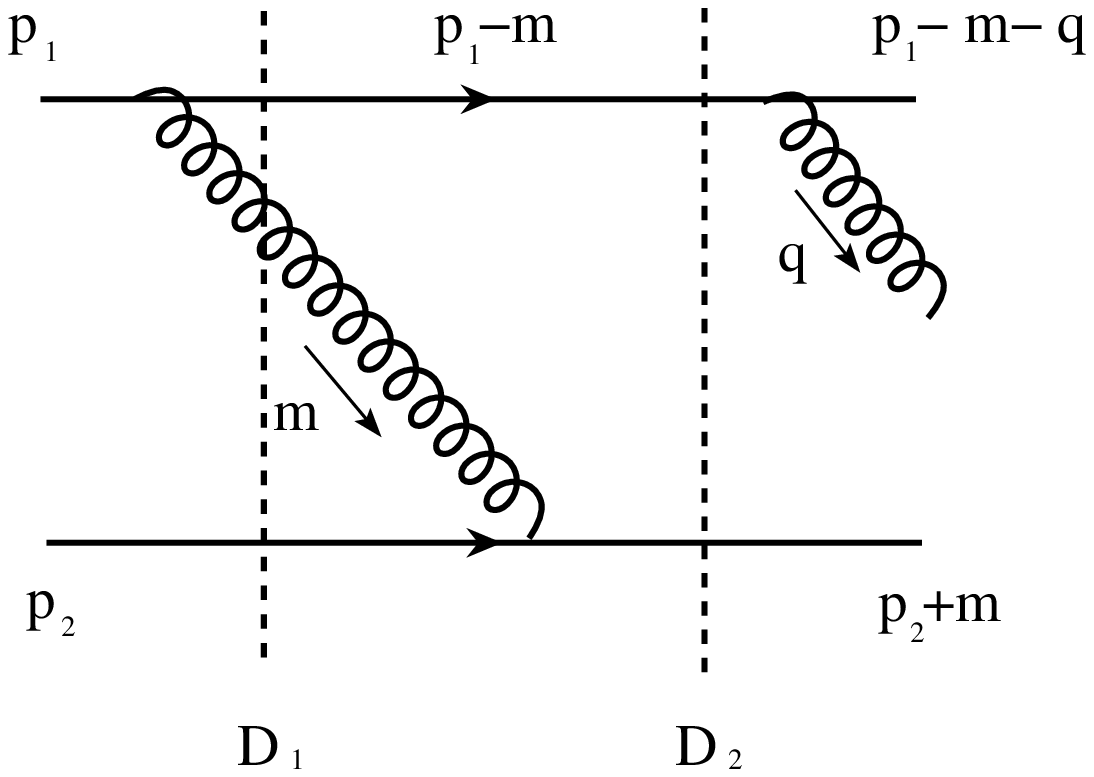,width=75mm}{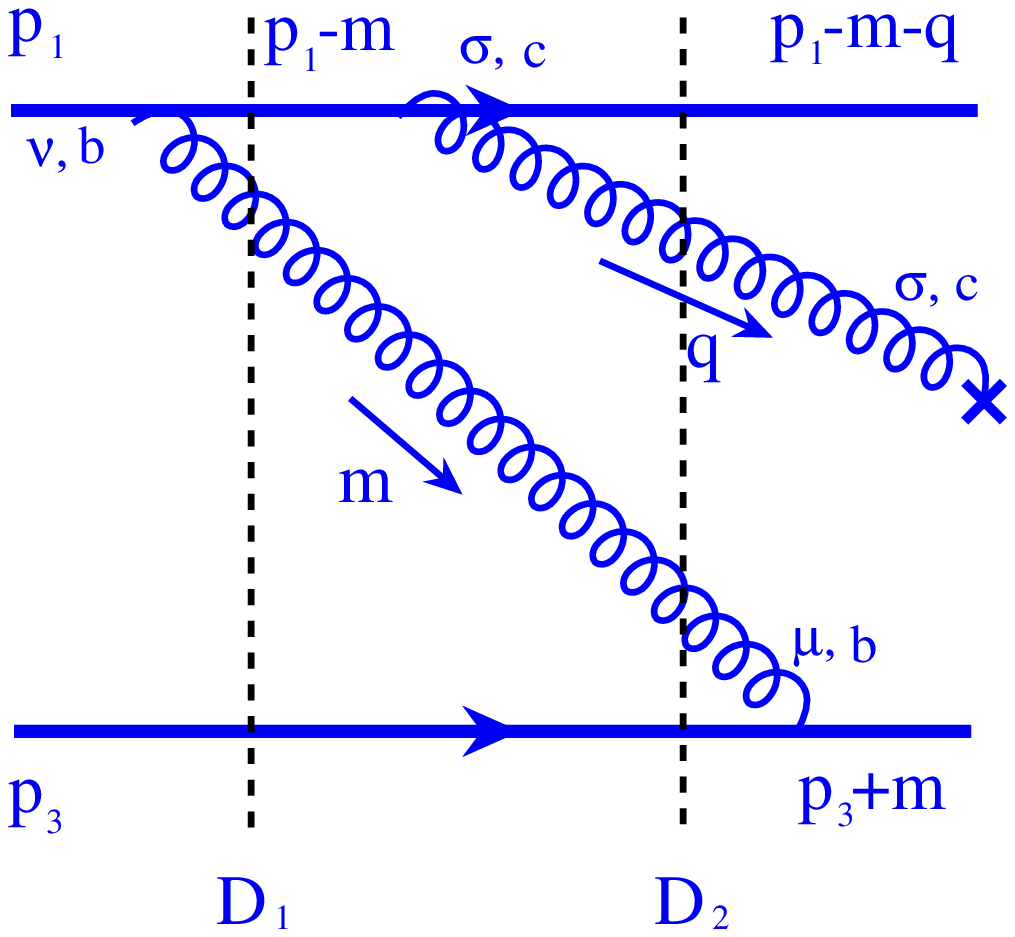,width=75mm}
  {A diagram contributing to the effective vertex.\label{Lip1}}
  {A diagram having relative $m_-$ suppression comparing to Fig. \protect\ref{Lip1}.\label{Lip2}}

\DOUBLEFIGURE[ht]{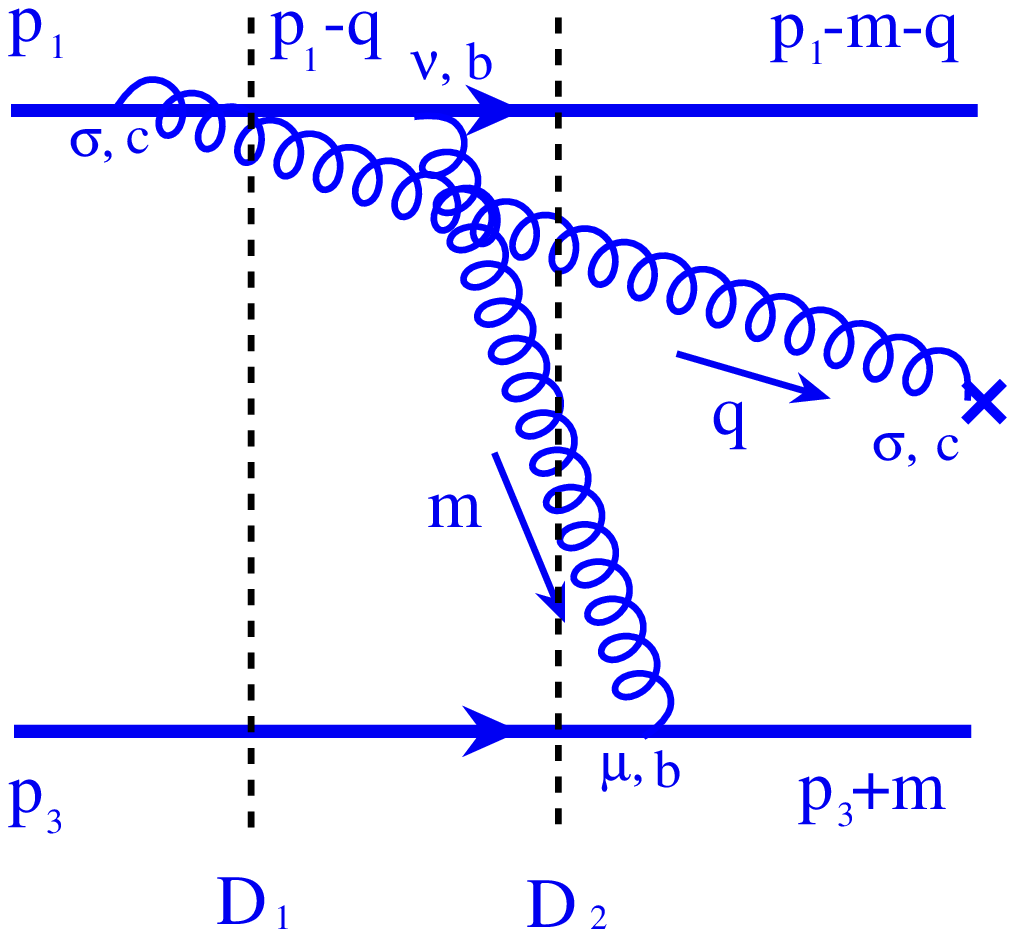,width=75mm}{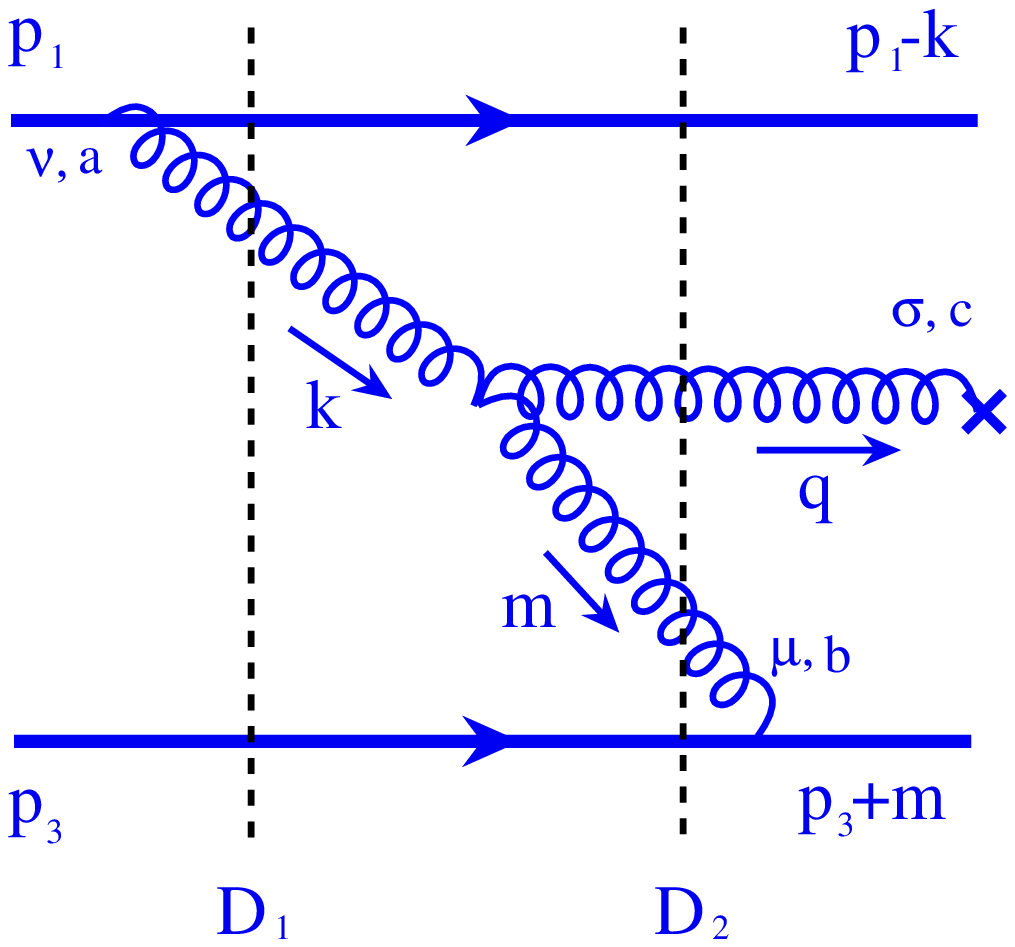,width=75mm}
  {A diagram contributing to the effective vertex.\label{Lip3}}
  {A diagram contributing to the effective vertex.\label{Lip4}}
Before starting our calculations we want to show that this kind of
diagrams indeed give the leading logarithm contribution. We consider
the total cross section shown in Fig. \ref{diag_6}.

We want to perform the integration over plus and minus  components
of all internal momenta leaving the transverse integration. It is
convenient to do the integration in terms of Sudakov parameters on
the complex plain. We assume the Regge limit, $p^\mu_{1}
p_{3\mu}\simeq p^\mu_{2} p_{3\mu}=s/2$ and forward scattering, i.e.
$Q_{\perp}=0$. Thus the internal momenta can be written as
\begin{center}
\bea   \label{sud}  k^\mu &=&\alpha_k p^\mu_1 + \beta_k
p^\mu_3+k_\perp^\mu \nonumber
\\
 l^\mu &=&\alpha_l p^\mu_1 + \beta_l p^\mu_3+l_\perp^\mu  \nonumber \\
 m^\mu &=& \alpha_m p^\mu_1 + \beta_m p^\mu_3+m_\perp^\mu \eea

 \end{center}

  The
total cross section have denominators that come from gluon
propagators \bea \label{propag}  && k^2=\alpha_k\beta_k s
-k_\perp^2  \\
&& m^2=\alpha_m\beta_m s -m_\perp^2 \\
&& l^2=\alpha_l\beta_l s - l_\perp^2 \\
&& (k-m)^2=(\alpha_k-\alpha_m)(\beta_k-\beta_m)s-(k-m)_\perp^2
\\
 && (l-Q)^2=(\alpha_l-\alpha_{Q})(\beta_l-\beta_{Q})s - l_\perp^2
  \\
    &&
(k+Q)^2=(\alpha_k+\alpha_{Q})(\beta_k+\beta_{Q})s-k_\perp^2 \\
 &&
(k-m+Q)^2=(\alpha_k-\alpha_m+\alpha_{Q})(\beta_k-\beta_m+\beta_{Q})s-(k-m)_\perp^2
 \eea
and the $\delta$-functions (from the cuts) of
 \bea \label{deltas}
(p_2-l)^2&=&-\beta_l s -l^2_\perp \label{p2-delta}\\
(p_1-k)^2&=&-\beta_k s -k^2_\perp   \\
(p_3+m)^2&=&\alpha_m s -m^2_\perp
 \eea
In forward scattering the transverse part of the transferred momenta
is zero ($Q_{\perp}=0$ ), from the on-shellness of the outgoing
quarks it can be seen that $\alpha_{Q}$ and $\beta_{Q}$ can be
neglected in further calculations.
 According to the Cutkosky rule the amplitude should be squared and integrated over all internal
momenta. If the integration is performed in terms of Sudakov
parameters it reads \bea \label{sudak-jacob} d^4k \; d^4l \; d^4m
=(\frac{s}{2})^3 d \alpha_k \; d \beta_k   \; d \alpha_l \; d
\beta_l  \; d \alpha_m \; d \beta_m  \; d^2 k_\perp \; d^2 l_\perp
\; d^2 m_\perp  \eea
The integration over transverse momenta  do not
bring the leading logarithm of energy and thus may be omitted in our discussion of the origin of the
$\ln s$ contribution.

First we do the integration with respect to $\beta_k$ which gives
$\beta_k=-k^2_\perp/s$ and factor $-1/s$ from $\delta^4((p_1-k)^2)$
(see Eq. \ref{p2-delta} ). Next the integration over $\beta_l$ is
performed, this gives at $\beta_l=- l^2_\perp/s$ and factor $-1/s$
from $\delta^4((p_2-l)^2)$. The $\delta$-function of $(p_3+m)^2$
gives $\alpha_m=+m^2_\perp/s$ and factor $1/s$. For $t$-channel
gluons the propagators  can be written as follows

\bea \label{t-channel-gluons}
 k^2&=&\alpha_k\beta_k s-k_\perp^2=\alpha_k(\frac{-k^2_\perp}{s})s
 -k_\perp^2\simeq -k_\perp^2 \\
   m^2&=&\alpha_m\beta_m s -m_\perp^2 =\frac{m^2_\perp}{s}\beta_m s -m_\perp^2 \simeq -m_\perp^2 \\
 l^2&=&\alpha_l\beta_l s - l_\perp^2=\alpha_l(\frac{-l^2_\perp}{s}) s -
 l_\perp^2 \simeq - l_\perp^2
 \eea
Similarly one finds that $(k+Q)^2\simeq -k_\perp^2$, $(l-Q)^2\simeq
-l_\perp^2$ .

 After this we proceed to
calculation of  other denominators are given by
\begin{center}
\bea \label{k-m}  (k-m)^2&=&(\alpha_k-\alpha_m)(\beta_k-\beta_m)s-(k-m)^2_\perp= \\
& &
(\alpha_k-\frac{m^2_\perp}{s})(-\frac{k^2_\perp}{s}-\beta_m)s-(k-m)^2_\perp=
\alpha_k\beta_m-(k-m)^2_\perp \nonumber
 \eea
 \end{center}
and the integration with respect to $\alpha_k$ is done on the
complex plain adding to denominator of Eq. \ref{k-m} some small
$i\epsilon$. The contour integration gives a residue at
$\alpha_k=[(k-m)^2_\perp-i\epsilon]/\beta_m s$ and the
 factor $1 / [\beta_m s]^2$ before the amplitude
 ($\beta^2_m$ squared is because of the  propagator of gluon $m-k-Q$). Next,
 $(k-m+l)^2$ gives $\alpha_l=[(k-m+l)^2_\perp-i\epsilon]/\beta_m$ and another
 factor if $1/\beta_m s$.

 We are left with some function of only transverse variables times
 $d\beta_m/\beta^3_m$. If we notice that the two triple gluon
 vertices in the center of Fig. \ref{diag_6} result into $\beta^2_m$
 in the numerator (see Eq. \ref{trigluon3} ) canceling two powers of
 $\beta_m$, than we end up with $d\beta_m/\beta_m$ which brings
 the logarithm of the energy.


 We start with three types of diagrams of
quark-quark scattering contributing to the effective vertex (see
Fig. \ref{Lip1}- \ref{Lip4}). This result will be useful for our
further considerations. The amplitude depicted of Fig. \ref{Lip1}
reads as

\begin{eqnarray} \label{fig1expr1}
g^3\frac{\bar{u}(p_1)}{\sqrt{p_{1+}}}\gamma^\nu t^b
\frac{u(p_1-m)}{\sqrt{p_{1+}-m_+}} \frac{-g_{\mu\nu}}{m_+}
\frac{1}{D_1} \frac{\bar{u}(p_1-m)}{\sqrt{p_{1+}-m_+}}\gamma^\sigma
t^c \frac{u(p_1-m-q)}{\sqrt{p_{1+}-m_+-q_+}}
\frac{\bar{u}(p_3)}{\sqrt{p_{3-}}}\gamma^\mu t^b
\frac{1}{D_2}\frac{u(p_3+m)}{\sqrt{p_{3-}+m_-}} \epsilon^q_{\sigma}
\end{eqnarray}

where $D_1$ and $D_2$ are given by

\bea \label{fig1den1}
  D_1 \,\,&=&\,\,p_{1-}+p_{3-}-\frac{(p_1-m)^2_\perp}{p_{1+}-m_+}-\frac{m^2_\perp}{m_+}-p_{3-}\simeq
-\frac{m^2_\perp}{m_+} \nonumber \\
D_2 &=&  p_{1-}+p_{3-}-\frac{(p_1-m)^2_\perp}{p_{1+}-m_+}-\frac{(p_3+m)^2_\perp}{p_{3+}+m_+}=
 \\
& &
\frac{(p_1-m-q)^2_\perp}{p_{1+}-m_+-q_+}+\frac{(p_3-m)^2_\perp}{p_{3+}-m_+}
-\frac{(p_1-m)^2_\perp}{p_{1+}-m_+}-\frac{(p_3+m)^2_\perp}{p_{3+}+m_+}
\simeq +\frac{q^2_\perp}{q_+} \nonumber
 \eea
In estimates of $D_2$ we used the fact that initial light cone
energy equals final light cone energy at high energies. Summing Eq.(\ref{fig1expr1})
over quark polarizations (see Ref. \cite{LEPAGE}) we have contribution only from
plus and minus components of $\gamma^\nu$ and  $\gamma^\mu$
respectively. Namely,
 \bea \label{gammaplus}
\frac{\bar{u}(p)}{\sqrt{p_+}}\gamma^\mu
\frac{u(p-m)}{\sqrt{p_+-m_+}}= n^\mu \eea where $n^\mu$  is defined
such that $n^\mu k_\mu=k_-$ for an arbitrary 4-vector $k^\mu$. Now
we can rewrite Eq.{\ref{fig1expr1}} as
\beq \label{fig1expr3}
t^b
t^c \otimes t^b g^3
n_{+}  n_{-}  \frac{q_-}{q^2_{\perp}} \frac{-m_+}{m^2_{\perp}} \frac{-1}{m_+} \epsilon_-
 =t^b
t^c \otimes t^b g^3
2\frac{q_+}{q_{\perp}^2}\frac{1}{m_{\perp}^2}\epsilon_-=t^b t^c \otimes
t^b
 4\frac{1}{m_{\perp}^2}\frac{q_{\perp} \epsilon_{\perp}}{q_{\perp}^2}
 \eeq

Here we used the transverse condition \bea\label{trans} q^\mu
\epsilon^q_\mu=\frac{1}{2}q_+\epsilon^q_- - q_\perp
\epsilon^q_\perp=0 \; \; \; \longrightarrow \;\;\; \epsilon_-=2
\frac{q_\perp \epsilon^q_\perp}{q_+} \eea

One should consider another diagram depicted in Fig. \ref{Lip2}, but
it can be readily seen that both of its denominators are of the
order of $D_1 \,\approx\,\,D_2\,\,\approx\, m_-=m^2_\perp/m_+$. Thus it is has an additional $m_-$
suppression comparing to the amplitude of Fig. \ref{Lip1} which will not give the logarithmic
contribution $\propto Y=\ln s$.
Next we consider a diagram shown in Fig. \ref{Lip3} given by
\begin{eqnarray} \label{fig3expr1}
g^3 \frac{\bar{u}(p_1)}{\sqrt{p_{1+}}}\gamma^{\sigma} t^c
\frac{u(p_1-q)}{\sqrt{p_{1+}-q_+}} \frac{1}{D_1}
\frac{\bar{u}(p_1-q)}{\sqrt{p_{1+}-q_+}}\gamma^\nu t^b
\frac{u(p_1-m-q)}{\sqrt{p_{1+}-m_+-q_+}}
\frac{\bar{u}(p_2)}{\sqrt{p_{3-}}}\gamma^\mu t^b
\frac{u(p_2+m)}{\sqrt{p_{3-}+m_-}} \frac{1}{D_2}\epsilon^q_{\sigma}
\end{eqnarray}
where $D_1$ and $D_2$ are given by

\bea \label{fig3den}
  D_1\,\,&=&\,\,p_{1-}+p_{2-}-\frac{(p_1-q)^2_\perp}{p_{1+}-q_+}-\frac{q^2_\perp}{q_+}-p_{2-}\simeq
-\frac{q^2_\perp}{q_+} \nonumber \\
D_2  &=& p_{1-}+p_{2-}-\frac{(p_1-m-q)^2_\perp}{p_{1+}-m_+-q_+}-p_{2-}\simeq
-\frac{m^2_\perp}{m_+}
 \eea
 Summing over quark polarizations we get the resulting expression
 for the amplitude shown in Fig. \ref{Lip3}
 \bea \label{fig3expr2}
t^c t^b \otimes t^b g^3
n_+n_-\frac{-q_+}{q_\perp^2}\frac{-m_+}{m_\perp^2}\frac{-1}{m_+}\epsilon_-
 =-t^b
t^c \otimes t^b g^3
2\frac{q_+}{q_\perp^2}\frac{1}{m_\perp^2}\epsilon_-=
 -t^b
t^c \otimes t^b g^3
4\frac{1}{m_\perp^2}\frac{q_\perp\epsilon_\perp}{q_\perp^2}
 \eea
 We see that the contribution of Fig. \ref{Lip3} is equal to that
 of Fig. \ref{Lip1} but opposite in sign. We also notice that
 they enter with different color matrices $t^a t^b\otimes t^b$ and $t^b t^a \otimes t^b$
 respectively. Using the basic relation for Lie group generators $[t^a,t^b]=i f^{abc}t^c$
 we can write the contribution from the two diagrams as

\bea \label{fig3plus1expr1}
 if^{abc}t^c \otimes t^b g^3 4\frac{1}{m_\perp^2}\frac{q_\perp\epsilon_\perp}{q_\perp^2}
 \eea
 The last diagram to be considered is shown in Fig. \ref{Lip4}
 and given by
\bea \label{fig4expr1}
i f^{abc}t^c \otimes t^b g^3
\frac{\bar{u}(p_1)}{\sqrt{p_{1+}}}\gamma^\mu\frac{\bar{u}(p_{1}-k)}{\sqrt{p_{1+}-k_+}}\frac{1}{D_1}
\frac{-g_{\nu \nu '}}{k_+}\Gamma^{\nu ' \mu' \sigma}\frac{-g_{\mu
\mu '}}{m_+}\frac{1}{D_2}
\frac{\bar{u}(p_3)}{\sqrt{p_{3-}}}\gamma^\mu\frac{\bar{u}(p_3-m)}{\sqrt{p_{3-}-m_+}}
\epsilon^q_{\sigma} \eea where $\Gamma^{\nu ' \mu' \sigma}$ is a
regular triple gluon QCD vertex.
\FIGURE[h]{
\begin{minipage}{55mm}{
\centerline{\epsfig{file= 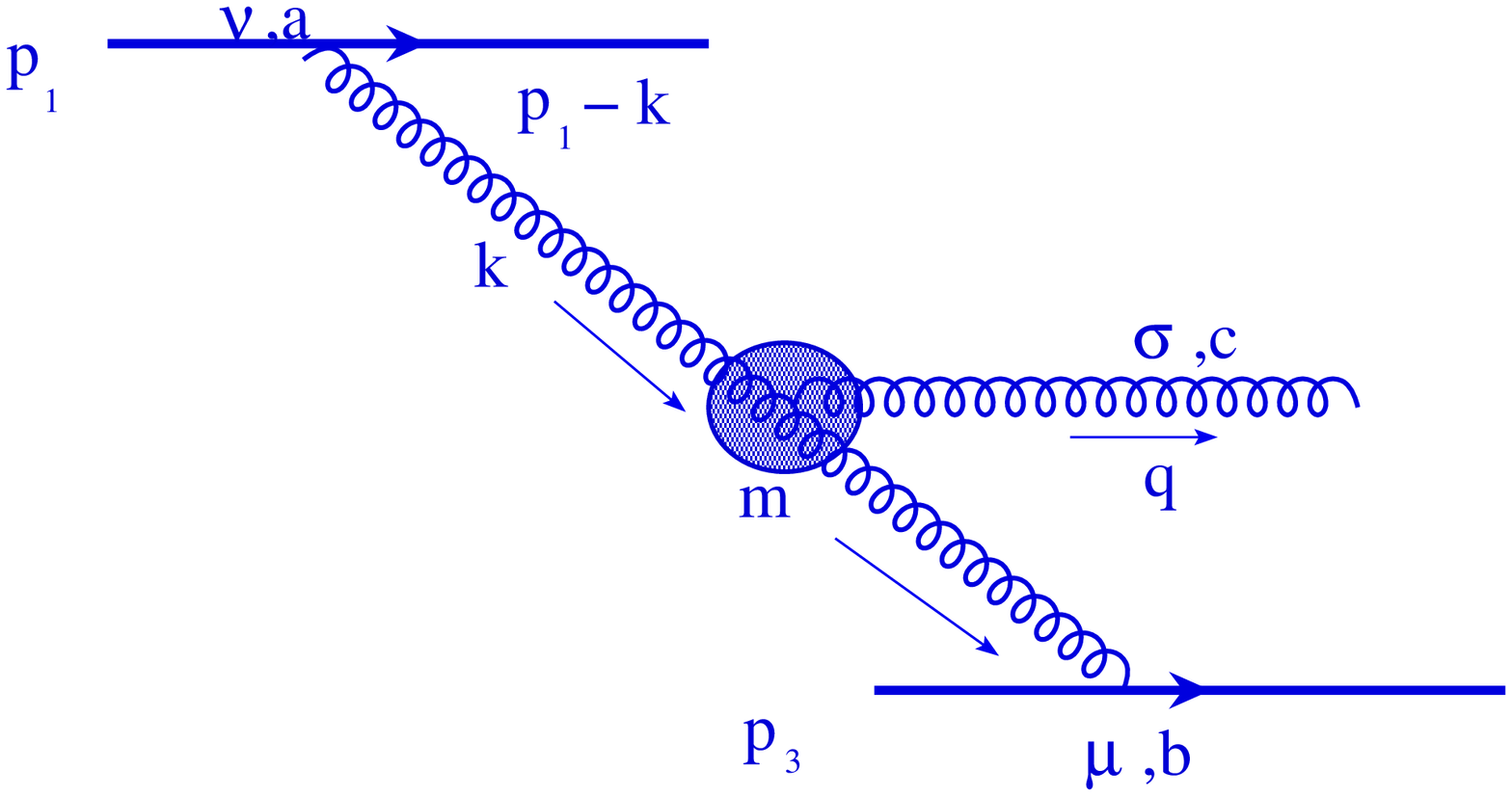,width=55mm}}
}
\end{minipage}
\caption{Schematic representation of the effective vertex. }
\label{Lip5}
}

 The light cone energy denominators $D_1$ and
$D_2$ are readily calculated
 \bea \label{fig4den}
 D_1&=&p_{1-}+p_{2-}-\frac{(p_1-k)^2_\perp}{p_{1+}-k_+}-\frac{k^2_\perp}{k_+}-p_{2-}\simeq
-\frac{k^2_\perp}{k_+} \nonumber \\
 D_2&=&p_{1-}+p_{2-}-\frac{(p_1-k)^2_\perp}{p_{1+}-k_+}
\frac{q^2_\perp}{q_+} -\frac{m^2_\perp}{m_+}-p_{2-}\simeq
-\frac{m^2_\perp}{m_+}
 \eea

The triple gluon vertex can be reduced as follows (see \cite{GLR})
\bea \label{trigluon}  \Gamma^{\nu ' \mu' \sigma} &=& g[g^{\nu ' \mu
'}(k+m)^\sigma+g^{\mu ' \sigma}(-m+q)^{\nu '}+g^{\nu'
\sigma}(-q-k)^{\mu '}] \nonumber \\ & =& [g^{\nu ' \mu
'}2k^\sigma+g^{\mu ' \sigma}(-m+q)^{\nu '}-g^{\nu' \sigma}2k^{\mu
'}]\simeq g^{\nu ' \mu '}2k^\sigma  \eea
 With these simplifications
Eq. \ref{fig4expr1} can be written as follows

\beq \label{fig4expr2}
 i f^{abc}t^c \otimes t^b g^3 n_+n_-\frac{-k_+}{k_\perp^2}\frac{-m_+}{m_\perp^2}\frac{-1}{m_+}
\frac{-1}{k_+}2k^\sigma
 \epsilon^q_\sigma= i f^{abc}t^c \otimes t^b g^3 4\frac{1}{m_\perp^2}
\frac{k_\perp\epsilon_\perp}{k_\perp^2}
 \eeq

 The resulting contribution from diagrams Figs.
 \ref{Lip1}-\ref{Lip4} is schematically shown in
 Fig.\ref{Lip5} and reproduces the effective Lipatov vertex
 \beq \label{effvert}
 i f^{abc}t^c \otimes t^b g^3 4\frac{1}{m_\perp^2}
\left(\frac{k_\perp\epsilon_\perp}{k_\perp^2}-\frac{q_\perp\epsilon_\perp}{q_\perp^2}\right)
 \eeq

We continue with the light cone perturbation theory and consider a
second quark emitting gluon $l$ that hits gluon $q$, as shown in
Fig. \ref{diag_5}. It is worth mentioning that the gluons $l$ and
$q$ should be coupled to the quark $1$ and quark $2$ lines via the
effective vertex found before. For simplicity, we consider only one
type of the diagrams contributing in this case and generalize the
result to all the possible couplings. There exists  a lot of ways to perform
the light cone energy cut when one adds second scattering quark. As
it was shown before that cuts double the same lower gluon are
suppressed in Regge limit. The amplitude shown in Fig. \ref{diag_5}
is written as follows
\FIGURE[hb]{\centerline{\epsfig{file=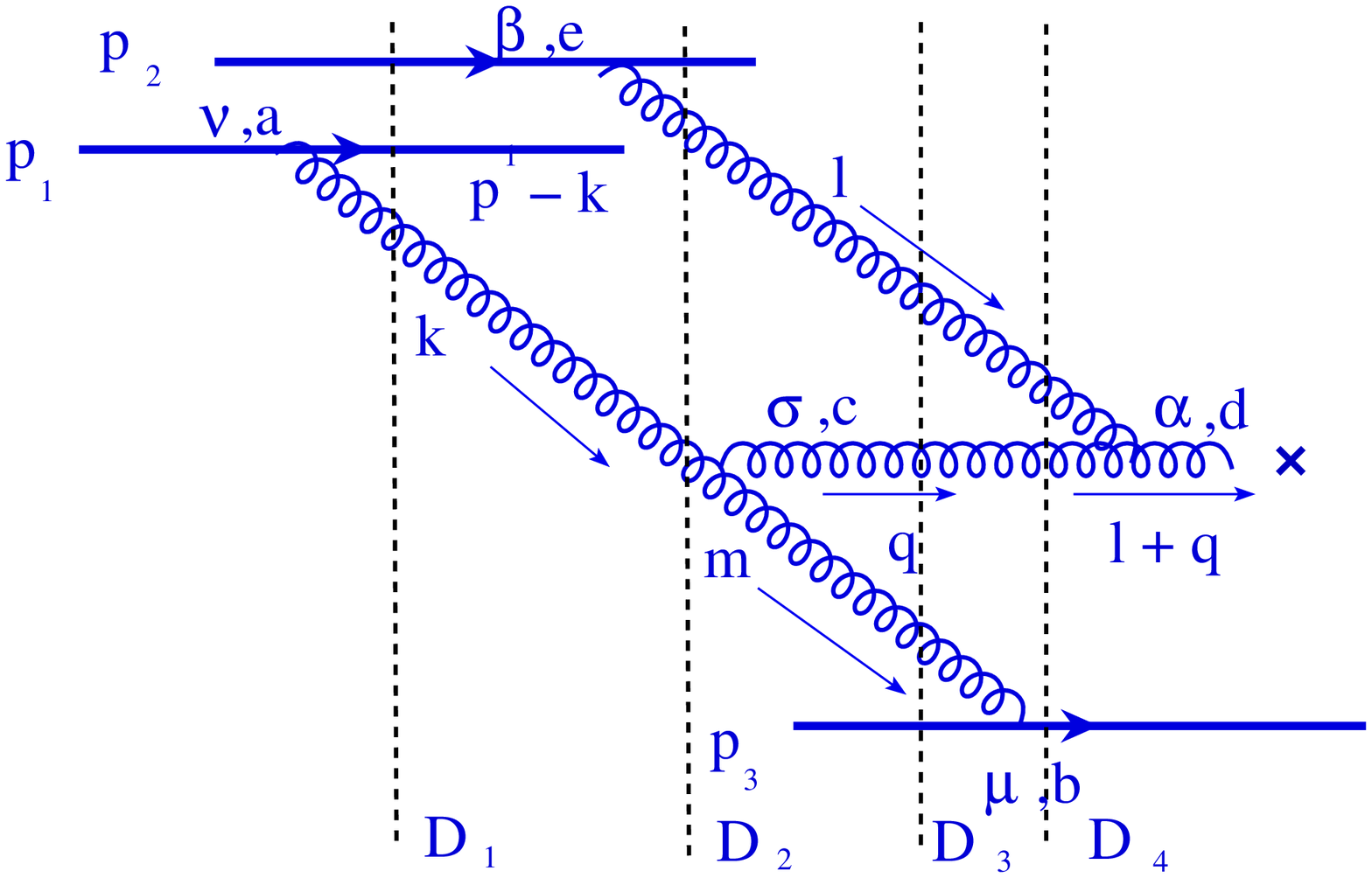,width=130mm}}
  \caption{One of the diagrams contributing to 2P $\rightarrow$ P  vertex calculated in
light cone perturbation theory}
\label{diag_5}
}

\bea \label{fig5exp1}
&& \frac{\bar{u}(p_1)}{\sqrt{p_{1+}}}\gamma^\nu
   \frac{u(p_1-k)}{\sqrt{p_{1+}-k_+}}  \frac{-g_{\nu \nu '}}{k_+}\frac{1}{D_1}
   \frac{\bar{u}(p_2)}{\sqrt{p_{2+}}}\gamma^\mu
      \frac{u(p_2-l)}{\sqrt{p_{2+}-l_+}}  \frac{-g_{\beta \beta '}}{l_+}\frac{1}{D_2}
  \nonumber \\
&& \Gamma^{\nu ' \mu ' \sigma}\frac{1}{D_3}\frac{\bar{u}(p_3)}{\sqrt{p_{3+}}}\gamma^\beta
   \frac{-g_{\mu \mu '}}{m_+}
   \frac{u(p_3-m)}{\sqrt{p_{3-}-m_-}}\frac{-g_{\sigma \sigma '}}{q_+}
   \frac{1}{D_4}
   \Gamma^{\sigma ' \beta ' \alpha} \epsilon^{q+l}_{\alpha}
   \eea

where the denominators $D_1$-$D_4$ are given by

\bea \label{fig5den}
D_1&\,\,=\,\,&p_{1-}+p_{2-}+p_{3-}-\frac{(p_1-k)^2_\perp}{p_{1+}-k_+}-\frac{k^2_\perp}{k_+}-p_{2-}-p_{3-}\simeq
-\frac{k^2_\perp}{k_+} \nonumber \\
D_2 &=& p_{1-}+p_{2-}+p_{3-}-\frac{(p_1-k)^2_\perp}{p_{1+}-k_+}-\frac{k^2_\perp}{k_+}
-\frac{(p_2-l)^2_\perp}{p_{2+}-l_+}-\frac{l^2_\perp}{l_+}-p_{3-}\simeq
-\frac{l^2_\perp}{l_+}\\
D_3
&=&p_{1-}+p_{2-}+p_{3-}-\frac{(p_1-k)^2_\perp}{p_{1+}-k_+}-\frac{q^2_\perp}{q_+}-\frac{m^2_\perp}{m_+}
-\frac{(p_2-l)^2_\perp}{p_{2+}-l_+}-\frac{l^2_\perp}{l_+}-p_{3-}\simeq
-\frac{m^2_\perp}{m_+} \nonumber \\
 D_4 &=& p_{1-}+p_{2-}+p_{3-}-\frac{(p_1-k)^2_\perp}{p_{1+}-k_+}-q_-
-\frac{(p_2-l)^2_\perp}{p_{2+}-l_+}-\frac{l^2_\perp}{l_+}-\frac{(p_3-m)^2_\perp}{p_{3+}-m_+}\simeq
-q_-\nonumber
 \eea
In Eq. \ref{fig5exp1} we omitted coupling constants and color
factors that will be easily restored at the end of the calculations.
 The amplitude Eq. \ref{fig5exp1} can be further
simplified as follows
\beq \label{fig5exp2}
 n^\nu
     \frac{-g_{\nu \nu '}}{k_+}\frac{-k_+}{k^2_\perp}
  n^\mu  \frac{-g_{\beta \beta '}}{l_+}\frac{-l_+}{l^2_\perp}
   \Gamma^{\nu ' \sigma
   \alpha'}\frac{-m_+}{m^2_\perp}
   \frac{-g_{\mu \mu '}}{m_+}
   \tilde{n}^\beta
   \frac{-g^{\sigma \sigma '}}{q_+ q_-}
   \Gamma^{\sigma ' \beta ' \alpha} \epsilon^{q+l}_{\alpha}
   \eeq
   where vector $\tilde{n}^\beta$ is defined as   $\tilde{n}^\beta
   k_\beta=k_-$ for any arbitrary vector $k^\beta$. The triple gluon
   QCD vertex $ \Gamma^{\nu ' \sigma
   \alpha'}$ is given in  Eq. \ref{trigluon}. The  vertex $\Gamma^{\sigma ' \beta '
   \alpha}$ in our case can be simplified as follows
   \bea \label{trigluon3} 
   \Gamma^{\sigma ' \beta '
   \alpha}=g^{\alpha \beta'} (l-q)^{\sigma'}+g^{\alpha \sigma'} (2q +l)^{\beta'}+ g^{\beta' \sigma'}(-q-2l)^\alpha
   \simeq  g^{\alpha \sigma'}(2q+l)^{\beta'}
   \eea
Using this we rewrite Eq. \ref{fig5exp2} as
 \bea \label{fig5exp3}
  +n^2 2\frac{1}{k^2_\perp}
    \frac{1}{l^2_\perp}
   \frac{1}{m^2_\perp}
   \frac{(2q+l)_- }{q_+ q_-}k^\alpha \epsilon^{q+l}_{\alpha}\simeq -2 \cdot 2\frac{1}{k^2_\perp}
    \frac{1}{l^2_\perp}
   \frac{1}{m^2_\perp}
   \frac{2}{q_+}k_\perp \cdot \epsilon^{q+l}_{\perp}
   \eea
   We note a plus component of $q$ in the denominator, this factor
   is not related to integration in transverse coordinates and leads
   to a logarithm of energy as was shown before for total cross
   section, thus it can be omitted at this stage and safely restored
   at the end of the calculations.
   Performing similar transformations for an amplitude of two gluons being emitted from quark (antiquark) lines
   (see Fig. \ref{Lip1}- \ref{Lip3}) we get
   \bea \label{fig5exp4}
  +2 \cdot 2\frac{1}{q^2_\perp}
    \frac{1}{l^2_\perp}
   \frac{1}{m^2_\perp}
  2q_\perp \cdot \epsilon^{q+l}_{\perp}
   \eea
   with the same color factor as of Eq. \ref{fig5exp3}.Thus, Fig. \ref{diag_5} gives
   \bea \label{fig5exp5}
  8\frac{1}{l^2_\perp}
   \frac{1}{m^2_\perp}
   \left(\frac{k_\perp^\alpha}{k_\perp^2}-\frac{q_\perp^\alpha}{q_\perp^2} \right)
   \eea
    One should add another one, where momenta $q$
   and $l$ are interchanged,    to this expression, then it gives an effective vertex
   for the scattering of two with one quarks (with color factor, powers of $g$ and plus-minus
components restored). Now we want to find the
   cross section that corresponds to two dipoles being scattered with
   a target. We use a mixed representation where we Fourier transform
   all transverse momenta to coordinates of dipoles (quarks). We
   assign transverse coordinates $x_i$ ($y_i$) to each
   quark (antiquark) line, and coordinate $z$ for the emitted gluon.
 The Fourier transform is performed separately for $d^2 k_\perp$ and $d^2 q_\perp$ integrations . We start
 with Eq. \ref{fig5exp3}
 \bea \label{fig5exp6}
 \int   \frac{d^2 k_\perp}{(2 \pi)^2} \frac{d^2 l_\perp}{(2 \pi)^2} \frac{d^2 m_\perp}{(2
 \pi)^2} &~&  e^{i (k-m+l)_\perp z}  \left(e^{i m_\perp x_3}-e^{i m_\perp y_3}\right)
 \left(e^{-i l_\perp x_2}-e^{-i l_\perp y_2}\right) \times \nonumber \\
 &\times &  \left(e^{-i k_\perp x_1}-e^{-i k_\perp y_1}\right)8\frac{1}{l^2_\perp}
   \frac{1}{m^2_\perp}
      \frac{k_\perp^\alpha}{k^2_\perp}
           \eea
 To simplify Eq. \ref{fig5exp6} we introduce the gluon propagator in
 the coordinate space
\beq \label{fig5exp7}
 D_G(x,x')= \int \frac{d^2 k_\perp}{(2 \pi)^2}
\frac{1}{k^2_\perp} e^{ik_\perp(x-x')}=-\frac{\ln\left(|x-x'|\mu\right)}{2\pi}
 \eeq
where $\mu$ is an arbitrary small mass that we need to send to $0$ at the end of our calculations,
 and the integral \beq
\label{fig5exp8} \int \frac{d^2 k_\perp}{(2 \pi)^2}
\frac{k_\perp^\alpha}{k^2_\perp}
e^{ik_\perp(x-x')}=\frac{i}{2\pi}\frac{(x-x')^\alpha}{(x-x')^2} \eeq
which will be interpreted later.
 In terms of Eq. \ref{fig5exp7}
 and Eq. \ref{fig5exp8} we can rewrite Eq. \ref{fig5exp6} as follows
\beq \label{fig5exp9} \frac{i}{2\pi}
\left[\frac{(z-x_1)^\alpha}{(z-x_1)^2}
-\frac{(z-y_1)^\alpha}{(z-y_1)^2}\right]\left(D_G(z,x_3)-D_G(z,y_3)\right)\left(D_G(z,x_2)-D_G(z,y_2)
\right)\eeq
In a similar way we rewrite Eq. \ref{fig5exp4} in the form of \bea
\label{fig5exp10}
 &&
 \int   \frac{d^2 q_\perp}{(2 \pi)^2} \frac{d^2 l_\perp}{(2 \pi)^2} \frac{d^2 m_\perp}{(2
 \pi)^2} e^{i (q+l)_\perp z}  \left(e^{i m_\perp x_3}-e^{i m_\perp y_3}\right)
 \left(e^{-i l_\perp x_2}-e^{-i l_\perp y_2}\right) \times \nonumber \\
 &&  \left(e^{-i q_\perp x_1}-e^{-i q_\perp y_1}\right)
 \left(e^{-i m_\perp x_1}-e^{-i m_\perp y_1}\right)8\frac{1}{l^2_\perp}
   \frac{1}{m^2_\perp}
      \frac{-q_\perp^\alpha}{q^2_\perp}=
      -\frac{i}{2\pi}
\left[\frac{(z-x_1)^\alpha}{(z-x_1)^2}
-\frac{(z-y_1)^\alpha}{(z-y_1)^2}\right]  \times \nonumber \\
&& \left(D_G(x_3,x_1)-D_G(y_3,x_1)-D_G(x_3,y_1)+D_G(y_3,y_1)\right)
                 \left(D_G(z,x_2)-D_G(z,y_2)\right)\eea
Eqs. \ref{fig5exp9}-\ref{fig5exp10} should be multiplied by the
wavefunctions of the original dipoles  $\psi^{(0)}(x_i, y_i)$, their
factors have transparent physical meaning, namely \bea
\label{fig5exp11} && \left[\frac{(z-x_1)^\alpha}{(z-x_1)^2}
-\frac{(z-y_1)^\alpha}{(z-y_1)^2}\right]  \hspace{1cm}
\mbox{describes the emission of gluon $z$ by dipole $(x_1,y_1)$ }
\nonumber \\   &&
\\ && \left(D_G(z,x_3)-D_G(z,y_3)\right) \hspace{1cm}
\mbox{describes the interaction of gluon $z$ with target dipole
$(x_3,y_3)$ }\nonumber
 \eea
It should be noted that  Eq. \ref{fig5exp9} should be multiplied by $2$
before adding to Eq. \ref{fig5exp10}. This could be explained as
follows, let us go back from large $N_c$ limit to $N_c=3$ and add a
color singlet corresponding to large $N_c$ limit. We want to
consider a combinatorial  factor for all possible emissions of two
gluons from dipole $(x_1,y_1)$ (see \fig{Col1}- \fig{Col4}) fixing
the colour in the final states. It is clearly seen from the picture that the upper
figure that corresponds to Eq. \ref{fig5exp3}, has as much as twice
options for two gluons to be
emitted.

\DOUBLEFIGURE[ht]{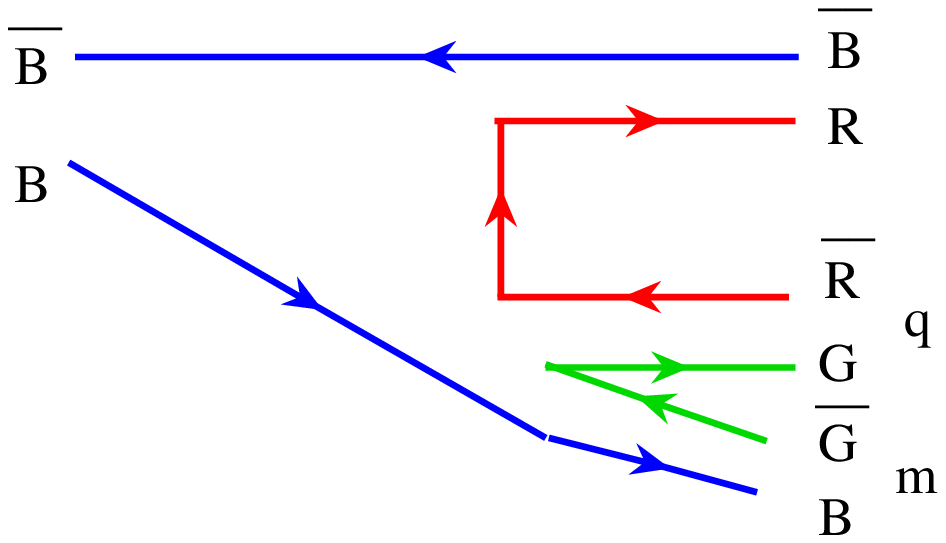,width=75mm}{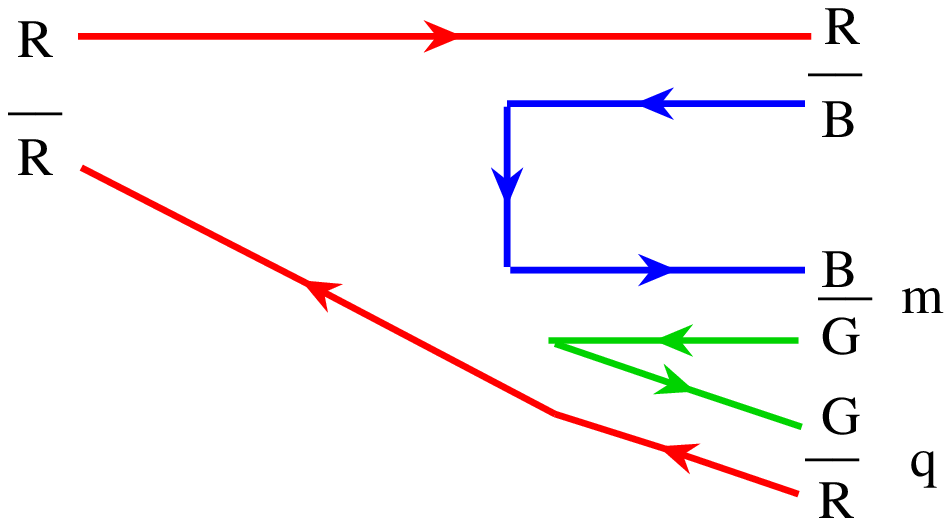,width=75mm}
{Colour structure of the reaction $q(blue) +
\bar{q}(antiblue)\,\rightarrow\,q \bar{q} + G(q + m) \rightarrow q
\bar{q} + G(q) + G(m)$. $ B,G,R(\bar{B},\bar{G},\bar{R})$  stand for  blue, green and red quarks (antiquarks), respectively. \label{Col1}}{Colour structure of the reaction $q(red) +
\bar{q}(antired)\,\rightarrow\,q \bar{q} + G(q + m) \rightarrow q
\bar{q} + G(q) + G(m)$.$ B,G,R(\bar{B},\bar{G},\bar{R})$ stand for  blue, green and red quarks (antiquarks), respectively.label{Col2}}

\DOUBLEFIGURE[ht]{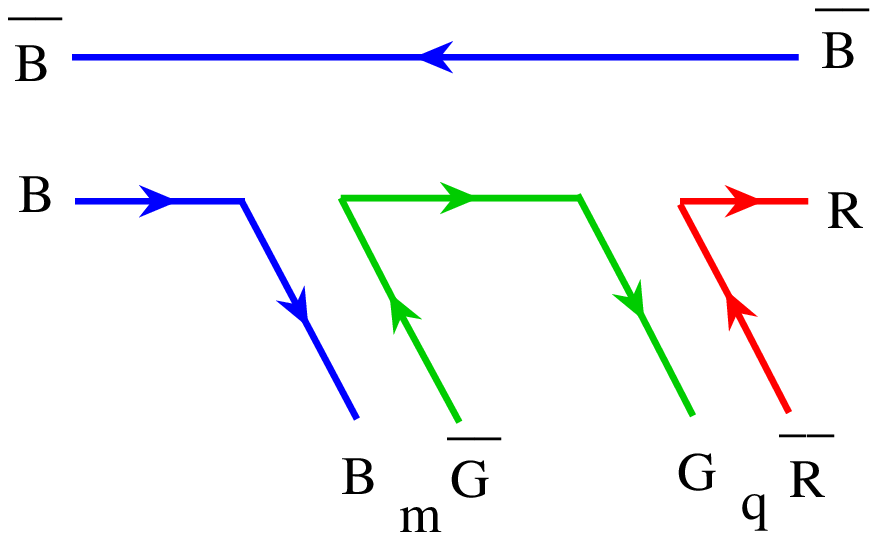,width=75mm}{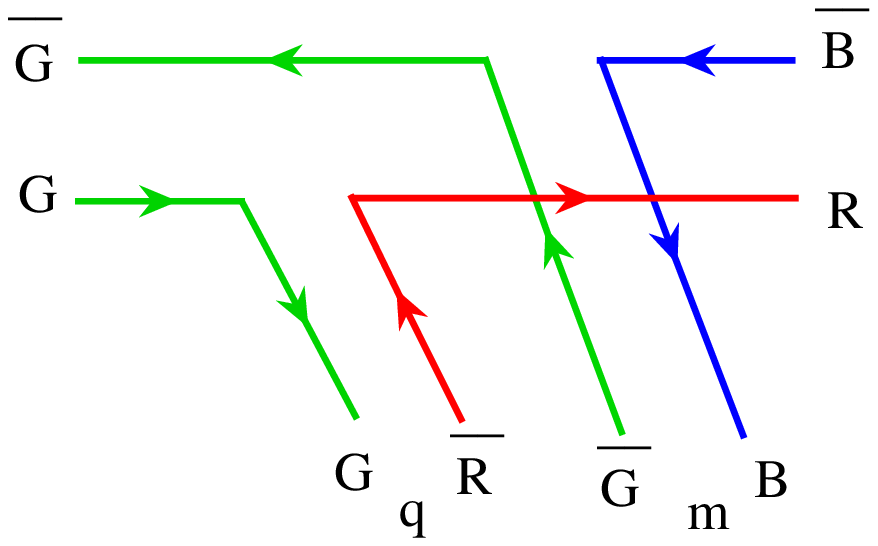,width=75mm}
{Colour structure of the reaction $q(blue) +
\bar{q}(antiblue)\,\rightarrow\, q
\bar{q} + G(q) + G(m)$. $ B,G,R(\bar{B},\bar{G} B,G,R(\bar{B},\bar{G},\bar{R}) $ stand for  blue, green and red quarks (antiquarks), respectively.\label{Col3}}{Colour structure of the reaction $q(green) +
\bar{q}(antigreen)\,\rightarrow\, q
\bar{q} + G(q) + G(m)$.  $ B,G,R(\bar{B},\bar{G},\bar{R})$  stand for  blue, green and red quarks (antiquarks), respectively.\label{Col4}}

Using this fact we may add Eq. \ref{fig5exp9} to Eq. \ref{fig5exp10}
\beq \label{fig5exp12}
 -\frac{i}{2\pi}
\left[\frac{(z-x_1)^\alpha}{(z-x_1)^2}
-\frac{(z-y_1)^\alpha}{(z-y_1)^2}\right]
\left(D_G(z,x_1;x_3,y_3)+D_G(z,y_1;x_3,y_3)\right)
                 \left(D_G(z,x_2)-D_G(z,y_2)\right)
\eeq
 where $D_G(z,x_1;x_3,y_3)$ stands for interaction of dipole
$(z,x_1)$ with the target. One should also add a term similar to Eq.
\ref{fig5exp12} with dipole $1$ interchanged with dipole $2$
($x_1,y_1\leftrightarrow x_2,y_2$). Now we are in position to square
the summed diagrams. It can be easily shown that all types of
interference terms vanish after the integration over $z$. Thus we
are left with each term squared and the final expression reads as
\bea
\label{fig5exp13}
&& |\psi^{(0)}(x_1,y_1)|^2
|\psi^{(0)}(x_2,y_2)|^2 |\psi^{(0)}(x_3,y_3)|^2 \frac{\alpha_s
N_c}{2 \pi^2} \times\\ && \int_0^Y dY'\int d^2z \{2\gamma
^{BA}(x_2,y_2|z)K(x_1,y_1|z)[\gamma^{BA}(x_1,z|x_3,y_3)+\gamma^{BA}(y_1,z|x_3,y_3)]
+(1\leftrightarrow 2) \} \nonumber
\eea where
$\gamma^{BA}(x,y|x',y')=(\frac{\alpha_s}{4\pi^2})^2G_0(x,y|x',y')$
with $G_0$ being the initial Green function given by Eq. \ref{G0},
and $K(x,y|z)$ is given by Eq. \ref{K}. $\gamma^{BA}(x_2,y_2|z)$
describes the interaction of a gluon with coordinate $z$ with dipole
$(x_2,y_2)$ (see Fig. \ref{amlfig}). It is equal to
 \beq \label{GAMMAZ}
\gamma^{BA}(x_1,y_1|z)=\frac{1}{8}\left(\frac{\alpha_s}{2\pi}\right)^2
\ln^2 \left[\frac{(x_1-z)^2}{(y_1-z)^2}\right] \eeq
\eq{fig5exp13} should be multiplied by the factor $\ln s$ which is originated from the integration
over the energy of produced gluon with the coordinate $z$.

One can see from \eq{fig5exp12} that we obtain the kernel of the BFKL equation for the interaction
of one dipole $(x_1,y_1)$ in \fig{v21pic}-d with the target dipole $(x,y)$ (see \fig{v21pic}-a).
Indeed, neglecting factor $D_G(x_2,z) - D_G(z,y_2)$ in \eq{fig5exp12} we have the following answer
taking derivative with respect to $Y = \ln s$ from $|A(\eq{fig5exp12})|^2$
\bea \label{A2}
\frac{d |A(\eq{fig5exp12})|^2}{d Y}&\,\,=\,\,& \\
 &= &\, \frac{\bas}{2 \pi}\int\,\,d^2 z\,|\Psi_0(x_2,y_2)|^2
\,K\left(x_1,y_1|z \right)\,\,\left(\,\gamma^{BA}( x_1,z | x_3,y_3)\,\,+\,\,\gamma^{BA}(z,y_1 |
x_3,y_3)\,\right)\nonumber
\eea
In \eq{A2} one can recognize the BFKL equation with the kernel $\frac{\bas}{2
\pi}\,\,K\left(x_1,y_1|z \right)$.

\subsection{Contribution to the generating functional}

 \eq{fig5exp13} gives the expression for the transition
$(x_1,y_1) \,+\,(x_2,y_2)\,\,\rightarrow\,\,(x_1,y_2)\,+\,(x_1,z)\,+\,(z,y_1)$ given by the diagrams
of \fig{v21pic}-c and \fig{v21pic}-d. This diagram gives a positive contribution but the sum of all
diagrams (see \fig{v21pic}, \fig{v211pic} and \fig{v212pic})  leads to a negative contribution due to
the AGK cutting rules (see \eq{AGK}). The advantages of this direct calculations are clear: first, we
obtain the vertex for $2 \to 1$ transition in more convenient form than it was calculated before (see
Refs.
\cite{IT,MSW,L3}). From \eq{fig5exp13} we see that this vertex is equal to
\beq \label{V21N}
\Gamma_{2 \to 1} \left( (x_1,y_1)  + (x_2,y_2) \to (x,y) \right) \,\,\,=\,\,\frac{\as N_c}{2
\pi^2}\,\,\tilde{K}\left(x_1,y_1;x_2,y_2|z \right)\,\,\,=\,\,
\eeq
$$
\,\,\frac{\as N_c}{2
\pi^2}\,\,\left\{ 2\,\gamma^{BA}\left(x_1,y_1|z \right)  \times
K\left(x_2,y_2;z \right) \,\,+\,\, x_1
\leftrightarrow x_2,  y_1
\leftrightarrow y_2 \right\}
$$
It should be stressed that \eq{V21N} leads to a vertex which is positive in the full phase space.

 The  second advantage is that  we can understand what process in terms of dipole these diagrams
describe. Indeed, at initial rapidity $Y$ we have two colour dipoles:$(x_1,y_1)$ and $(x_2,y_2)$.
The diagram of \fig{amlfig}-a  describes the transition of these two intitial dipoles which goes in two
initial
\beq \label{STAGE}
(x_1,y_1)
\,+\,(x_2,y_2)\,\,\rightarrow\,\,(x_1,y_1)\,+\,(x_2,z)\,+\,(z,y_2)\,\, \rightarrow
\,\,(x_1,y_2)\,+\,(x_2,z)\,+\,(z,y_1);
\eeq
and the target dipole $(x,y)$
interacts with dipoles: $(x_2,z)$ and $(z,y_2)$ during the first stage of the process  or with
$(x_2,z)$ and $(z,y_2)$ in the last stage.

Therefore, in terms of the functional of \eq{N} these diagrams can be written as
\beq \label{V21FN}
- \,\h\,\Gamma_{2 \to 1} \left( \eq{V21N} \right)\,\left( \gamma(x_1,z) \,\,+\,\,\gamma(z,y_1)
\,\,+\,\, \gamma(x_2,z)\,\,+\,\,\gamma(z,y_2) \right)\,\frac{\delta}{\delta
\gamma(x_1,y_1)}\,\frac{\delta}{\delta \gamma(x_2,y_2)}
\eeq
The form of \eq{V21FN} suggests that this term can be originated by the following term in the
Hamiltonian $\chi[u]$  (see \eq{ZEQ} and \eq{chi}) :
\beq \label{V21FZ}
\h\,\Gamma_{2 \to 1} \left( \eq{V21N}
\right)\,\left[u(x_1,z)\,u(z,y_1)\,\,+\,\,u(x_2,z)\,u(z,y_2) \right]\,\,\frac{\delta}{\delta
u(x_1,y_1)}\,\frac{\delta}{\delta u(x_2,y_2)}
\eeq
The Feynman diagrams that correspond to this term are shown in \fig{feymdi}-b and \fig{feymdi}-c.
However, \eq{V21FZ} cannot be correct since it leads to a positive  two dipoles to two dipole
amplitude.    Indeed, for  two dipoles to two dipole
amplitude we have the same three diagrams of \fig{v21pic}, \fig{v211pic} and  \fig{v212pic}
types. The sum of these diagrams gives the negative sign (see Ref. \cite{BLAGK} where this problem was
studied in  details ).

\FIGURE[ht]{
\centerline{\epsfig{file=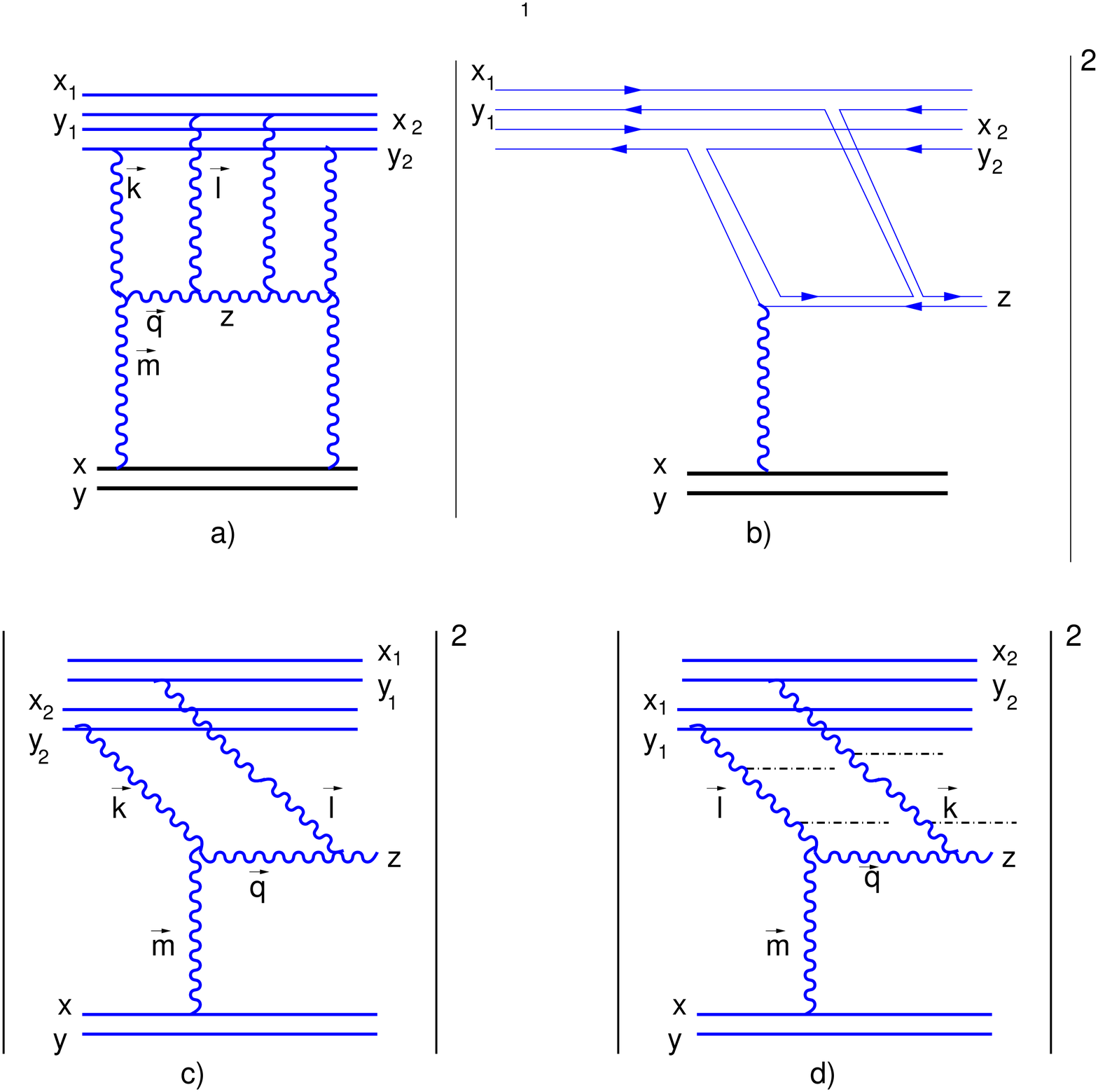,width=120mm}}
\caption{The diagram of the 2 Pomeron into one Pomeron transition and its dipole interpretation:
diagonal contribution.}
\label{v21pic}}

\FIGURE[ht]{
\centerline{\epsfig{file=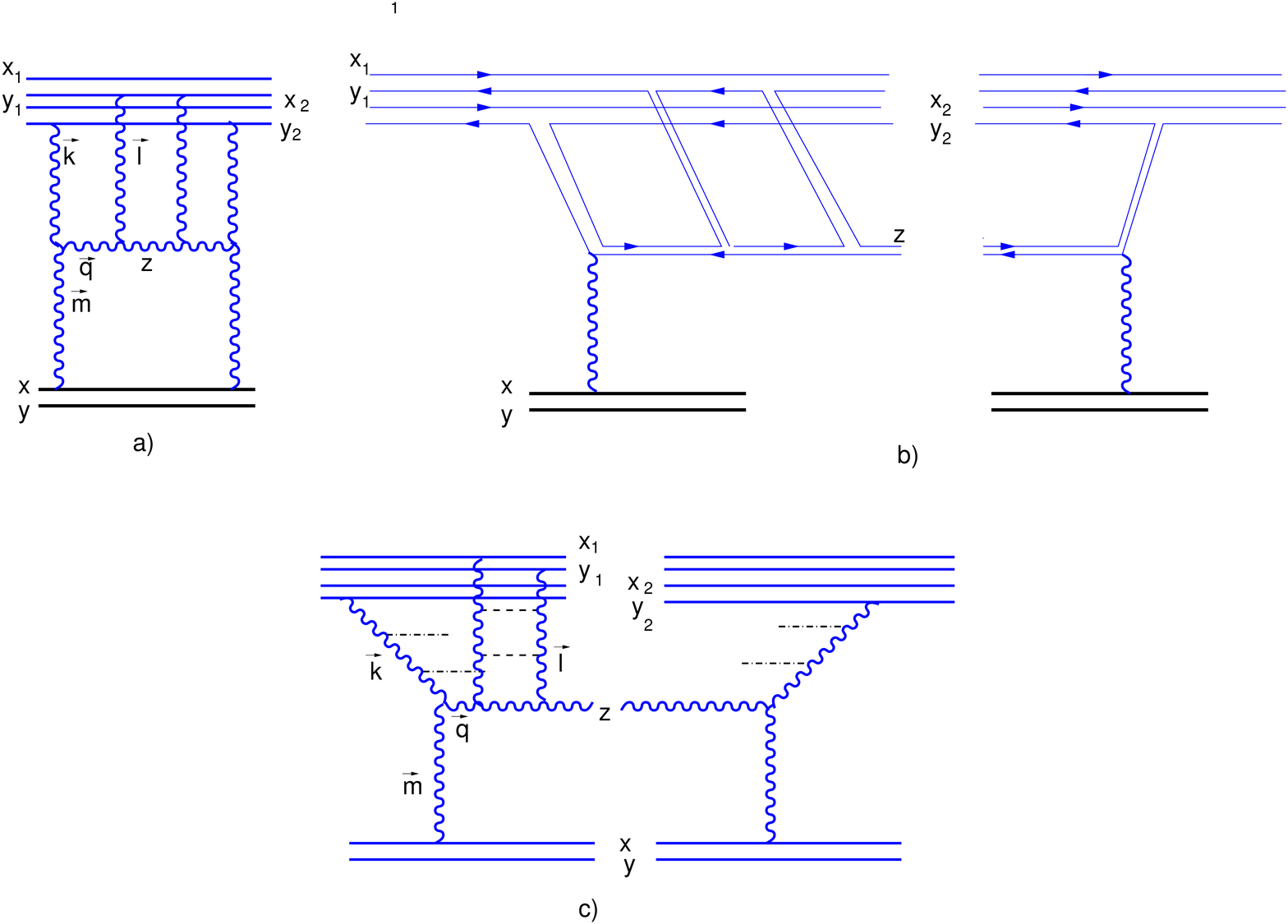,width=120mm}}
\caption{The diagram of the 2 Pomeron into one Pomeron transition and its dipole interpretation:
non-diagonal contribution with  elastic rescattering of the dipole $x_1,y_1$).}
\label{v211pic}}

\FIGURE[ht]{
\centerline{\epsfig{file=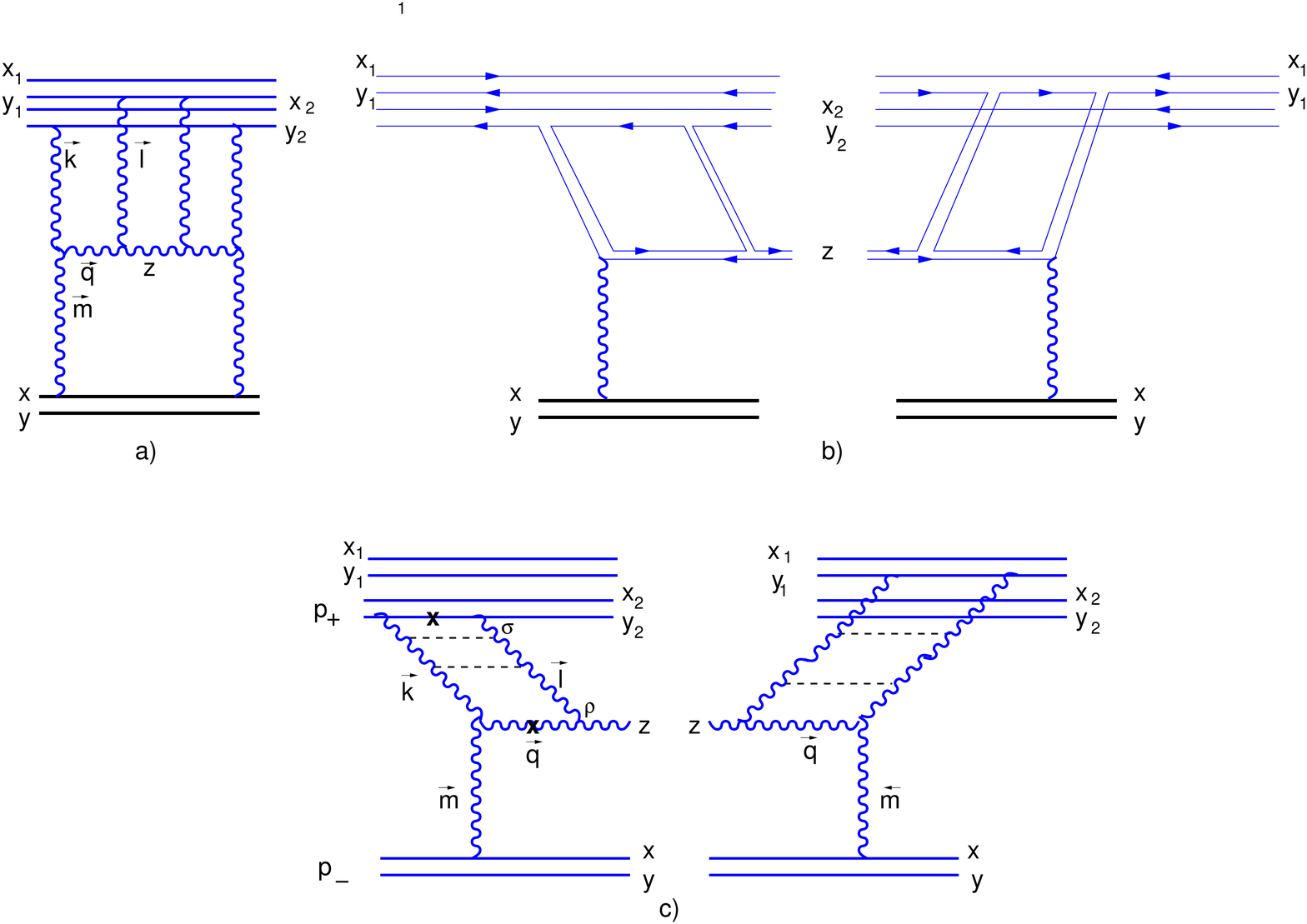,width=120mm}}
\caption{The diagram of the 2 Pomeron into one Pomeron transition and its dipole interpretation:
non-diagonal contribution with the interference between the decays of two dipoles.}
\label{v212pic}
}

Repeating the same calculation which we made, summing \fig{v21pic},\fig{v211pic} and \fig{v212pic},
one obtain for $2 \to  2$  dipole amplitude the following answer
\beq \label{V22N}
-\h \,\Gamma_{2 \to 1} \left( \eq{V21N} \right)\,\left\{ \gamma(x_1,z) (\gamma(z,y_1)
\,\,+\,\, \gamma(x_2,z)\,\,+\,\,\gamma(z,y_2)) \,+ \right.
\eeq
$$
\left.
\,\gamma(x_2,z) (\gamma(z,y_1)\,+\,\gamma(z,y_2))\,+\,
\gamma(z,y_1)\gamma(z,y_2) \right\}\,\frac{\delta}{\delta
\gamma(x_1,y_1)}\,\frac{\delta}{\delta \gamma(x_2,y_2)}
$$

\eq{V21N} and \eq{V22N} suggest that all dipoles that have been produced in both stages of the process
(see \eq{STAGE}) interact with the target. It means that the contribution to $ \chi[u]$ (see \eq{ZEQ}
and \eq{chi}) looks as follows
\beq \label{V21FZF}
-\h\,\Gamma_{2 \to 1} \left( \eq{V21N} \right)\,u(x_2,z)\,u(z,y_2)\,u(x_1,y_2)\,u(z,y_1)
\,\frac{\delta}{\delta u(x_1,y_1)}\,\frac{\delta}{\delta u(x_2,y_2)}
\eeq

\eq{V21FZF} describes the process of decay of two dipoles into four  dipoles\footnote{As far as we know
A. Kovner was the first who discussed the possibility that two Pomeron to one Pomeron merging is
related to the decay process in the dipole approach.}.   As one can see from \eq{P1} - \eq{P3} each
process for the dipole transition
generates two terms in Markov's chain which lead in \eq{P1} - \eq{P3} to  the positive contribution
which represents the increase in the number of dipoles due
to the elementary process ( birth term)and a negative contribution which describes a decrease of  the
number of
dipoles
as a result of the elementary process (death term).  The death term which corresponds to the birth
term given by  \eq{V21FZF} has the form
\beq \label{V21FZD}
-\h\,\int\,\,d^2 z\,\,\Gamma_{2 \to 1} \left( \eq{V21N}\right)\, u(x_1,y_1)\,\,u(x_2,y_2)\,
\,\frac{\delta}{\delta u(x_1,y_1)}\,\frac{\delta}{\delta u(x_2,y_2)}
\eeq
where
\beq \label{V21INT}
\int\,\,d^2 z\,\,\Gamma_{2 \to 1} \left( \eq{V21N}\right)\,\,=\,\,
\eeq
$$
\frac{\as N_c}{2
\pi^2}\,\,\left\{ 2\,\left( \gamma^{BA}\left(x_1,y_1|x_2 \right) \,,\,+\,\,\gamma^{BA}\left(x_1,y_1|y_2
\right) \right) \times
\int\, d^2 z\,\,K\left(x_2,y_2;z \right) \,\,+\,\, x_1
\leftrightarrow x_2,  y_1
\leftrightarrow y_2 \right\}
$$
It is easy to see that for the case when two dipoles are the same, \eq{V21FZD} describes the diagram of
\fig{death}

\FIGURE[ht]{
\centerline{\epsfig{file=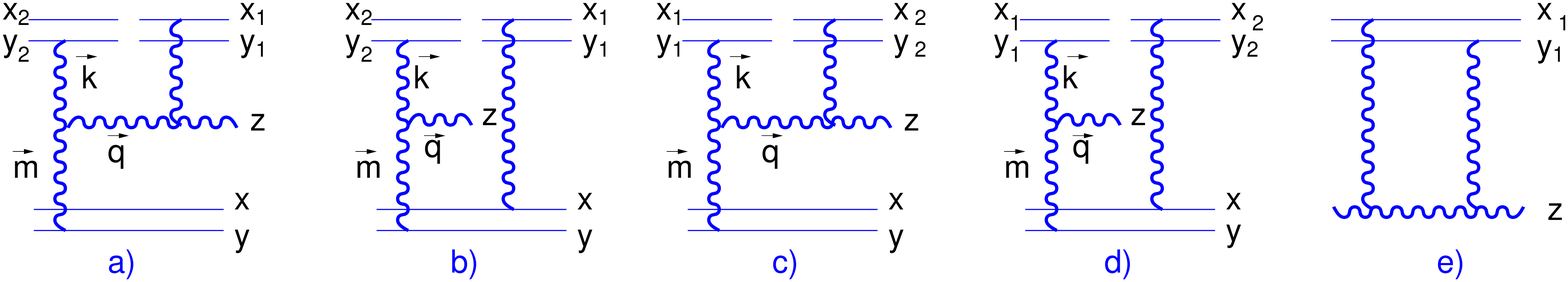,width=120mm}}
\caption{The diagrams for the amplitude of two dipole with one dipole interaction with emission of
one extra gluon.}
\label{amlfig}}

\FIGURE[ht]
{\centerline{\epsfig{file=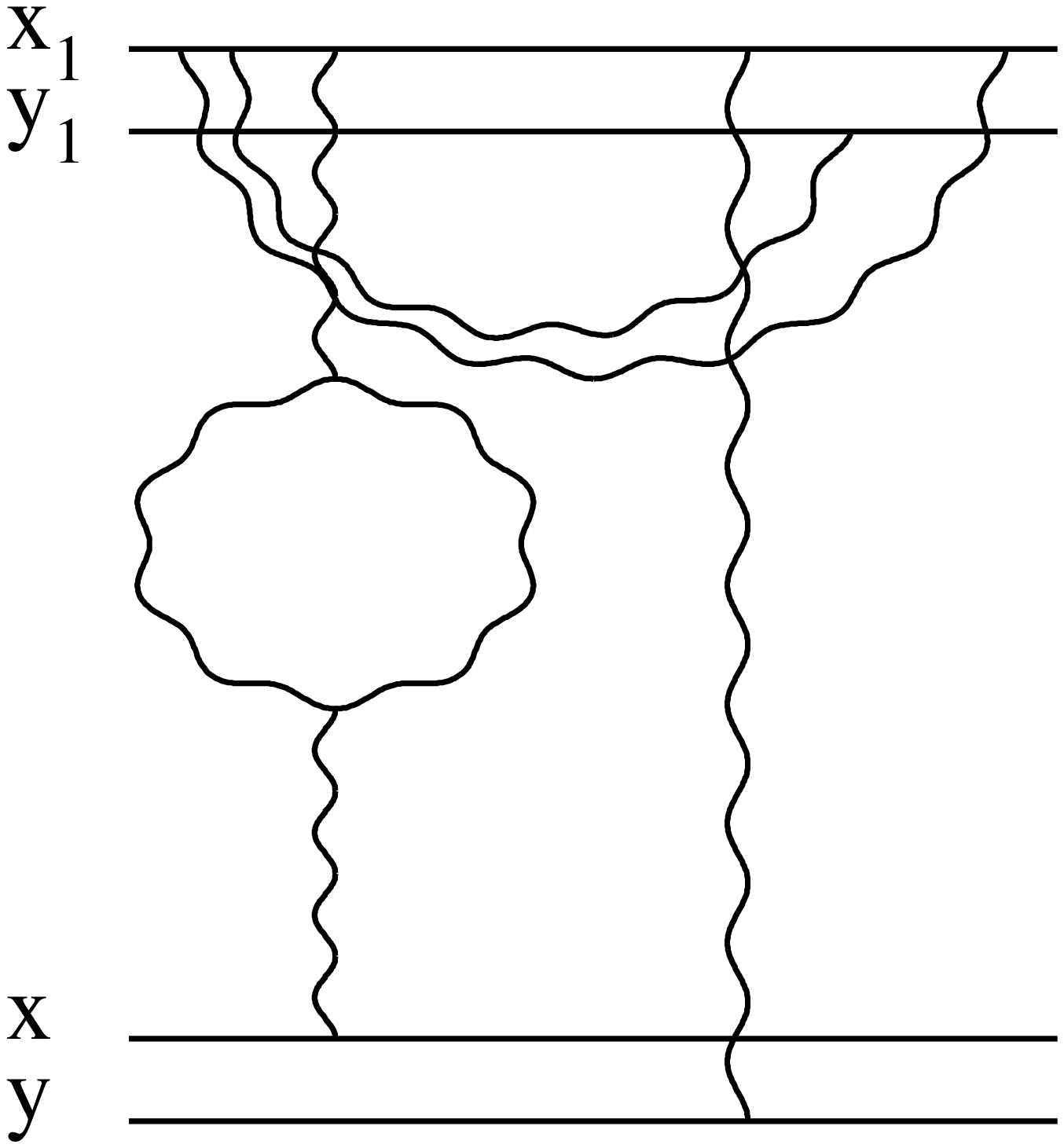,width=80mm}}
\caption{The diagrams for the amplitude which describes the death term in the equation for the
generating functional in the case when dipoles $(x_1,y_1)$ and $(x_2,y_2)$ are equal.
}
\label{death}}

Finally, the equation for $\chi[u]$ (see \eq{chi}) has the form
\bea
\chi[u]\,&=&\,\,\frac{\bas}{2 \pi}\left\{\int d^2 x\, d^2 y \,d^2 z K\left(x,y|z \right)\,\left(
- u(x,y)\,\,+\,\,u(x,z)u(z,y) \right)\,\frac{\delta}{\delta u(x,y)} \right. \label{chif}\\
 &-& \left.\h\, \int\,\prod^2_{i=1}\, d^2 x_i\, d^2 y_i \,d^2 z \,\tilde{K}\left(x_1,y_1;x_2,y_2|z
\right)\,\,\right.\nonumber \\
 &\times& \,\left.\left( u(x_1,z)\,u(z,y_1)\,u(x_2,z)\,u(z,y_2)
\,\,-\,\,u(x_1,y_1)\,u(x_2,y_2) \right)\,
\frac{\delta}{\delta u(x_1,y_1)}\,\frac{\delta}{\delta u(x_2,y_2)} \right\} \nonumber\\
& + & \frac{\bas}{2\,N^2_c \pi}\h\, \int\,\prod^2_{i=1}\, d^2 x_i\, d^2 y_i \,d^2 z\,\left(
K(x_1,y_2|z) \,+\,K(x_2,y_1|z)\right)\, \nonumber \\
 & \times&
\left(1 - u(x_1,y_2))\,(u(x_2,y_1) -
u(x_2,z)\,u(z,y_1)\right)\frac{\delta}{\delta u(x_1,y_1)}\,\frac{\delta}{\delta u(x_2,y_2)}\nonumber
\eea
where the last term is the transition of two dipoles to three dipoles taken from Ref. \cite{L3}.
We introduce this term  for completeness of presentation as well as to demonstrate that  a  part of
terms
for the Pomeron interactions induced by the second term in \eq{chif}, have a contribution which is
suppressed by the extra power of the QCD coupling in comparison with the same interactions generated by
the last term in \eq{chif}.

\subsection{Generating functional in the toy model}
In this subsection we return to the simple toy model to investigate the specific properties of the
generating functional given by \eq{chif}. The equation for the generating functional in this model has
the form
\beq \label{NFTM}
\frac{\partial Z}{\partial Y}\,\,=\,\,-\,\Gamma(1 \to 2)\,u\,(1 - u)\,\frac{\partial Z}{\partial
u}\,\,+\,\,\left\{\Gamma(2 \to 1)\,u^2\,(1 - u^2) \,+\,\Gamma(2 \to 3)\,u\,( 1 - u)^2 \right\}
\frac{\partial^2 Z}{(\partial u)^2}
\eeq

The asymptotic solution to this equation can be found solving the equation with $\frac{\partial
Z}{\partial Y} = 0$. In semi-classical approach for which $ Z(Y=\infty, u) = e^{\Phi(u)}$ with
$\Phi_{u,u} \ll
\Phi^2_{u}$ $\Phi(u)$ is equal to
\beq \label{PHITM}
\Phi_u(u)\,\,=\,\,\frac{\kappa_1}{u(1 + u) \,+\,\kappa_2 (1 - u)}
\eeq
where
\beq \label{KAPPA}
\kappa_1\,\,=\,\,\frac{\Gamma(1 \to 2)}{\Gamma(2 \to 1)}\,\,\propto\,\frac{1}{\as^2}\,\gg\,\,1\,;
\,\,\,\,\,\,
\,\,\,
\,\,\,
\kappa_2\,\,=\,\,\frac{\Gamma(2 \to 3)}{\Gamma(2 \to 1)}\,\,\propto\,\frac{1}{N^2_c\,\as^2}\,\gg\,\,1\,;
\eeq

From \eq{PHITM}
\beq \label{ZSOLTM}
Z(\infty,u)\,\,=\,\,\left( \frac{(\kappa_2 - u)(  \frac{2}{\kappa_2})}{(\kappa_2 - 1)( 1 +
\frac{2}{\kappa_2} -u)} \right)^{\frac{\kappa_1}{\kappa_2}}\,\,
\eeq
This solution satisfies the normalization condition:$Z(\infty,u=1)=1$. One can see that the main
role that plays   the second term in \eq{NFTM} is to regularize the singularity at $u \to 1$ which
occurs if we neglect this term. It should be stressed that the vertex for $2 \to 1$ transition is positive for $u \to 1$.   \eq{ZSOLTM} suggests that the most important region of $u$ for the
solution is $u \to 1$ where \eq{NFTM} has a simple form
\beq \label{NFTMS}
\frac{\partial Z}{\partial{\cal Y}}\,\,=\,\,-\,\,\,u\,(1 - u)\,\left(\,\kappa_1\,\frac{\partial Z}{\partial
u}\,\,-\,\,\,\,L(u)
\frac{\partial^2 Z}{(\partial u)^2} \right)
\eeq
 where ${\cal Y} \,=\,\,\Gamma(1 \to 2)\,Y$ and
\beq \label{L}
L(u)\,\,=\,\,\,u\,(1 + u)\,\,+\,\,\,\kappa_2\,(1 - u)\,\,=\,\,\left\{\begin{array}{c}
2 \,\,\,\,\,\,\,\,\mbox{for} \,\,(1 - u)\,<\,\,2/\kappa_2\,;\\ \\
\kappa_2\,(1 - u)\,\,\,\,\,\,\,\,\mbox{for} \,\,(1 - u)\,\geq \,\,2/\kappa_2\,;\end{array} \right.
\eeq
First, we go to Mellin transform
\beq \label{MT}
Z({\cal Y}; u)\,\,=\,\,\int^{a + i \infty}_{a - i \infty}\,\,\frac{d \,\omega}{2 \pi i}\,\,\,e^{\omega
\,{\cal
Y}}\,\,Z(\omega, u)
\eeq
and equation for $Z(\omega, u)$ has the form
\beq \label{ZOMU}
\frac{\omega}{u\,(1 - u)}\,=\,\,\,- \Phi_u\,\,+\,\,\frac{1}{\kappa_1}\,\,L(u)\,\, \Phi^2_u
\eeq
where,  searching the  semi-classical solution, we assume that $Z(\omega,u) = \exp\left(\Phi(\omega,u)
\right)$.

\eq{ZOMU} has two roots:
\beq \label{ROOTS}
\Phi^{\pm}_u\,\,=\,\,\frac{\kappa_1}{2\,\,L(u)}\,\left\{\,\, 1\,\,\pm\,\,\sqrt{1 \,\,+\,\,\frac{4
\,\,\omega\,\,L(u)}{\,\,\kappa_1\,\,u\,(1 - u)\,}}\,\,\right\}
\eeq
The initial condition is $Z({\cal Y} =0,u)= u$. The general solution to \eq{ZOMU} has the following
form
\beq \label{SCTM1}
Z\left({\cal Y} ,u\right)\,\,=\,\,=\,\,\int^{a + i \infty}_{a - i \infty}\,\,\frac{d \,\omega}{2 \pi
i}\,\,\,e^{\omega
\,{\cal
Y}}\,\left\{\,e^{\Phi^+(\omega,u)}\,\phi^+(\omega)\,\,+\,\,e^{\Phi^-(\omega,u)}\,\phi^-(\omega)\right\}
\eeq
where functions $\phi^+$ and $\phi^-$ should be found from  the initial condition at ${\cal Y} =0$.

The asymptotic behaviour of the solution is determined by the region of small $\omega$ . One can see
from \eq{SCTM1} that the first term in this equation at small $\omega$ has the form of
\beq \label{SCTM2}
Z^{(+)}({\cal Y},u)\,\,=\,\,e^{\Phi({\cal Y} = \infty,u)}\,\,\int^{a + i \infty}_{a - i
\infty}\,\,\frac{d
\,\omega}{2 \pi
i}\,\,\,e^{\omega
\,{\cal
Y}}\,\left\{\,e^{\frac{\kappa_1}{\kappa_2}\,\omega \,\ln\left(\frac{u}{1 -u}\right)}
\right\}\,\phi^+(\omega)
\eeq
and has a chance top approach the asymptotic solution discussed above. The second term for $\omega \ll
1$ leads to

\beq \label{SCTM3}
Z^{(-)}({\cal Y},u)\,\,=\,\,\int^{a + i \infty}_{a - i \infty}\,\,\frac{d \,\omega}{2 \pi
i}\,\,\,e^{\omega
\,{\cal
Y}}\,\left\{\,e^{-\,\frac{\kappa_1}{\kappa_2}\,\,\omega \,\ln\left(\frac{u}{1 -u}\right)}
\right\}\,\phi^-(\omega)
\eeq
and its behaviour depends on the type of singularity at $\omega = 0$ in function $\phi^-(\omega)$. For
$\phi^-(\omega) = 1/\omega$ this solution generates a constant contribution. We will try to find the
solution considering $\phi^-(\omega) = 0$.

Therefore, we want to find $\phi^+(\omega)$ from the initial condition
\beq \label{SCTM4}
Z^{(+)}({\cal Y} = 0,u)\,\,=\,\,e^{\Phi({\cal Y} = \infty,u)}\,\,\int^{a + i \infty}_{a - i
\infty}\,\,\frac{d
\,\omega}{2 \pi
i}\,\,\,\,e^{ \Phi^+(\omega,u)} \,\phi^+(\omega)\,\,=\,\,u
\eeq

The main observation is that $\Phi^+(\omega,u)$ (see \eq{ROOTS}) in \eq{SCTM4} can be taken in the
form
\beq \label{SCTM5}
\Phi^{+}(\omega,u)\,\,=\,\,\int^u_0\,\,d u'\,\,\frac{\kappa_1}{2\,\,L(u')}\,\left\{\,\,
1\,\,+\,\,\sqrt{\frac{4
\kappa_1\,\,\omega\,\,
L(u')}{\,u'\,(1 - u')\,}}\,\,\right\}
\eeq
since $\frac{4 \,
\omega\,\,L(u)}{\,\kappa_1\,u\,(1 - u)\,}\,\gg\,1$ either because $u$ is close to unity or $\omega
\,\approx\,\frac{\kappa_1}{\kappa_2}\,\,\gg\,\,1$. Using \eq{L} we can take the integral over $u'$ in
\eq{SCTM5}
and reduce $\Phi^{+}$ to the form
\bea
\Phi^{+}(\omega,u)\,\,\,&=&\,\,\sqrt{\frac{\kappa_1\omega}{2}\,(1 -
u)}\,\,\,\,\Theta\left(\frac{2}{\kappa_2}\,-
\,(1
-u)\,
\right)  \label{SCTM6}\\
\,\,&+&\,\, \left\{\,-\, \frac{\kappa_1}{2\,\kappa_2} \,\ln ( 1 - u)\,\,+\,\,\sqrt{\frac{ \omega
\kappa_1}{\kappa_2}}\,\ln\left( \frac{1 + \sqrt{u}}{1- \sqrt{u}} \right)\, \right\}\,\,\,\,\Theta\left(
\,(1 -u) -
\frac{2}{\kappa_2} \,\right) \nonumber
\eea

It easy to see that
we can choose to satisfy the initial condition of \eq{SCTM4} the function $\phi^{+}(\omega)$ in the
following form
\beq \label{PHI+}
\phi^{+}(\omega)\,=
\,\frac{1}{\omega} \,\,+\,\,
\eeq
$$
+\,\,
\frac{\Gamma\left(\frac{\alpha}{2 } - \sqrt{\alpha\,\omega}\right)
\Gamma\left(\frac{\alpha}{2 } +
\sqrt{\alpha\,\,\omega}\right)}{\Gamma\left( \alpha \right)}\,\,{}_2F_1\left(-2,
\frac{\alpha}{2 } - \sqrt{\alpha\,\,\omega}, \alpha, 2
\right)\,\,-\,\,\frac{\Gamma\left(\alpha\, - \sqrt{\alpha\,\,\omega}\right)
\Gamma\left(\alpha\, + \sqrt{\alpha\,\,\omega}\right)}{\Gamma\left(2\,\alpha \right)}
$$
where $\alpha = \kappa_1/\kappa_2$.

Indeed, for $(1 - u)\,\leq 2/\kappa_2$ only the first  term in $\phi^+(\omega)$ contribute leading
to $Z({\cal Y}=0,u)\,=\,1$. This result is the same as $Z({\cal Y}=0,u)\,=\, u$ within our accuracy
since $ u = 1 + O(1/\kappa_2)$.

 The rest of \eq{PHI+} gives a small contribution of the order
of $e^{ - \frac{\kappa_1}{\kappa_2}}$.
For $(1 - u)\,\geq 2/\kappa_2$ the contribution of the first term cancels the contribution of the last
one, while the second term leads to  $Z({\cal Y}=0,u)\,=\, u$. Each $\Gamma$ function has a series of
poles at $\sqrt{\alpha\,\omega}\,\,=\,\,\alpha/2 + l$ ( $l$ is integer number ) with the residues
$(-1)^l/l!$. The sum over $l$ for the first term in ${}_2F_1$  looks  as
\beq \label{SCTM7}
(1 - u)^{- \frac{\alpha}{2}}\,\sum^{\infty}_{l =0}\,
\frac{(-1)^l}{l!}\,\frac{\Gamma \left( \alpha\,+\,l \right)}{\Gamma \left(
\alpha\,\right)}\,\tau^{\frac{\alpha}{2}}\,\tau^{l}\,\,=\,\,(1 - u)^{ - \frac{\alpha}{2}}
\frac{\tau^{\frac{\alpha}{2}}}{(1 + \tau)^\alpha}
\eeq
with $\tau \,\,\equiv\,\,\frac{ 1 + \sqrt{u}}{1 - \sqrt{u}}$.
Taking into account all three terms in ${}_2F_1$ we have
\beq \label{SCTM8}
Z^{(+)}({\cal Y}=0,u)\,\,=(1 - u)^{ - \frac{\alpha}{2}}\frac{\tau^{\frac{\alpha}{2}}}{(1 +
\tau)^\alpha}\,\frac{(1 - \tau)^2}{(1 + \tau)^2}\,\,=\,\,u
\eeq

The behaviour at large ${\cal Y}$ in \eq{SCTM1} comes from the region of small $\omega$. The first term
stems from the pole in $\omega$ which leads to the asymptotic solution. The energy corrections comes
from the singularities at negative  $\omega$ in $\Gamma$ functions. Using the following formula (see
{\bf 8.326(2)} in Ref. \cite{RY})
\beq \label{SCTM9}
\frac{\Gamma\left(\frac{\alpha}{2 } - \sqrt{\alpha\,\omega}\right)
\Gamma\left(\frac{\alpha}{2 } +
\sqrt{\alpha\,\,\omega}\right)}{\Gamma\left( \alpha \right)\,\Gamma\left( \frac{\alpha}{2}
\right)}\,\,=\,\,\prod^\infty_{k = 0} \,\frac{( \frac{\alpha}{2 } + k)^2}{
( \frac{\alpha}{2 } + k)^2 \,+\,\alpha\,\omega}
\eeq
we can see that the solution steeply approaches the asymptotic one having corrections that are
proportional to $e^{ - \frac{\alpha}{2} {\cal Y}}$.

\eq{PHI+} shows nicely the general property of the Sturm-Liouville equation \cite{KAMKE}: the discrete
spectrum at negative $\omega$. It should be mentioned  that our equation can be reduced to
Sturm-Liouville equation in a general case. We also would like to stress that the semi-classical
approach is valid since from \eq{ROOTS} one can see that $1 \,\ll\,\Phi_{u u}
\,\approx\,\kappa_1\,\ll\,\Phi^2_{u}
\,\approx\, \kappa^2_1$.

Concluding this discussion we would like to summarize that we found the analytical solution in the
entire phase space of our variable $Y$ and $u$.

 Coming back to discussion of Pomeron interaction, we  see that \eq{NFTM} leads to a different interactions
 of the Pomerons if we  write the equation for the
functional $N$ of \eq{N}. The number of Pomeron vertices that will be generated by  \eq{NFTM} is greater
than in the previous version of the generating functional $Z$ suggested in Ref. \cite{L3} . Indeed, the
equation for $N$ looks as
\beq \label{NTM}
\frac{\partial N}{\partial Y}\,\,=
\eeq
$$
=
\,\,\,\Gamma(1 \to 2)\,\gamma (1 - \gamma)\,\frac{\partial
Z}{\partial
\gamma}\,\,+\,\,\left\{\Gamma(2 \to 1)\,\gamma\,( 1 - \gamma)^2\,(2 - \gamma) \,+\,\Gamma(2 \to 3)\,
\gamma^2 (1 - \gamma) \right\}
\frac{\partial^2 N}{(\partial \gamma)^2}
$$

One can see that \eq{NTM} the following vertices for Pomeron interactions:
\bea
\mbox{(Pomeron intercept)} \,\,P \to P  & \,\,\,=\,\,\,&  \Gamma(1 \to 2)
\label{P1PTM}\\
\,P \to 2P & \,\,\,=\,\,\,& -  \,\Gamma(1 \to 2) \label{P2PTM}\\
2P \to \,P  & \,\,\,=\,\,\,& 2\,\Gamma(2 \to 1) \label{2PPTM}\\
2P \to 2P  & \,\,\,=\,\,\,&  \Gamma(2 \to 3) \, -\, 5\, \Gamma(2 \to 1) \,\,\approx\,\,
\Gamma(2 \to 3) \label{2P2PTM}\\
2P \to 3P & \,\,\,=\,\,\,& - \, \Gamma(2 \to 3) \,+\, 4\, \Gamma(2 \to 1)\approx\,\,
\Gamma(2 \to 3) \label{2P3PTM}\\
2P \to 4P & \,\,\,=\,\,\,& - \, \Gamma(2 \to 1) \label{2P4PTM}
\eea

One can see that $\Gamma(2 \to 1)$ leads only to small corrections for $2P \to 2P$ and for $ 2P \to 3P$
transitions but generates correctly the $2P \to P$ vertex.

It should be stressed that \eq{NFTM} leads to quite different noise term in \eq{EQRT} which looks as follows
\beq \label{TMFINEQ}
\frac{\partial \,\,\Phi(Y)}{\partial Y}\,\,\,=\,\,\Gamma(1 \to 2)\,\left( \Phi(Y)\,\,\,-\,\,\,\,\,\Phi^2(Y)
\right)
\,\,+\,\,\zeta(Y)
\eeq
with the noise term
\beq \label{FINNOISE}
<|\zeta(Y) |> \,\,=\,\,0\,;
\eeq
$$
<|\zeta(Y)\,\,\zeta(Y')|>\,\,
=\,\,2\,\left(\,\Gamma(2 \to 1)\Phi^2(Y)\left\{ 1 - \Phi^2(Y)\right\}\,\,+\,\,\Gamma(2 \to 3)\,\,\Phi(Y)
\left\{1 - \Phi(Y)\right\}^2 \right) 
\,\delta(Y - Y')\,\,.
$$
The noise term in \eq{FINNOISE} is determined by the $2 \to 1$ transition only at $1 - \Phi(Y) \approx 
\as^2\,\ll\,1$, 
while in the entire kinematic  region outside this band the dynamics is due to the second term in 
\eq{FINNOISE}.

\subsection{A practical way to find solution: Monte Carlo simulation.}

In this section we consider the BFKL Pomeron calculus in the form of \eq{S0ISMR} which leads to the simplest
approach in the framework of the generating functional aopproach. This approach is not onlapproachplest but
also it is free from all troubles related to the negative contribution for the process of $2P \to P$
transition. Repeating procedure discussed in section 3.2 one can see that  \eq{S0ISMR} for action leads to the
following equation for the generating functional $Z(Y,[u(b,k)])$ defined as
\beq \label{ZMR}
Z\left(Y\,-\,Y_0;\,[u(b_i,k_i)] \right)\,\,\equiv\,\,
\eeq
$$
\equiv\,\,\sum_{n=1}\,\int\,\,
P_n\left(Y\,-\,Y_0;\,b_1, k_1; \dots ; b_i, k_i; \dots ;b_n, k_n
 \right) \,\,
\prod^{n}_{i=1}\,u(b_i, k_i) \,d^2\,x_i\,d^2\,k_i
$$
where $u(b_i,k_i)$ are  arbitrary functions.
The equation has the form

\begin{equation}\label{ZEQMR}
\frac{\partial \,Z\,\Lb Y-Y_0; [\,u(b,k)\,]\Rb}{
\partial \,Y}\,\,= \,\,\chi\,[\,u(b,k)\,]\,\,Z\,\Lb Y- Y_0; [\,u(b,k)\,] \Rb
\end{equation}
with
\begin{eqnarray}
\chi[u]\,\,&=&\,\, \,\int\,d^2\,b\, d^2\, k\, \left( \frac{\bas}{2 \pi}
 \left(- \int\,d^2\,k'\, K(k,k') u(b,k')\,\frac{\delta}{\delta u(b,k')} \,+\,u(b,k) \,u(b,k)\,
\,\frac{\delta}{\delta u(b,k)} \right)\,- \right. \label{chimr} \\
 & &\left. - \left( \frac{4\,\pi^2 \bas}{N_c} \right)^2\,\frac{\bas}{2 \pi}
\left( u(b,k)  \,u(b,k) \,-\, u(b,k) \right) \,\,\frac{1}{2} \,\frac{\delta^2}{\delta
u(b,k)\,\delta
u(b,k)} \right)\,;
\label{VE21MR}
\end{eqnarray}

\eq{chimr} and \eq{VE21MR} show that the $2P \to 1P$ transition can be written as the two dipole to one dipole
merging with a positive probability. Therefore, \eq{ZEQMR} has a very simple probabilistic interpretation
which can be written as the following Markov's chain:
\begin{equation}  \label{PMR}
\frac{\partial \,P_n(Y;\dots;b_i=b,k_i=k; \dots;b_n=b,k_n=k)}{\partial Y}\,\,\,=
\end{equation}
\begin{eqnarray}
 &=&\,\,  \frac{\bas}{2 \pi}\,\sum_{i} \left(P_{n-1}(Y; \dots;b_i=b,k_i=k; \dots;b_{n-1},k_{n
-1})\,- \right.\nonumber\\
 & &\left. -\,\,\int \,d^2\,k' \,K(k,k')\,\,P_n
(Y;\dots; b_i=b,k_i=k';
\dots;x_n=b,k_n=k)\right) \label{PMR1} \\
 &+ & \,\, \left( \frac{4\,\pi^2 \bas}{N_c} \right)^2\,\frac{\bas}{2 \pi} \sum_{i > j} \left( P_{n+1}(Y;
\dots;b_j=b,k_j=k; \dots ;
b_i=b,k_i=k;\dots; b_n=b,k_n=k ) -\right.\nonumber\\
 & &\left.\,- \, P^{}_n(Y;\dots;b_j=b,k_j=k;\dots b_i=b,k_i=k; \dots; b_n=b,k_n=k;) \right);  \label{PMR2}
\end{eqnarray}

This set of equations can be solved numerically and it gives a practical way to discuss the influence of the Pomeron loops on the solution for the scattering amplitude at high energies.

\section{High energy amplitude in QCD: solution to the functional equation.}
\label{sec:HEA}
In this section we will find the solution to the general equation of \eq{ZEQ} using the strategy
suggested in Ref. \cite{L4}. This strategy directly  follows from the solution for the toy model and
consists of three steps:

{\bf 1.} \,\,\,\, We  find an asymptotic solution to the functional equation \ref{ZEQ} with zero
l.h.s. using the semi-classical approach. In other words looking for solution in the form
$Z\left(Y=\infty,[u_i]\right) = \exp \left\{ \Phi\left(Y=\infty,[u_i]\right)\right\} $ assuming that
$$
\frac{\delta}{ \delta u_i}\,\frac{\delta}{ \delta u_k}
\Phi\left(Y=\infty,[u_i]\right)\,\,\ll\,\,\frac{\delta}{ \delta u_k}
\Phi\left(Y=\infty,[u_i]\right)\frac{\delta}{ \delta u_i} \Phi\left(Y=\infty,[u_i]\right)
$$
In searching this solution we can neglect the second tern in \eq{ZEQ} but take into account the decay
of two dipoles into three dipoles\footnote{This term is not written in \eq{ZEQ} but one can find it
in Ref.\cite{L3,L4}.}.

{\bf 2.} \,\,\,\, Search for the function $u(x,y)=u_0(x,y)$ which satisfies the equation:
\bea
&\frac{1}{\kappa_2}\,\int\,\,\prod^2_{i=1}\,d^2\,x_i\,d^2\,y_i\,\,d^2\,z\,\, 2
\gamma^{BA}\left(x_1,y_1|z\right)
\,K(x_2,y_2;z)\,\left( \gamma_0(x_2,z)\,\,+\,\,\gamma_0(z,y_2)\,-\,\gamma_0(x_2,y_2) \right)\,\,=&
\nonumber \\
& & \nonumber \\
&=
\int\,\,\prod^2_{i=1}\,d^2\,x_i\,d^2\,y_i\,\,d^2\,z\,\,K(x_2,y_1;z)\,\,
\,\gamma_0(x_2,y_1)\,\Lb \gamma_0(x_1,z)\,+\,  \gamma_0(z,y_2)\,-\,\gamma_0(x_2,y_1) \Rb&
\label{GAMMA0}
\eea
where we assume that $\gamma_0(x,y) = 1 - u_o(x,y)\,\, \propto\,\,\frac{\bas}{N^2_c}\,\,\ll \,\,1$.
\eq{GAMMA0} is the condition that the term in \eq{ZEQ} that is responsible for $2 \to 3$ transition of
dipoles, is
equal to the second term in \eq{ZEQ} which describes the decay of two dipoles into four dipoles.

{\bf 3.} \,\,\,\,  Specify the asymptotic solution $Z\left(Y=\infty,[u_i]\right)$ from the
normalization
condition
\beq \label{HE1}
Z\left(Y=\infty,[u_i=u_0]\right)\,\,=\,\,1
\eeq

{\bf 4.} \,\,\,\,  Search the solution in the form $ \Phi\left(Y,[u_i]\right)\,\,=\,\,
\Phi\left(Y=\infty,[u_i]\right)\,\,+\,\,\Delta \Phi\left(Y,[u_i]\right)$ and assuming that
$\left(\frac{\delta \Delta
\Phi\left(Y,[u_i]\right)}{\delta u} \right)^2$ are small and can be neglected.  In doing so we obtain
a linear equation which we need to solve.

\subsection{Asymptotic solution}
The asymptotic solution to \eq{ZEQ} with the term describing the transition of two dipoles to three
dipoles has been found in Ref. \cite{L4} and we discuss it here for the sake of completeness of our
presentation. This solution has the form
\beq \label{AS1}
\Phi(Y=\infty,[u_i])\,\,=\,\,- \,\alpha \,\ln \Lb
\frac{\int\,\,d^2\,x\,d^2\,y\,\,\Theta(x,y)\,\,\gamma(x,y)}{\int\,\,d^2\,x\,d^2\,y\,\,\Theta(x,y)}
\Rb\,
\eeq
where $\alpha = \frac{\kappa_1}{\kappa_2}$ with  $\kappa_1$ and $\kappa_2$ are defined in \eq{KAPPA}
and $\Theta$ is a function which is equal to
1 for $|\x| < R$ and $|\y| < R$ and  $\Theta = 0$
for $|\x| > R$ and $|\y| > R$ where $R$ is the
largest scale in the problem. From \eq{AS1} one can see that ${\cal F}$ is defined as
\beq \label{AS2}
{{\cal F}}(\infty,[\gamma_i])\,\,\equiv\,\,\int\,\,d^2\,x\,d^2\,y\,\,\Theta(x,y)\,\,\gamma(x,y)
\,\,\,\mbox{and} \,\,\,N\,\,=\,\,\int\,\,d^2\,x\,d^2\,y\,\,\Theta(x,y)
\eeq
Substituting $Z(Y,[u_i]) = \exp\Phi(Y,[u_i])$ with $\Phi(Y=\infty,[u_i])$ given by \eq{AS1}
we have
$$
\kappa_1\,\alpha\,\int\,d^2 \,x\,\,d^2\,y\,\,d^2\,z\,\,K(x,y;z)\,\Lb \,\gamma(x,y)
\,-\,\gamma(x,z)\,\gamma(z,y) \Rb\,\frac{ \Theta(x,y)}{{\cal
F}}\,= \,\kappa_1\,\alpha\,\int\,\,d^2 \,x_1\,\,d^2\,y_1\,\,d^2 \,x_1\,\,d^2\,y_2\,\,d^2\,z\,
$$
\beq \label{AS3}
K(x_1,y_2;z)\,\,\Lb \gamma(x_1,y_2)\,\gamma(x_2,y_1)\,
\,-\,\gamma(x_1,z)\,\gamma(z,y_2)\,\gamma(x_2,y_1)
\Rb
\frac{
\Theta(x_1,y_1)}{{\cal F}}\,\frac{
\Theta(x_2,y_2)}{{\mathcal F}}
\eeq
Canceling common factors, multiplying by ${\cal F}$ both sides of \eq{AS3}  and substituting ${\cal
F}\,\,=$\\ $\int\,d^2\,x'\,d^2\,y'\,\,\gamma(x',y')$ we obtain
the following equation
\beq \label{AS4}
\int\,d^2 \,x\,\,d^2\,y\,\,d^2 \,x'\,\,d^2\,y'\,\,d^2\,z\,\,K(x,y;z)\,\gamma(x',y')\,\,\Lb
\,\gamma(x,y)
\,-\,\gamma(x,z)\,\gamma(z,y) \Rb\,=\,
\eeq
$$
=\,\,\int\,\,d^2 \,x_1\,\,d^2\,y_1\,\,d^2 \,x_1\,\,d^2\,y_2\,\,d^2\,z\,
\,\,
K(x_1,y_2;z)\,\,\Lb \gamma(x_1,y_2)\,\gamma(x_2,y_1)\,
\,-\,\gamma(x_1,z)\,\gamma(z,y_2)\,\gamma(x_2,y_1) \Rb
$$
One can see that if we denote in the r.h.s. of this equation $x_1 \to x$,  $y_2 \to y$, $x_2 \to x'$
and
 $ y \to y'$, \eq{AS4} becomes the identity. This means that \eq{AS1} is the asymptotic solution of
\eq{ZEQ}.

\begin{boldmath}
\subsection{$ \gamma_0(x,y)$}
\end{boldmath}
In this subsection we will find $ \gamma_0(x,y)$ as the solution to \eq{GAMMA0} in more direct way
than in Ref. \cite{L4} using the explicit form of the vertex for $2 \to 4$ dipole decay (see
\eq{V21N}). If we change the notation  in the r.h.s. of \eq{GAMMA0}, namely $x_1 \leftrightarrow
x_2$ and $y_1 \leftrightarrow
y_2$ one can see that
\beq \label{GAMMA0SOL}
\gamma_0(x,y) \,\,=\,\,\frac{2}{\kappa_2}\,\,\gamma^{BA}(x,y|z)
\eeq
is the solution to this equation. It looks strange that $\gamma_0(x,y)$ has three arguments
instead of two. However, the explicit form of $\gamma^{BA}(x,y|z)$ in \eq{GAMMAZ} shows that this
function is actually a function of only two variable $x - z$ and $z - y$. We can change all
integration variable shifting them by $z$ without inducing any change in the integral. It means that
$\gamma_0(x,y)$ is equal
\beq \label{GAMMA0SOL1}
 \gamma_0(x,y)\,\,\,=\,\, \,\frac{1}{8\,N^2_c}\,\ln^2\frac{x^2}{y^2}
\eeq
\eq{GAMMA0SOL1} gives the same $ \gamma_0(x,y)$ as was derived in Ref.\cite{L4} from rather lengthy
discussions.

Using \eq{GAMMA0SOL1} we can build the solution
 that satisfies the modified initial condition of \eq{HE1}
($Z(Y,[\gamma_i=\gamma_{0,i}]) \,=\,1$). This solution  has the  form
\beq \label{ZF23}
Z(Y,[\gamma_i]) \,\,=\,\,\Lb \,\frac{{\cal F}( \infty;[\gamma_i{x,y}])}{{\cal F}( \infty;
[\gamma_{0i}(x,y)]) }
\,\Rb^{- \alpha}
\eeq
\subsection{ Approaching the asymptotic solution}
Substituting $ \Phi\left(Y;[\gamma_i(x,y)]\right)\,\,=\,\,\Phi\left(Y =
\infty;[\gamma_i(x,y)]\right)\,\,+\,\,\Delta \Phi\left(Y;[\gamma_i(x,y)]\right)$ we obtain a linear
equation
for
$\Delta \Phi$ if we assume that
\bea
\frac{\delta \Delta \Phi\left(Y;[\gamma_i(x,y)]\right)}{\delta \gamma(x,y)}\,\,& \ll & \,\,\frac{\delta
\Phi\left(Y=\infty;[\gamma(x,y)]\right)}{\delta \gamma_i(x,y)}; \label{AAS1} \\
\,\,\,\,\,\,\,&\mbox{and} &\,\,\,\,\,\,\,\nonumber\\
\frac{\delta^2 \Phi\left(Y=\infty;[\gamma_i(x,y)]\right)}{ \delta \gamma(x_1,y_1) \delta
\gamma(x_2,y_2)}
\,\,&\ll &\,\,\frac{\delta
\Phi\left(Y=\infty;[\gamma(x,y)]\right)}{\delta \gamma_i(x,y)}\,\frac{\delta \Delta
\Phi\left(Y;[\gamma_i(x,y)]\right)}{\delta \gamma(x,y)}  \nonumber
\eea

We can use the solution for the toy model (see section 4.4) to show that these assumptions are
reasonable but we have to check \eq{AAS1} after finding the solution for $\Delta \Phi$. It should be
mentioned that the second equation is valid since
$\frac{\delta^2 \Phi\left(Y=\infty;[\gamma_i(x,y)]\right)}{ \delta \gamma(x_1,y_1) \delta
\gamma(x_2,y_2)}\,=\,0$ for the asymptotic solution.

The  equation for $\Delta \Phi(Y;[\gamma_i])$  has the form \cite{L4}:
\bea
\frac{ \partial\, \Delta \Phi(Y;[\gamma_i])}{\partial\,{\cal Y}}\,\,\,&=&\,\,-\,\int
\,\,d^2\,x\,d^2\,y
\,d^2\,z\,\,\,K(x,y;z)\,\Lb \,\gamma(x,y)\,\,-\,\,\gamma(x,z)\,\gamma(z,y)\,\Rb \frac{\de\, \Delta
\Phi(Y;[\gamma_i])}{\de\,\gamma(x,y)}\,\,+ \nonumber \\
&+& \frac{1}{\kappa_1}\,\int\,\prod^2_{i=1}\,d^2\,x_i\,d^2\,y_i\,\,d^2\,z\,\, 2
\gamma^{BA}\left(x_1,y_1|z\right)\,
\,K(x_2,y_2;z)\,\left( \gamma(x_2,z) + \gamma(z,y_2) - \gamma(x_2,y_2)
\right) \nonumber \\
 &\times& \frac{\delta^2 \Phi\left(Y=\infty;[\gamma_i(x,y)]\right)}{ \delta
\gamma(x_1,y_1) \delta
\gamma(x_2,y_2)} \,\,
 +\,\,2\,\frac{1}{{\cal
F}(\infty,
[\gamma_{0i}])}\,\int\,d^2\,z\,d^2\,x_1\,d^2\,y_1\,d^2\,x_2\,d^2\,y_2\,\,\,K(x_1,y_2;z)\,\,
\nonumber\\
\ &\times& \Lb
\,\gamma(x_1,y_2)\,\gamma(x_2,y_1)\,\,-\,\,\gamma(x_1,z)\,\gamma(z,y_2)\,\gamma(x_2,y_1)\,\Rb
\,\frac{\de \Delta\Phi(Y;[\gamma_i])}{\de\,\gamma(x_1,y_1)} \label{AAS2}
\eea
\eq{AAS2} is the  Liouville-type equation with the  source term which is given by the two dipoles to four
dipoles decay. It is easy to reduce this equation to the  Liouville equation without the  source term and
even linearize it if we assume that $\Delta \gamma({\cal
Y};x,y) \,\equiv\,  \gamma({\cal Y};x,y) -  \gamma_0({\cal Y};x,y)\,\,\ll\,\gamma_0({\cal Y};x,y)$
where $\gamma_0$ is the solution to \eq{GAMMA0} and using the explicit form of
$\Phi(Y=\infty,[\gamma_i])$. The equation has the form
\beq \label{AAS3}
\frac{ \partial\, \Delta \Phi(Y;[\Delta \gamma_i])}{\partial\,{\cal Y}}\,\,\,=
\eeq
$$
\,\,\int\,\,d^2\,z \,K(x,y;z)\,\Lb \Delta\,\gamma({\cal Y}; x,z) \,+\,\Delta\,\gamma({\cal Y};
z,y)\,-\,\Delta\,\gamma({\cal Y}; x,y) \Rb \,\frac{ \partial\, \Delta \Phi(Y;[\Delta
\gamma_i])}{\partial\,\gamma(x,y)}
$$
which   can be  solved   assuming that $\Delta
\Phi(Y;[\Delta \gamma_i])\,=\,\Delta
\Phi([\Delta \gamma_i(Y;x_i,y_i)])$ (see Refs. \cite{L3,L4} for details). For function
$\Delta \gamma(Y;x,y)$
we
obtain the
equation
\beq \label{AAS4}
\frac{\partial\,\,\Delta\,\gamma({\cal Y}; x,y)}{\partial\,\,{\cal Y}}\,\,=\,\,
-\,\,\int\,\,d^2\,z \,K(x,y;z)\,\Lb \Delta\,\gamma({\cal Y}; x,z) \,+\,\Delta\,\gamma({\cal Y};
z,y)\,-\,\Delta\,\gamma({\cal Y}; x,y) \Rb
\eeq
This equation is just the BFKL equation but with the negative sign in front of l.h.s.

The initial condition for $\Delta
\Phi([\gamma(Y;x_i,y_i)])$ stems from the equation $ Z\Lb Y = 0,[u_i] \Rb $, that leads to
\beq \label{AAS5}
\exp\left\{ \Phi( Y=\infty,[\gamma_i])\,\,+\,\,\Delta
\Phi( Y=0,[\gamma_i])\right\} = 1 - \gamma(Y=0,x,y)
\eeq
or
\bea
\Delta \Phi({\cal Y} = 0;[\gamma_i])\,\,&=&\,\,\ln\frac{(1 - \gamma({\cal Y} =0 ;
x,y))}{(1 - \gamma_)({\cal Y};
x,y))}\,\,+\,\,\alpha \,\ln
\Lb
\frac{\int\,d^2\,x\,d^2\,y\,\Theta(x,y)\,\,\gamma({\cal Y} =0 ; x,y)}{{\cal F} ({\cal Y}=\infty;
[\gamma_{0,i}])}
\Rb\,\nonumber\\
&\,\approx &
\, -\,\Delta \gamma({\cal Y} =0 ;
x,y)  \,\,-\,\,\alpha\,
\frac{\int\,d^2\,x\,d^2\,y\,\Theta(x,y)\,\,\Delta\,\gamma({\cal Y} =0 ;
x,y)}{{\cal F}
({\cal Y}=\infty; [\gamma_{0,i}])} \label{AAS6}
\eea

The solution to \eq{AAS4} can be easily found using the Green function of the BFKL equation (see
\eq{BFKLGF}), namely,
\beq \label{AAS7}
\Delta\,\gamma({\cal Y}; x,y)\,\,=\,\,\int\,d^2\,x'\,d^2\,y'\,\tilde{G}\left(x,y; Y|x',y';Y=0 \right)\,\Delta
\gamma(x',y')
\eeq
where $\Delta \gamma(x',y')$ is an arbitrary function and $\tilde{G}\left(x,y; Y|x',y';Y=0 \right)$ is
the Green function of \eq{AAS4} which satisfies the initial condition (see Eq. (120) of Ref. \cite{LI})
\beq \label{AAS8}
 \tilde{G}\left(x,y; Y|x',y';Y \right)\,\,=\,\,\delta^{(2)}(\vec{x} -
\vec{x}^{\,'})\,\delta^{(2)}(\vec{y} - \vec{y}^{\,'})\,\,+\,\,\delta^{(2)}(\vec{x} -
\vec{y}^{\,'})\,\delta^{(2)}(\vec{y} - \vec{x}^{\,'})
\eeq
Such a Green function is equal to
\beq \label{AAS9}
\tilde{G}\left(x,y; Y|x',y';Y=0 \right)\,\,=\,\,\Theta(Y) \,\times
\eeq
$$
\times \sum^{\infty}_{n = - \infty}\,\int\,\frac{d \nu}{\pi^4}\,d^2\,x_0\,e^{ - \omega(n,\nu)\,Y}\,
\,\frac{ \nu^2 + n^2/4}{(x - x')^2\,(y - y')^2}\,E(x,y;x_0|\nu)\,E^*(x',y';x_0|\nu)
$$
where all ingredients of \eq{AAS9} are defined by \eq{BFKLE} - \eq{BFKLOM}.
One can see that $\Delta\,\gamma({\cal Y}; x,y)\,\,$  decreases as $\exp\left( - \omega(n=0,\nu=0)\,Y
\right)$ at large values of rapidity $Y$.

Using \eq{AAS7} and  \eq{AAS5} we find the functional $\Delta \Phi(Y;[\delta \gamma_i])$

\beq \label{FINDPHI}
\Delta \Phi(Y ;[\Delta \gamma_i])\,\,=  \,\,
\, -\,\Delta \gamma(Y;x,y)  \,\,-\,\,\alpha\,
\frac{\int\,d^2\,x\,d^2\,y\,\,\,\Delta\,\gamma(Y;
x,y)}{{\cal F}( Y=\infty; [\gamma_{0,i}])} \nonumber
\eeq
with $\Delta \gamma$ given by \eq{AAS7}
while the final answer for the generating functional $Z$ has the form
\beq\label{ZFIN}
Z\left(Y,[\Delta \gamma_i]\right)\,\,=\,\,\exp \left\{ \Phi(Y = =\infty;[ \gamma_0(x,y) + \Delta
\gamma(Y;x,y)] ) \,\,+\,\,\Delta \Phi(Y ;[\Delta \gamma(Y;x,y)]) \right\}
\eeq

Using \eq{ZFIN} we can easily calculate the behaviour of the scattering amplitude at high energies as
it has been suggested in Ref. \cite{L4}.


\section{High energy scattering amplitude: semi-classical approach}
\label{sec:HEASC}
\subsection{General equations for scattering amplitude in semi-classical approach.}
The method suggested in the previous sections leads to the exact solution at  the ultra high energies,
but it does not look very practical to reconstruct the behaviour of the scattering amplitude in the
entire kinematic region starting from low energies. It is especially clear in comparison with the
remarkable progress in the searching of the solution for the mean field approach (Balitsky-Kovchegov
equation). For the Balitsky-Kovchegov
equation we have developed both analytical \cite{LT,IIM} and numerical\cite{THEORVRSDATA} approaches
which
give us  the effective way for calculation of the scattering amplitude in the accessible kinematic
region of energies. We firmly believe that the method of the generating functional will lead to a
practical application in spirit of Ref. \cite{MS}, namely, to creation of the new Monte Carlo codes
which will allow to solve the equations for the scattering amplitude as well as to give some
predictions to the inclusive observable expanding the  region of application of this approach.

In this section we develop the semi-classical approach to the amplitude determined by the BFKL Pomeron
Calculus in the functional integral form of \eq{BFKLFI} or \eq{SM2}. It is well known that the
semi-classical approach is related to the fields $\Phi$ and $\Phi^+$ that satisfy the equation
\beq \label{HESC1}
\delta S \,\,=\,\,0
\eeq
 Making variation of $S$  given by \eq{S0} - \eq{SE}  with respect to both fields
we obtain (see also  Ref. \cite{BRN})
\bea
 \,\,\,\,\,& & \frac{\partial N(x,y;Y')}{\partial\,Y'}\,\,= \label{SCEQ1}\\
 &=&\,\,\frac{\bas}{2
\pi}\,\int\,d^2\,z\,K\left(x,y|z
\right)\,\left(N(x,z;Y') \,+\,N(z,y;Y')\,-\,N(x,y;Y')\,- N(z,y;Y')\,N(z,y;Y') \right) \nonumber
\\
 &-& \,\,\frac{1}{(2 \pi)^4}\,\frac{\bas}{ \pi}\,\int\,\,G_0\left(x,y;Y'|x',y';Y'\right)\,\frac{d^2 x'\,d^2\,y'}{( x' -
y')^4}\,K \left(x',y'|z \right)\,\,d^2\,z\,N^+\left(z,y', Y - Y'\right)\,N(x',z;Y')\,\nonumber\\
 & & \frac{\partial N^+(x,y;Y - Y')}{\partial\,Y'}\,\,=\,\, \label{SCEQ2} \\
&=&\,\,\frac{\bas}{2 \pi}\,\int\,d^2\,z\,K\left(x,y|z
\right)\,\left(N^+(x,z;Y - Y') \,+\,N^+(z,y;Y - Y')\,-\,N^+(x,y;Y - Y')\,- \right. \nonumber \\
&-&\left.
\, \, N^+(z,y;Y -
Y')\,N^+(z,y;Y
- Y')
\right) \nonumber\\
 &-& \,\,\frac{1}{(2 \pi)^4}\,\frac{\bas}{ \pi}\,\int\,\,G_0\left(x,y;Y'|x',y';Y'\right)\,\frac{d^2 x'\,d^2\,y'}{( x' -
y')^4}\,K \left(x',y'|z \right)\,\,d^2\,z\,N\left(z,y'; Y'\right)\,N^+(x',z;Y - Y')\,\nonumber
\eea
where we use \eq{LG} and the new normalization for $\Phi$ and $\Phi^+$,
namely,$\Phi(x,y;Y)\,=\,N(x,y;Y)/4 \pi \as$ and $\Phi^+(x,y;Y)\,=\,N^+(x,y;Y)/4 \pi \as$, which
corresponds to the normalization of the scattering amplitude. The mean field approximation follows from
\eq{SCEQ1} and \eq{SCEQ2} if we neglect the last terms in these equations which describes the two
Pomeron to one Pomeron merging. The boundary conditions for these equations are (see \eq{SE})
\beq \label{SCBC}
N(x,y;Y'=Y_0)\,\,=\,\,4\, \pi\, \as\,\,
\tau_{tar}(x,y);\,\,\,\,\,\,\,\,\,\mbox{and}\,\,\,\,\,\,\,\,\,N^+(x,y;Y - Y'=Y_0)\,\,=\,\,4 \pi\,
\as\,\,
\tau_{pr}(x,y)
\eeq
From \eq{SCBC} one can see when the mean field approximation could work. It should be either large
$\tau_{tar}$ as in the case of scattering off nuclei for which $\tau_{tar}\,\propto\, A^{\frac{1}{3}}$
and
$N(x,y;Y'=Y_0)\,\propto\, \as  A^{\frac{1}{3}} \,\approx 1$\cite{K}; or the size of the target dipole
is
much
large than the size of the projectile and , therefore $\alpha_{S,tar} \,\gg\,\alpha_{S,pr}$
\cite{GLR}. In these
both cases
the interaction of the initial field $N(x,y;Y'=Y_0)$  leads to  large contribution and the last term
can be neglected.

\subsection{Linear approximation to the general equation and Mueller-Shoshi band.}

 As it was shown \cite{GLR,BALE,MP,MUTR} that we can find the typical scale of the process (so called
saturation momentum  $Q_s$ without finding the explicit solutions of non-linear equations (see
\eq{SCEQ1} and \eq{SCEQ2}). The equation for $Q_s$ is actually a value of $(x - y)^2 \approx 1/Q^2_s$
at which
\beq \label{QSEQ1}
N^{BFKL}(r^2 = ( x - y)^2 =r^2_s=1/Q^2_s ;Y') \,\,\,\approx\,\,\,1,
\eeq
 where $N^{BFKL}$ is the
solution to the linear equation (see \eq{SCEQ1} and \eq{SCEQ2} but without non-linear contributions).

This equation has been solved (see Refs. \cite{GLR,MUTR}) and the following expressions for the
saturation momentum has been derived

\bea
Q^2_s(Y'; \mbox{fixed} \,\,\, \as)\,\,&=&\,\,Q^2_s(Y_0)\,\exp  \left(
\,\frac{\omega(\gamma_{cr})}{1 -
\gamma_{cr}}\,\,(Y' -
Y_0)\,\,\right)\,; \label{QSFC}\\
Q^2_s(Y;\mbox{running}\,\,\, \as)\,\,&=&
\,\,Q^2_s \left( Y_0 \right)\,\,\exp \left( \sqrt{\frac{8\,N_c
\,\chi(\gamma_{cr})}{b\,\gamma_{cr}}}\,(\,\sqrt{Y'}\,-\,\sqrt{Y_0}
)\,\,\right)\,; \label{QSRC}
\eea

where the value of $\gamma_{cr}$ can be found
from the equation \cite{GLR,MUTR}:
\beq \label{GAMMACR}
\frac{\omega(\gamma_{cr})}{ 1 - \gamma_{cr}} \,\,\,=\,\,\,-
\,\,\frac{d \omega(\gamma_{cr})}{ d \gamma_{cr}}\,,
\eeq
and
\beq \label{CHILO}
\omega(\gamma)\,\,=\,\,\bas\,\chi(\gamma)\,\,=\,\,\bas\,\left(2\,\psi(1)
\,\,-\,\,\psi(\gamma)\,\,-\,\,\psi(1 -
\gamma)\,\,\right)
\eeq

 In the mean field approach based on the Balitsky-Kovchegov equation for dipoles which size are
less
than the $1/Q_s$ we can safely use the linear BFKL equation to obtain the scattering amplitude. As one
can see from \eq{SCEQ1} and \eq{SCEQ2} that we need to demand that the amplitude $N^+(x,y;Y -Y')$
should be smaller than unity ($ N^+(x,y;Y -Y') \,<\,1$). This condition means that the dipole size $x -
y$ should be smaller that a new saturation scale $Q_s(Y - Y')$ which we can obtain from the same
equations \eq{QSFC} and \eq{QSRC} but changing $Y_0 \to Y $ and $Q^2_s \left(Y'= Y_0
\right)\,\to\,Q^2_s \left(Y - Y'= Y_0 \right)$.

In \fig{qsd} we plot the kinematic region (`band') in which we can replace the system of
non-linear
equations of \eq{SCEQ1} and \eq{SCEQ2} by the BFKL linear equation. Two lines in this picture show the
two saturation momenta: $Q_s(Y')$ ans $Q_s(Y - Y')$ in the case of fixed QCD coupling, for the
amplitude of two dipole scattering. One dipole has the value of rapidity $Y' = Y_0$ and  the size $x_0
- y_0 = R_0$ while the second dipole has rapidity $Y$ and its size is equal to $R\,\,\ll\,R_0$.
Therefore, for $Q_s(Y')$ the value of $Q_s(Y'=Y_0) \,\approx\, 1/R^2_0$ and for  all sizes of dipoles
smaller than $1/Q_s(Y')$ we can neglect the non-linear contribution that are proportional to
$N^2(r,Y')$. This condition can be written as follows
\beq \label{C1}
\ln (R^2_0/r^2)\,\geq\,\,\frac{\omega(\gamma_{cr})}{1 -
\gamma_{cr}}\,\,(Y' -Y_0)
\eeq
\FIGURE[ht]{
\centerline{\epsfig{file=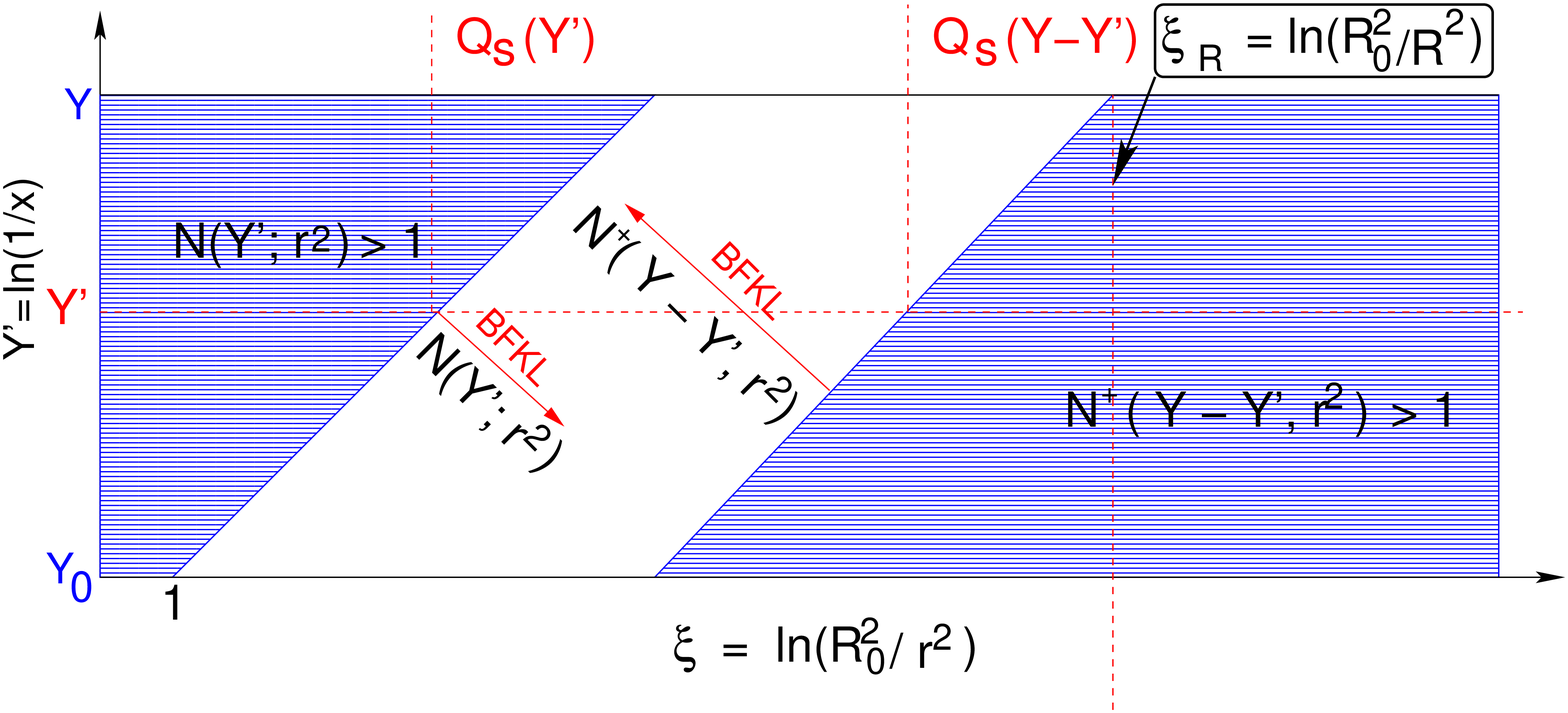,width=130mm}}
\caption{The kinematic region, where we can replace the non-linear system of equations given by
\eq{SCEQ1} and \eq{SCEQ2} (see also \eq{BANDYY'} and \eq{BANDY}), by the linear BFKL equation.$R$
and $R_0$ are the sizes of projectile and target dipoles, $R \,\ll\,R_0$).}
\label{qsd}}

These sizes are to the right of the left line in \fig{qsd}. For $Q_s(Y - Y') $ we have a different
normalization, namely $Q_0(Y -Y'=Y_0)=1/R^2$ and instead of \eq{C1} we have
\beq \label{C2}
\ln (r^2/R^2)\,\geq\,\,\frac{\omega(\gamma_{cr})}{1 -
\gamma_{cr}}\,\,(Y - Y' -Y_0) \,\,\,\,\,\mbox{or}\,\,\,\,\,\ln
(R^2_0/r^2)\,\,\leq\,\,\ln(R^2_0/R^2)\,\,\,-\,\,\frac{\omega(\gamma_{cr})}{1 -
\gamma_{cr}}\,\,(Y - Y' -Y_0)
\eeq

Finally, the BFKL equation can describe the scattering amplitude in the band shown in \fig{qsd} which
has the form
\bea
\ln\left(R^2_0 Q^2_s(Y - Y')\right)\,\,\,\,\,\,\,&\geq& \,\,\,\,\,\ln
(R^2_0/r^2)\,\,\,\,\,\geq \,\,\,\,\,\,\,\,\,\,\,\ln\left(R^2_0
Q^2_s( Y')\right) \,\,\,\mbox{where} \label{BAND} \\
\ln\left(R^2_0 Q^2_s(Y - Y')\right)\,\,\,\,\,\,\,&=&
\,\,\,\,\,\ln(R^2_0/R^2)\,\,\,-\,\,\frac{\omega(\gamma_{cr})}{1 -
\gamma_{cr}}\,\,(Y - Y' -Y_0)\,;\,\label{BANDYY'}\\
\ln\left(R^2_0
Q^2_s(Y')\right)\,\,\,\,\,\,\,\,\,\,\,\,\,\,\,\,\,\,&=&\,\,\,\,\,\,
\,\frac{\omega(\gamma_{cr})}{1 -
\gamma_{cr}}\,\,(Y' -Y_0)\,; \label{BANDY}
\eea

 This
result has been derived by Mueller and Shoshi (see Ref. \cite{MUSH}) using quite a different approach
based on the completeness relation for the dipole number density (or in other words, a unitarity
constraints in $t$-channel). In this approach the same conclusions stem from the semi-classical
equation for the scattering amplitude.

The most interesting result of Ref. \cite{MUSH} is the appearance of a new saturation scale which
has the form
\beq \label{QSNEW}
\ln\Lb R^2_0\,\tilde{Q}^2_s(Y) \Rb\,\,=\,\,\ln \Lb R^2_0\,Q^2_s(Y)\Rb\,\,+\,\,c_s \frac{\pi^2
\bas\,\chi^{''}_{\gamma, \gamma}(\gamma_{cr})(1 -
\gamma_{cr})^2}{8\,\ln^3(1/\bas)}\,\,=\,\,\bas\,C_s
\,Y
\eeq where $c_s$ is a constant which is determined by the exact value of the amplitude where the
saturation is declared to occur. Constant $C_s$ is equal to
\beq \label{CS}
C_s\,\,\,\equiv\,\,\frac{\chi(\gamma_{cr})}{1 - \gamma_{cr}}\,\,+\,\,\,c_s \frac{\pi^2
\,\chi^{''}_{\gamma, \gamma}(\gamma_{cr})(1 - \gamma_{cr})^2}{8\,\ln^3(1/\bas)}
\eeq

\subsection{Solution in the saturation region}

In the system of \eq{SCEQ1} and \eq{SCEQ2} we have used the relation of \eq{LG}. However, we need
to return to the general form of the action (see \eq{SI})  and to a general form of \eq{SCEQ1} and
\eq{SCEQ2}. This general form can be written as follows
\bea
 \,\,\,\,\,& & \frac{\partial N(x,y;Y')}{\partial\,Y'}\,\,= \label{SCEQG1}\\
 &=&\,\,\frac{\bas}{2
\pi}\,\int\,d^2\,z\,K\left(x,y|z
\right)\,\left(N(x,z;Y') \,+\,N(z,y;Y')\,-\,N(x,y;Y')\,- N(z,y;Y')\,N(z,y;Y') \right) \nonumber
\\
 &-& \,\frac{1}{(2 \pi)^4}\,\frac{\bas}{ \pi}\,\int\,\,G_0\left(x,y;Y'|x',y';Y'\right)\,\frac{d^2 x'\,d^2\,y'}{( x' -
y')^4}\,K \left(x',y'|z
\right)\,\,d^2\,z\,\left(L_{z,y'} \,N^+\left(z,y',Y-Y'\right)\right)\,N(x',z;Y')\,\nonumber\\
 & & \frac{\partial N^+(x,y;Y - Y')}{\partial\,Y'}\,\,=\,\, \label{SCEQG2} \\
&=&\,\,\frac{\bas}{2 \pi}\,\int\,d^2\,z\,K\left(x,y|z
\right)\,\left(N^+(x,z;Y - Y') \,+\,N^+(z,y;Y - Y')\,-\,N^+(x,y;Y - Y')\,- \right. \nonumber \\
&-&\left.
\, \, N^+(z,y;Y -
Y')\,N^+(z,y;Y
- Y')
\right) \nonumber\\
 &-& \,\,\frac{1}{(2 \pi)^4}\,\frac{\bas}{ \pi}\,\int\,\,G_0\left(x,y;Y'|x',y';Y'\right)\,\frac{d^2 x'\,d^2\,y'}{( x' -
y')^4}\,K \left(x',y'|z \right)\,\,d^2\,z\,\left( L_{z,y'}N\left(z,y';
Y'\right)\right)\,N^+(x',z;Y - Y')\,\nonumber
\eea

One can see that in the kinematic  region where (see \fig{qsat})
\beq \label{QSAT}
\frac{r^2}{R^2_0} \,Q^2_s(Y')\,\,<\,\,1\,; \,\,\,\,\,\,\,\,\,\,\,\,\,\,\,\,\,\,\mbox{and}
\,\,\,\,\,\,\,\,\,\,\,\,\,\,\,\,\,\,\frac{R^2}{r^2} \,Q^2_s(Y - Y')\,\,<\,\,1
\eeq
both amplitudes $N\left(r, R_0;Y'\right)$ and $N\left(R,r; Y - Y'\right)$ are deeply in the
saturation region. We can find the solution of \eq{SCEQG1} and \eq{SCEQG2} in this region
replacing both $N$ by $N = 1 +\Delta N$ and noticing that constant $N^+$ in \eq{SCEQG1}, as well
as
constant $N$ in \eq{SCEQG2}, does not give a contribution we see that the asymptotic solution is
$N = 1$ and for $\Delta N$ we have the following equation
\beq \label{DEN}
 \frac{\partial \left(L_{x,y}  \Delta N(x,y;Y') \right)}{\partial\,Y'}\,\,=
\eeq
$$
\,\,- \frac{\bas}{2
\pi}\,\int\,d^2\,z\,K\left(x,y|z\right)\, \left(L_{x,y}  \Delta N(x,y;Y') \right)\,\,-\,\,
\frac{\bas}{2
\pi}\,\int\,d^2\,z\,K\left(x,y|z\right)\left(L_{x,z}  \Delta N(x,z;Y') \right)
$$
In \eq{DEN} we neglect all contributions of the order $(\Delta N)^2$, $ \Delta N\,\Delta N^+$ and
$(\Delta N^+)^2$.

\eq{DEN} has a solution if we assume that $\Delta N$ has a geometrical scaling \cite{LT}, namely,
$\Delta N$ is a function of one variable
\beq \label{ZV}
\zeta\,\,\,=\,\,\,\ln \left(\frac{r^2}{R^2_0}\,\tilde{Q}^2_s(Y') \right)\,\,=\,\,\xi\,\,-\,\,\ln
\left(R^2_0\,\tilde{Q}^2_s(Y')\right)
\eeq
where $\tilde{Q}_s$ is given by \eq{QSNEW}.
\FIGURE[h]{
\centerline{\epsfig{file= 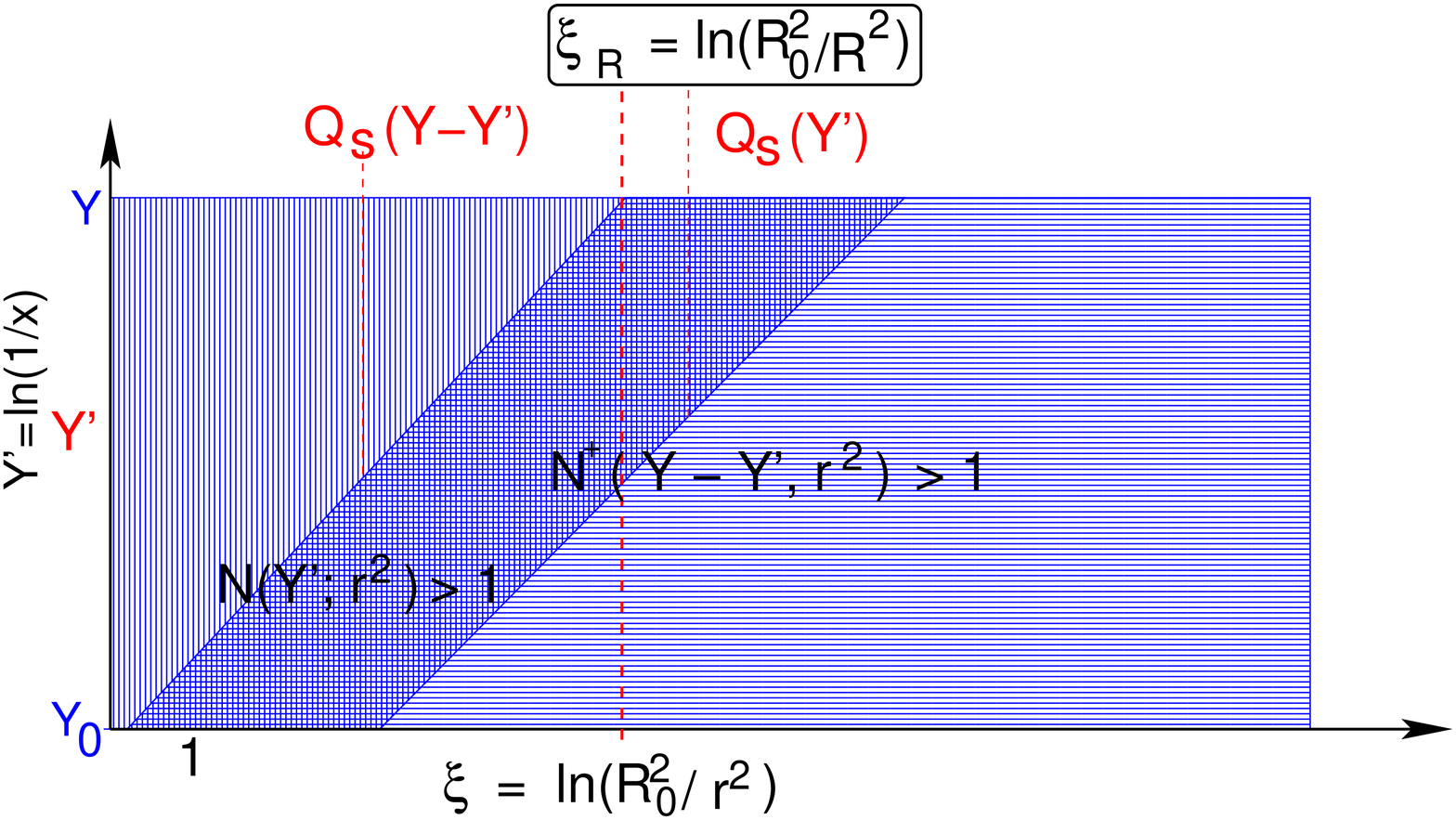,width=100mm}}
\caption{The kinematic region where both amplitudes $N(r^2,R^2_0;Y')$ and $N(r^2,R^2; Y - Y')$ are
in the saturation region. }
\label{qsat}}

Using \eq{QSFC} we can rewrite \eq{DEN} in the form
\beq \label{DEN1}
C_s \frac{\Delta \tilde{N}(\zeta)}{d
\zeta}\,\,=\,\,-\,\,\zeta\,\Delta \tilde{N}(\zeta)\,\,-\,\,\int_{1/\tilde{Q}_s(Y')}\,\frac{d^2
z}{\pi}\,K(x,y;z)\,\,\Delta \tilde{N}(x,z;Y')
\eeq
where $\tilde{N} \,\,=\,\,L_{x,y} N(x,y;Y)$.
In Ref. \cite{LT} it is shown that the main contribution in the saturation region stems from
logarithmic integration, namely,
\beq \label{DLAK}
\int_{1/\tilde{Q}_s(Y')}\,\,\frac{d^2
z}{\pi}\,\,K(x,y;z)\,\,\,\,\longrightarrow\,\,\,\,\int^{(x - y)^2}_{1/\tilde{Q}_s(Y')}\,\,\frac{d
z^2}{( x
- z)^2}\,\,\,\,+\,\,\,\, \int^{(x - y)^2}_{1/\tilde{Q}_s(Y')}\,\,\frac{d z^2}{( y
- z)^2}
\eeq
\eq{DLAK} reduces \eq{DEN1} to
\beq \label{DEN2}
C_s \,\frac{\Delta \tilde{N}(\zeta)}{d
\zeta}\,\,=\,\,-\,\,2\,\,\zeta\,\Delta \tilde{N}(\zeta)\,\,- \,\int^{\zeta}\,d \zeta'\Delta
\tilde{N}(\zeta')
\eeq

Introducing a new function $ \Delta {\cal N} (\zeta)\,\,=\,\,\int^{\zeta}\,d \zeta'\Delta
\tilde{N}(\zeta')$ we can rewrite \eq{DEN2} in the form
\beq \label{DEN3}
C_s\,\Delta {\cal
N}"_{\zeta,\zeta}\,\,=\,\,-\,\,2\,\,\zeta\,\Delta {\cal N}"_\zeta (\zeta)\,\,- \,\,\Delta {\cal N}
\eeq
the solution of this equation is equal to (see {\bf 9.255(2)} in  Ref. \cite{RY}
\beq \label{SOLDEN}
\Delta {\cal N}(\zeta)\,\,=\,\,\exp\left\{- \frac{\zeta^2}{2\, C_s}\,\right\}\,D_{- \frac{1}{2}}
\left( \sqrt{\frac{2}{C_s}}\,\zeta \right)
\eeq
 where $D_p(t)$ is the parabolic cylinder function (see {\bf 9.240 - 9.255} in Ref. \cite{RY}).
\eq{SOLDEN} leads to
\beq \label{SOLDEN1}
\Delta {\cal N}(\zeta)\,\,\,\longrightarrow\,\,\exp\left( - \frac{\zeta^2}{C_s} \right)
\eeq

In semi-classical approach \eq{SOLDEN1} leads to  $\Delta N\, \propto\, \exp\left( - \,\frac{1 -
\gamma_{cr}}{\omega(\gamma_{cr})}
\,\zeta^2 \right)$ in the case if we neglect the difference between $Q_s$ and $\tilde{Q}_s$ (see
\eq{QSNEW}.  This result is very surprising. Indeed, the approach to the asymptotic
solution in the case of the Balitsky-Kovchegov equation gives $\Delta N \propto\,exp\left( -
\,\frac{1 -
\gamma_{cr}}{2\,\omega(\gamma_{cr})}
\,\zeta^2 \right)
$ \cite{LT} which is slower than \eq{SOLDEN1}. The estimates
 based on the Iancu - Mueller approach \cite{IM,KOLE}  gave even smaller coefficient in the
exponent than for the Balitsky-Kovchegov equation. The solution of the functional equation (see
previous section and Ref. \cite{L4}) also gives a coefficient which is in two times smaller than
in the Iancu - Mueller approach.  In the last two approaches the $t$-channel unitarity has been
heavily used. Certainly, the BFKL Pomeron Calculus satisfies the $t$-channel unitarity
constraints, but we cannot guarantee that the semi-classical approach respects $t$ - channel
unitarity.

The second problem with the  solution of \eq{SOLDEN1} could be our assumption that the solution in
the saturation region has the same geometrical scaling behaviour as the solution to the
Balitsky-Kovchegov equation. In Ref. \cite{MUSH} it is shown that the geometrical scaling is not
correct assumption in the vicinity of the saturation scale in the Mueller-Shoshi band.
What happens inside the saturation region we have not studied yet.
 To find out whether the geometrical scaling behaviour is an
inherent property of the system we need to solve \eq{SCEQ1} and \eq{SCEQ2} in the vicinity of two
saturation scales taking into account the non-linear corrections.  We will do this in further
publications.

\section{Conclusions}
We show in this paper that the BFKL Pomeron Calculus in the kinematic region given
by \eq{REPC} has two equivalent descriptions: (i)\,\,one is the generating functional  which  gives a  clear
probabilistic
interpretation of the processes of high energy scattering and   provides also a Hamiltotian-like
description of the
system
of interacting dipoles; (ii)\,\,the second is the Langevin equation for directed percolation (see \eq{DPG}). We are able to prove this equation for the dipoles in the momentum represetation if the impact parametrepresentationlarger than
$b \,\gg 1/k$ where $k$ is the transverse momentum of a dipole.

 In other words, the BFKL Pomeron Calculus can be considered as an alternative description
of the statistical system of dipoles with different kinds of interactions between them.

One of the  results of this paper is the understanding the two Pomerons to one Pomeron merging in the
framework of the dipole approach. It turns out that the Feynman diagrams of \fig{v21pic}, \fig{v211pic}
and \fig{v212pic} describe the process given by \eq{STAGE}. In this process two dipoles are produced
in
the first stage and two other dipoles in the second stage of the processes. All these four dipoles can
interact with the target leading to \eq{V21FZF} for the contribution to the generating functional.
 At first sight,  \eq{V21FZF} contradicts the probabilistic treatment due to the negative
sign in front of this term in \eq{ZEQ}. Indeed, it seems strange that two dipoles decay into four
dipoles with negative amplitude since this amplitude looks rather as probability of this decay.
We showed in section 4 that this sign minus has a very simple explanation and stems from the fact that
the decay of two dipoles into four actually goes in two stages. During the first one, one dipole decays
into two with positive probability. During the second stage of the process the second dipole scatters
off the produced two dipoles. This rescattering does not produce the new dipoles but change the two
dipoles, produced during the first stage,  to two new dipoles.  This rescattering process goes in the
Born
approximation of perturbative QCD but it generates almost a  pure imaginary scattering amplitude.
This fact results in extra minus sign. The calculations of section 4 suggest the educated guess: all
processes of merging for Pomerons  ( say $nP \to mP$ with $m <n$,  for example such as $3 P \to 2 P$)
will lead to the term in
\eq{ZEQ} and \eq{chi} of the following type:
\beq \label{GT1}
\prod_i^{2 n} u(x_i,y_i)\,\prod^n_l \frac{ \delta}{\delta u(x_l,y_l)}
\eeq

It should be stressed that the vertex in front of \eq{GT1} can be found directly from the JIMWLK approach as
\cite{JIMWLK} as it was demonstrated in Ref. \cite{KOLU}. Therefore, the evolution equations (Markov's chain) for the generating functional together with the JIMWLK approach lead to the selfconsistent and complete theory of high energy interaction in QCD.

It should be stressed that the negative sign does not change the fact that the high energy amplitude
can be written in terms of probabilities to find the given number of dipoles of definite sizes (see
\eq{P}). It was shown in section 3.3 that the equations for generating functional can be rewritten as
the path integral for the partition function using which we can introduce the thermodynamic potential
to describe the system of dipoles.

It is shown that the question about negative amplitude does not arise  if we treat the system of dipoles in the momentum representation. Markov's chain of equation for this system is written in the paper (see \eq{PMR})  and can be considered as a practical way to find a solution in accessible range of energies.

The  result of this paper is the solution to the linear functional equation for the generating
functional in the  high energy region. We found the semi-classical solution for the toy model in the
entire kinematic region while we were able to calculate the asymptotic solution and
the first correction
to it at high energy in the general case of interacting dipoles. This solution shows the two BFKL Pomerons to one BFKL Pomeron merging is essential only in the limited region of the phase space where the resulting vertex is positive.

The third result of this paper is the first attempt to solve the semi-classical equations for the
BFKL Pomeron calculus. We confirm the Mueller-Shoshi band in $\ln(1/x)$ and $\xi = \ln( R^2_0/r^2)$
plane where the solution to the linear BFKL equation can describe the scattering amplitude which follows from the
set of equation in the semi-classical approach.
We found the solution deeply in the saturation region where the scattering amplitude approaches
unity ($ N \,\rightarrow \,1$). It turns out that approach is steeper for our case than for the
mean field approximation.

 Being elegant and beautiful the BFKL
Pomeron Calculus has a clear
disadvantage: it  lacks   theoretical ideas what kind of Pomeron interactions we
should take into account and why.  Of course, the Feynman diagrams in leading
$\ln(1/x)$ approximation of perturbative QCD allow us, in principle, to calculate all
possible Pomeron
interactions but, practically, it is very hard job. Even if we  calculate these
vertices we need to understand what set of vertices we should take into account for
calculation of the scattering amplitude.  In this paper we show that the the BFKL Pomeron calculus together with the JIMWLK approach  allows us to build the selfconsistent and complete theory of the high energy scattering in QCD.   However, we lack the formalism how to extract the multi Pomeron vertices from the JIMWLK approach
 as well as the general approach which will allow us to take into account all possible multi Pomeron vertices.
 This is the reason why we need to
develop a more general formalism. Fortunately, such a formalism has been
built and it is known under the abbreviation JIMWLK-Balitsky approach
\cite{MV,JIMWLK,B}. In this approach we are able to calculate all vertices for
Pomeron interactions as it was demonstrated in Ref. \cite{KOLU} and it solves the
first part of the problem: determination of all possible Pomeron interactions.
However, we need to understand what vertices we should take into account for
calculation of the scattering amplitude. We hope that a further progress in going
beyond of the BFKL Pomeron Calculus (see Refs. \cite{KOLU,HIMS}) will lead to such
a development of the BFKL Pomeron Calculus that we will have a consistent
theoretical
approach. Hopefully this approach will be simpler than Lipatov's effective action
\cite{LIEF} which is not easier to solve than the full QCD Lagrangian.

It is well known that the mean field approach to our problem which includes only one dipole  to two
dipoles decay in the master equation (see \eq{ZEQ}) has been studied quite well both analytically
\cite{LT} and numerically \cite{THEORVRSDATA}. We firmly believe that  the probabilistic interpretation
gives a practical method
for creating a Monte Carlo code in spirit of the approach suggested in Ref. \cite{MS}. This code will
allow us to find a numerical solution to the problem and to consider the inclusive
observables. This extension is very desirable since the most experimental data
exist for these observables. We also hope that the semi-classical approach  for searching the solution
to the master equation will be useful  for developing  both analytical and numerical  approaches which
will result in predictions for the LHC.

\section*{Acknowledgments:}
We want to thank Asher Gotsman, Larry McLerran, Dima Kharzeev, Alex Kovner, Misha Lublinsky and Uri
Maor for very useful
discussions on the subject
of this paper. Our special thanks go to Jose Guilherme Milhano, who draw our attention to Refs. \cite{HH,DMS}
and the discussions with whom on the subject of this paper were very useful and instructive.

 This research was supported in part  by the Israel Science Foundation,
founded by the Israeli Academy of Science and Humanities and by BSF grant \# 20004019.
\appendix
\begin{boldmath}
\section{ Calculation of $G_0(x_1,x_2|x'_1,x'_2)$.}  \label{sec:A}
\end{boldmath}
The solution of the BFKL equation is given by \cite{LI}
 
 \beq \label{sol_pr}
G(x_1,x_2;x^{'}_1,x^{'}_2|\omega)\,=\, 
\sum_{n=-\infty}^{+\infty}\int_{-\infty}^{+\infty}
\frac{(\nu^2+n^2/4) d\nu}{[\nu^2+(n-1)^2/4][\nu^2+(n+1)^2/4]}
\frac{G_{\nu \mu}(x_1,x_2,x^{'}_1,x^{'}_2)}{\omega-\omega(\nu, \mu)}
 \eeq

where $G_{\nu \mu}(x_1,x_2,x^{'}_1,x^{'}_2)$ is the Mellin transform of 
\eq{BFKLGF} and
$x_i$ are two-dimensional vectors in complex coordinates 
  \begin{eqnarray}\label{xcomp_pr}  
  x_i=x_{i,x}+i x_{i,y} & &  x^*_i=x_{i,x}-i x_{i,y}
  \end{eqnarray} 
The function $\omega(\nu, \mu)$ is the eigen value of the BFKL equation
given by  \eq{BFKLOM}.

The four-point Green function is presented in terms of the hypergeometric 
functions \cite{LI,NP}
  \begin{eqnarray}\label{4Green_pr}
  &  G_{\nu \mu}(x_1,x_2,x^{'}_1,x^{'}_2)=C_1 x^h x^{*\tilde{h}}
F(h,h,2h;x) F(\tilde{h},\tilde{h},2\tilde{h};x^*) & \nonumber \\
&+C_2 x^{1-h} x^{* 1- \tilde{h}}
F(1-h,1-h,2-2h;x) F(1-\tilde{h},1-\tilde{h},2-2\tilde{h};x^*) &
  \end{eqnarray}
with $h=\frac{1}{2}+i \nu +\frac{n}{2}$, $\tilde{h}=\frac{1}{2}+i \nu
-\frac{n}{2}$, and $x$ is the anharmonic ratio 
  
  \beq \label{anhar_rat}
    x= \frac{x_{12} x_{1' 2'}}{x_{1 1'} x_{22'}}
  \eeq
Coefficients  $C_1$ and $C_2$ are given by \cite{LI} 
  \begin{eqnarray}\label{C1C2_pr}
    C_1=\frac{b_{n,-\nu}}{2 \pi^2} \hspace{2cm} 
    C_2=\frac{b_{n,\nu}}{2 \pi^2}  
  \end{eqnarray}
with 
  \begin{eqnarray}\label{bnu_pr}
  b_{n,\nu}=\pi^3 2^{4i\nu}
\frac{\Gamma(-i\nu+(1+|n|)/2) \Gamma(i\nu+|n|/2)}{\Gamma(i\nu
+(1+|n|)/2)\Gamma(-i\nu+|n|/2)}
  \end{eqnarray}
As we have discussed, the high energy asymptotic behaviour stems from $n=0$ term in  
\eq{sol_pr}. The initial condition for \eq{BFKLGF} at $Y=Y_0$  is given by the 
following expression
\beq\label{sol1_pr}
G_0(x_1,x_2;x'_1,x'_2)\,\,=
\eeq
$$
\,\,\int\,\frac{d \omega}{2 \,\pi 
\,i} \int_{-\infty}^{+\infty}\,\,
\frac{\nu^2 d\nu}{(\nu^2+1/4)^2}
\frac{G_{\nu \mu}(x_1,x_2,x'_1,x'_2)}{\omega - \omega(\nu,n=0)}\,=\,\frac{1}{\omega}
\int_{-\infty}^{+\infty}\,\,
\frac{\nu^2 d\nu}{(\nu^2+1/4)^2}\,\,G_{\nu \mu}(x_1,x_2,x'_1,x'_2)
$$
This integral can be taken by closing contour of integration over
singularities of the integrand.

The function 
\begin{eqnarray}\label{nu2_pr}
\frac{\nu^2 }{(\nu^2+1/4)^2}
\end{eqnarray}
has two poles at $\frac{i}{2}$ and $-\frac{i}{2}$.
The four-point Green function of  \eq{4Green_pr} consists of two terms, one
with $C_1(x x^*)^{\frac{1}{2}+i\nu}$ and the other one $C_2(x
x^*)^{\frac{1}{2}-i\nu}$.
 For small $|x|$ this terms should be integrated
closing contour in  upper and lower semi-planes respectively. 
The resulting contour in the lower  semi-plane runs
 anticlockwise
and thus the value of the contour integral enters with a minus sign.

The terms
could be expanded in the vicinity of their poles. Let us consider the
first term. We expand the function $C_1$ in the vicinity of $\frac{i}{2}$
  \begin{eqnarray}\label{C1exp_pr}
  C_1 = \frac{\pi}{2}2^{-4i\nu}
\frac{\Gamma(-i\nu)}{\Gamma(-i\nu+\frac{1}{2})\Gamma(1+i\nu)} 
   \frac{\Gamma(i\nu+\frac{3}{2})}{(i\nu+\frac{1}{2})}
  \end{eqnarray}
The hypergeometric function can be written as a sum
  \begin{eqnarray}\label{hypersum_pr}
   F(a,b,c;x)=1+\frac{\Gamma(c)}{\Gamma(a)\Gamma(b)}\sum^{\infty}_{n=1}
   \frac{\Gamma(a+n)\Gamma(b+n)}{\Gamma(c+n)}\frac{x^n}{n!}
  \end{eqnarray} 
In the case of
$F(h,h,2h;x)$ the
singularity of $\Gamma$ function at $\frac{i}{2}$ can factorized out the
sum
  \begin{eqnarray}\label{hyperln_pr}
   F(h,h,2h;x)=1+\frac{\Gamma(2h)}{\Gamma(h)\Gamma(h)}\sum^{\infty}_{n=1}
   \frac{\Gamma(h+n)\Gamma(h+n)}{\Gamma(2h+n)}\frac{x^n}{n!}\nonumber\\
  \simeq 1+\frac{1}{2\Gamma(h)}\sum^{\infty}_{n=1}
   \frac{x^n}{n}=1-\frac{1}{2} \frac{ln(1-x)}{\Gamma (i\nu + \frac{1}{2} )}
  \end{eqnarray}
At this stage the first term of the integrand of \eq{sol1_pr} can be
 written as
   \beq\label{1term_pr}
\eeq
$$
\frac{1}{\omega}\frac{\nu^2}{(i\nu+\frac{1}{2})^3(i\nu-\frac{1}{2})^2} 
\frac{\pi}{2}  2^{-4i\nu} \frac{\Gamma(-i\nu)}{\Gamma(-i\nu+\frac{1}{2})\Gamma(1+i\nu)} 
 (xx^*)^{\frac{1}{2}+i\nu}
\left(1-\frac{1}{2}\frac{ln(1-x)}{\Gamma(\frac{1}{2}+i\nu)}\right)
\left(1-\frac{1}{2}\frac{ln(1-x^*)}{\Gamma(\frac{1}{2}+i\nu)}\right) 
 $$
 
It is clearly seen that the term of zero order in $ln|1-x|$  has a 
third order pole at $\nu=\frac{i}{2}$; the term of first order in $ln|1-x|$
has a second order pole at $\nu=\frac{i}{2}$, and, the term of second order in
$ln|1-x|$ has a simple pole at $\nu=\frac{i}{2}$. The contributions of those
terms in the contour integral are found to be 
   \beq\label{C1zero_pr}
   \frac{i \pi^2}{\omega}[-2\,\,+
\,\, ln(xx^*)(-4 + ln(xx^*) )]
   \eeq
   \beq\label{C1first_pr}
   -\frac{ \pi^2}{2\omega}
[-2\,\,+\,\,\gamma\,\, + \,\, ln(xx^*)]ln[(1-x)(1-x^*)]
   \eeq
    \beq\label{C1second_pr}
  - \frac{i\pi^2}{4\omega}ln(1-x)ln(1-x^*)
   \eeq
respectively.

In a similar way we may expand the second of the integrand in vicinity 
of its pole at $\nu=-\frac{i}{2}$, namely

\beq \label{2term_pr}
\eeq
$$
\frac{1}{\omega}\frac{\nu^2}{(\frac{1}{2}-i\nu)^3(i\nu+\frac{1}{2})^2} 
\frac{\pi}{2}  2^{4i\nu} \frac{\Gamma(i\nu)}{\Gamma(i\nu+\frac{1}{2})\Gamma(1-i\nu)} 
 (xx^*)^{\frac{1}{2}-i\nu}
\left(1-\frac{1}{2}\frac{ln(1-x)}{\Gamma(\frac{1}{2}-i\nu)}\right)
\left(1-\frac{1}{2}\frac{ln(1-x^*)}{\Gamma(\frac{1}{2}-i\nu)}\right) 
$$
  The integration is performed on the lower semicircle and  results in overall minus sign 
  of the integral. The contributions corresponding to \eq{C1zero_pr}, 
\eq{C1first_pr} 
and \eq{C1second_pr} are 

\beq\label{C2zero_pr}
-\frac{i \pi^2}{\omega}[-\,\,2\,\,+\,\,
 ln(xx^*)(-4 + ln(xx^*) )]
   \eeq
   \beq \label{C2first_pr}
   -\frac{ \pi^2}{2\omega}
[-2\,\,+\,\,\gamma \,\,+\,\,  ln(xx^*)]ln[(1-x)(1-x^*)]
   \eeq
\beq \label{C2second_pr}
  + \frac{i\pi^2}{4\omega}ln(1-x)ln(1-x^*)
   \eeq
   respectively.

Comparing the contributions we note that that of zero and second order \\ in $ln[(1-x)(1-x^*)]$ are 
exactly canceled out, and we are left with 
\beq \label{finalCont_pr}
   -2\frac{ \pi^2}{2\omega}
[-2\,\,+\,\,\gamma \,\,+\,\,  ln(xx^*)]ln[(1-x)(1-x^*)]
   \eeq 
   For small $|x|$ this can written as
   \beq \label{final2_pr}
   -\frac{4\pi^2}{\omega}\hspace{0.15cm}ln|x|\hspace{0.15cm}ln|1-x|
   \eeq 
   or
   \beq\label{final02_pr}
  +\frac{4\pi^2}{\omega}\hspace{0.15cm}ln\frac{1}{|x|}\hspace{0.15cm}ln|1-x|
   \eeq 
   Going  back to complex vector representation of x and rewriting \eq{final3_pr}  as
   \beq \label{final3_pr}
G_0(x_1,x_2;x'_1,x'_2)\,\,=\,\,
  +\frac{4\pi^2}{\omega}\hspace{0.15cm}ln \left|\frac{x_{11'}x_{22'}}{x_{12'}x_{1'2}}\right|  
   \hspace{0.15cm}ln \left|\frac{x_{11'}x_{22'}}{x_{12}x_{1'2'}}\right|
   \eeq
we see that we    reproduce the result of \cite{LI}. Therefore, we demonstrated that    
\eq{final3_pr} gives a correct initial condition for searching the scattering 
amplitude at high energies restricting ourselves by the one term in \eq{BFKLGF} with 
$n =0$. 

\section{ The path integral formalism for the generating  functional.} \label{sec:B}
We want to develop a path integral formalism similar to that we
found for the toy model, but where the probabilities to find $n$
dipoles depend of dipole sizes. In our notation we denote by Latin
index rapidity interval, and Greek indices relate to a size of
dipole.

As in Chapter 3.3  we introduce the creation and annihilation operators
\begin{center}\label{appB_1}
\begin{eqnarray}
\hat{a}(q)=\frac{\delta}{\delta u(q)} \hspace{2cm} \hat{a}^{\dagger}(q)=u(q)
\end{eqnarray}
\end{center}
with commutation relations $[\hat{a}(q),\hat{a}^{\dagger}(q')]=\delta(q - q')$ at fixed $Y$. The expression for the coherent states in this case takes form of
\begin{center}
\begin{eqnarray}
|\phi(q)>=e^{\phi(q)\hat{a}^{\dagger}(q)-\phi(q)}|0>
\end{eqnarray}
\end{center}
with
\begin{center}
\begin{eqnarray} \label{appB_11}
\hat{a}(q')|\phi(q)>=\phi(q)|\phi(q)> \delta(q -q')
\end{eqnarray}
\end{center}

First we consider discrete dipole sizes $q_\alpha=(L/N)\alpha$, where $L$ is a maximal possible dipole size, $N$
a number of intervals of $L$ discretization, and $\alpha$ is an integer number running from $0$ to $N$.
In this case the commutation relations become $[\hat{a}(q_\alpha),\hat{a}^{\dagger}(q_{\beta})]=\delta _{\alpha,\beta}$.

The unit operator can be written in terms of the coherent states
\begin{center}
\begin{eqnarray}
\hat{I}=\prod_{\alpha}\int \frac{d\phi^*(q_\alpha) d\phi(q_\alpha)}{i\pi } e^{ -\phi^*(q_{\alpha})\phi(q_\alpha)+\phi(q_\alpha)+\phi^*(q_\alpha)}  |\phi(q_\alpha)><\phi(q_\alpha)|
\end{eqnarray}
\end{center}

The operator $\mathcal{H}$ defined in Eq.~(\ref{chi}) can be written as
\begin{center}
\begin{eqnarray} \label{appB_2}
\mathcal{H}[\hat{a}^{\dagger},\hat{a}] = &-& \sum_{\beta} \sum_{\gamma} \sum_{\lambda} [V_{1\rightarrow 2}
(q_\beta\rightarrow q_\gamma+q_\lambda) \{-\hat{a}^{\dagger}(q_{\beta}) +\hat{a}^{\dagger}(q_{\gamma})\hat{a}^{\dagger}(q_{\lambda}) \} \hat{a}(q_{\beta}) \nonumber \\
&-& V_{2 \rightarrow 1} (q_\gamma+q_\lambda \rightarrow q_\beta) \{\hat{a}^{\dagger}(q_{\gamma})\hat{a}^{\dagger}(q_{\lambda}) -\hat{a}^{\dagger}(q_{\beta})\} \frac{1}{2} \hat{a}(q_{\gamma})\hat{a}(q_{\lambda}
)]
\end{eqnarray}
\end{center}
Following the logic of Chapter  3.3  we consider a matrix element
\begin{center}
\begin{eqnarray}
\left\{\prod_{\alpha'}<\phi_{j+1}(q_{\alpha'})|\right\} (1+\mathcal{H}\Delta Y )\left\{|\prod_{\alpha}\phi_{j}(q_{\alpha})>\right\}
\end{eqnarray}
\end{center}
First, look at the second term of the Hamiltonian
\begin{center}
\begin{eqnarray}\label{appB_4}
\left\{\prod_{\alpha'}<\phi_{j+1}(q_{\alpha'})|\right\}
 \sum_{\beta} \sum_{\gamma} \sum_{\lambda} [-V_{1\rightarrow 2}
(q_\beta\rightarrow q_\gamma+q_\lambda)\hat{a}^{\dagger}(q_{\gamma})\hat{a}^{\dagger}(q_{\lambda})\hat{a}(q_{\beta})]
 \left\{\prod_{\alpha}|\phi_{j}(q_{\alpha})>\right\} \nonumber \\
\left\{\prod_{\alpha'}<\phi_{j+1}(q_{\alpha'})|\right\}
 \sum_{\beta} \sum_{\gamma} \sum_{\lambda} [-V_{1\rightarrow 2}
(q_\beta\rightarrow q_\gamma+q_\lambda)
\phi^*_{j+1}(q_{\gamma})
\phi^*_{j+1}(q_{\lambda})
\phi_{j}(q_{\beta})
]
 \left\{\prod_{\alpha}|\phi_{j}(q_{\alpha})>\right\}
\end{eqnarray}
\end{center}
In Eq.~(\ref{appB_4}) we used the property of the coherent states given by  Eq.~(\ref{appB_11}).

In the continuous limit $\delta_{\alpha,\beta}$ is replaced by $\delta(q_\alpha-q_\beta)$, and
$\prod_{\alpha}d\phi^*(q_\alpha)d\phi(q_\alpha)$  by  functional integration $\int \mathcal{D} \phi^* \mathcal{D} \phi$.

From here we see that rest of the calculations is similar to that of Chapter  3.3 and we end up with the expression
for a matrix element of an operator $A$ between states of initial $Y_0$ and final rapidity $Y$
\begin{center}
\begin{eqnarray}\label{appB_3}
<Y|A|Y_0>  \sim \int \mathcal{D}\Phi^+ \mathcal{D}\Phi A(Y) e^{S}
\end{eqnarray}
\end{center}
where
\begin{eqnarray}\label{appB_5}
 S\, &=& \,\int \left( \int\Phi^+(q)  \frac{d}{dY} \Phi(q)dq +\mathcal{H} ( \Phi^+ +1, -\Phi ) \, \right) dY
\end{eqnarray}
with the Hamiltonian given by

\begin{eqnarray}
\mathcal{H} &=& \nonumber
\\
 &+&\int d^4 q_0 d^4 q_1 d^4 q_2 [V_{1\rightarrow 2}
(q_0\rightarrow q_1+q_2) \{-\Phi^+(q_{0}) +\Phi^+(q_{1})+\Phi^+(q_{2})+ \Phi^+(q_{1})\Phi^+(q_{2}) \} \Phi(q_{0}) \nonumber \\
&-& V_{2 \rightarrow 1} (q_1+q_2 \rightarrow q_0) \{
\Phi^+(q_{1})\Phi^+(q_{2})+\Phi^+(q_{1}) +\Phi^+(q_{2})-\Phi^+(q_{0})
\} \frac{1}{2}
\Phi(q_{1})\Phi(q_{2})
 ]
\end{eqnarray}

\section{The exact calculations in the semi-classical approach} \label{sec:C}

At this appendix we calculate exactly exponent in  semi-classical solution and compare it with it approximations at
 small and large values of $\omega$, which ware used for calculation of corresponding weight functions.

Remind that \eq{ROOTS}

\beq 
\Phi^{\pm}_u\,\,=\,\,\frac{\kappa_1}{2\,\,L(u)}\,\left\{\,\, 1\,\,\pm\,\,\sqrt{1 \,\,+\,\,\frac{4 
\,\,\omega\,\,L(u)}{\,\,\kappa_1\,\,u\,(1 - u)\,}}\,\,\right\}
\eeq

\noindent Therefore exponent in  semi-classical solution $\Phi^{\pm}$ will be defined by the following integrals:

\beq \label{ROOTS_INT}
\Phi^{\pm}(u) \,\,= \,\, \int_{0}^{u} \frac{\kappa_1}{2\,\,L(u')}\,\left\{\,\, 1\,\,\pm\,\,\sqrt{1 \,\,+\,\,\frac{4 
\,\,\omega\,\,L(u')}{\,\,\kappa_1\,\,u'\,(1 - u')\,}}\,\,\right\} \, du' \, = \, I_1 \, + \, I_2 
\eeq

\noindent where

\bea 
I_1 \, &=& \,\, \int_{0}^{u} \frac{\kappa_1}{2\,\,L(u')} \, du' \nonumber \\ 
&=& \,\, \frac{2}{\sqrt{6 \kappa_2 - 1 - \kappa_2^2}} \cdot \left[ \arctan \left( \frac{1 - \kappa_2 + 2u}{\sqrt{6 \kappa_2 - 1 - \kappa_2^2}} \right) \, - \, \arctan \left( \frac{1 - \kappa_2}{\sqrt{6 \kappa_2 - 1 - \kappa_2^2}} \right) \right] \nonumber \\ 
\eea

\bea 
I_2 \,\,&=& \,\, \pm \, \frac{\kappa_1}{2} \int_{0}^{u} \,\frac{du'}{L(u')} \,\sqrt{1 \,\,+\,\,\frac{4 
\,\,\omega\,\,L(u')}{\,\,\kappa_1\,\,u'\,(1 - u')\,}} \nonumber \\ 
&=& \,\, \pm \, \frac{\kappa_1}{2} \int_{0}^{u} \,\frac{du'}{u'(1+u') + \kappa_2(1-u')} \,\sqrt{1 \,\,+\,\,\frac{4 
\,\,\omega \cdot [u'(1+u') + \kappa_2(1-u')]}{\,\,\kappa_1\,\,u'\,(1 - u')\,}} \nonumber \\ 
&=& \,\, \pm \,\, [ \, II_2(u) - II_2(0) \, ]
\eea

\noindent where $II_2$ defined as following indefinite integral:

{
\footnotesize

\bea 
II_2(t) \,\, &=& \, t \,\, \sqrt{ - \,\, \frac{(A^+ \, + \, 8 \omega \frac{\kappa_2}{\kappa_1} (1-t)) \cdot (A^- \, + \, 8 \omega \frac{\kappa_2}{\kappa_1} (1-t))}{4^3 \omega^2 \frac{\kappa_2}{\kappa_1^2}}} \,\, \sqrt{1 + 4 \omega \, \frac{\frac{\kappa_2}{\kappa_1} + \frac{1 - \kappa_2}{\kappa_1}t + \kappa_2 t^2}{t(1-t)}}\nonumber \\ 
&& \, \cdot \{ \left( 4 \omega \sqrt{ \frac{1}{\kappa_1^2} - 6 \frac{\kappa_2}{\kappa_1^2} + \frac{\kappa_2^2}{\kappa_1^2} } \right) EF \left[ i \cdot \arcsin h \left( \sqrt{\frac{4 \omega}{A^-} \cdot \frac{\kappa_2}{\kappa_1} \cdot \frac{t}{1-t}} \right), \,\, \frac{A^-}{A^+}  \right] \nonumber \\ 
&& - \, EPi \left[ - \, \frac{A^-}{\frac{4 \omega}{\kappa_1} \cdot (- 1 - \kappa_2 + \sqrt{1 - 6 \kappa_2 + \kappa_2^2})}, \,\, i \cdot \arcsin h \left( \sqrt{\frac{8 \omega}{A^-} \cdot \frac{\kappa_2}{\kappa_1} \cdot \frac{1-t}{t}} \right), \,\, \frac{A^-}{A^+}  \right] \nonumber \\ 
&& + \, EPi \left[ \frac{A^-}{\frac{4 \omega}{\kappa_1} \cdot (1 + \kappa_2 + \sqrt{1 - 6 \kappa_2 + \kappa_2^2})}, \,\, i \cdot \arcsin h \left( \sqrt{\frac{8 \omega}{A^-} \cdot \frac{\kappa_2}{\kappa_1} \cdot \frac{1-t}{t}} \right), \,\, \frac{A^-}{A^+}  \right] \} \nonumber \\ 
&& \cdot \frac{1}{ \left[ -t(1-t) - 4 \omega \, \frac{\frac{\kappa_2}{\kappa_1} + \frac{1 - \kappa_2}{\kappa_1}t + \kappa_2 t^2}{t(1-t)} \right] \cdot \sqrt{1 - 6 \kappa_2 + \kappa_2^2} \,\,\, \sqrt{\frac{4 \omega}{A^-} \cdot \frac{\kappa_2}{\kappa_1} \cdot \frac{t}{1-t}}} \nonumber \\ 
\eea
}

In the last relation we used the following notations:

{
\footnotesize

\bea 
A^\pm &\equiv& 1 + \frac{4 \omega}{\kappa_1} + 4 \omega \frac{\kappa_2}{\kappa_1} \; \pm \; \sqrt{ - 128 \omega \frac{\kappa_2}{\kappa_1^2} + \left(1 + \frac{4 \omega}{\kappa_1} + 4 \omega \frac{\kappa_2}{\kappa_1} \right)^2 } \nonumber \\ 
EF(\varphi, m) &\equiv&  \int_{0}^{\varphi} \frac{d \theta}{\sqrt{1 - m \cdot \sin^{2} (\theta)}} \; = \; Elliptic \; integral \; of \; the \; first \; kind  \nonumber \\ 
EPi(n, \varphi, m) &\equiv&  \int_{0}^{\varphi} \frac{d \theta}{(1 - n \cdot \sin^{2} (\theta)) \cdot \sqrt{1 - m \cdot \sin^{2} (\theta)}}  \; = \; Elliptic \; integral \; of \; the \; third \; kind  \nonumber \\ 
&& \nonumber \\ 
&&  \nonumber \\ 
\eea

}

We can see that in spite the fact that $\Phi^{\pm}$ can be calculated explicitly (in radicals), it is too complicated 
in order to use it for the calculations of the weight functions $\phi^\pm(\omega)$ in \eq{SCTM2} and \eq{SCTM3}. 
Nevertheless, this exact calculation very fruitful for two purposes: (I) using exact form of $\Phi^{\pm}$ we can 
estimate how good our approach for small and large $\omega$ (or in another words how good we estimated spectrum of our 
solution), (II) once we know our spectral functions $\phi^\pm(\omega)$ we can use this exact solution for the following 
calculations. It is important to note that $II_2(t)$ looks as imaginary  functions, but on fact the imaginary
 part is 
closely related to regularization of singularity at $u=0$, thus it is essential in some since.
 The plots below illustrates how good our approach for small and large $\omega$, which we used for calculations of $\Phi^{\pm}$.

\FIGURE[h]{
	\centerline{\epsfig{file=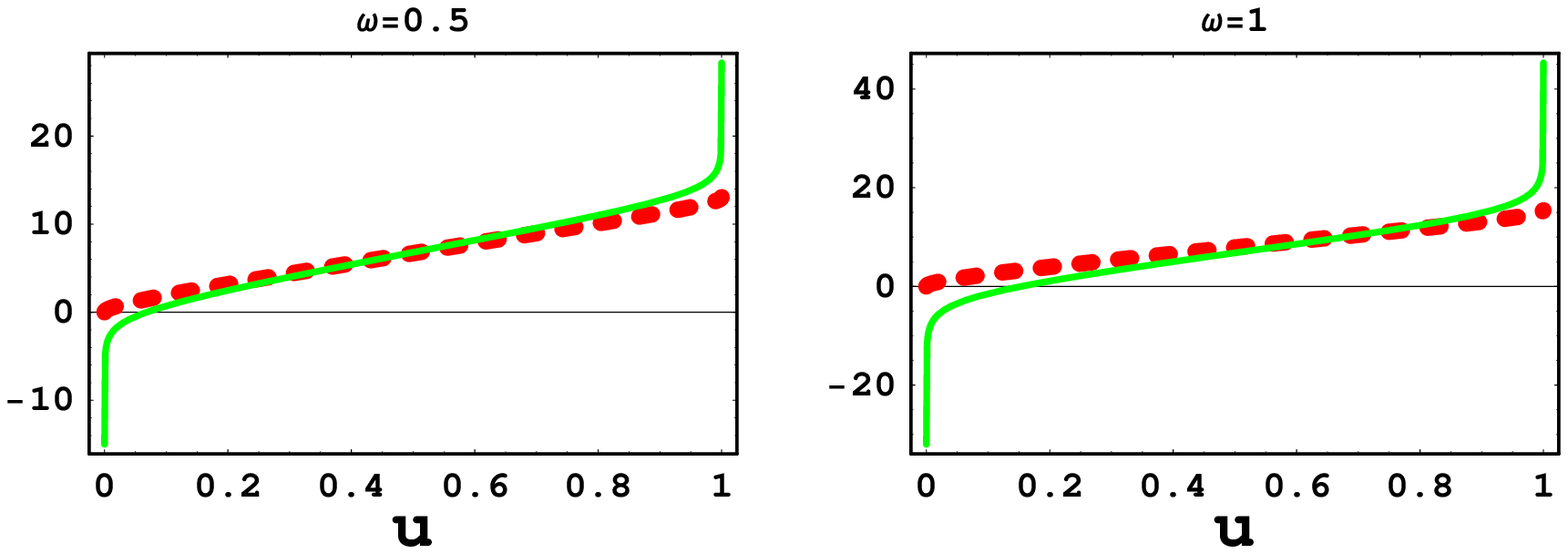,width=15cm, height=5cm}}
		\caption{At this plot we compare positive branch of our exact result (thick doted line) with for $\phi(\omega)$ 
with approximation for small values of $\omega$ (i.e. $\omega > \frac{\kappa_1}{\kappa_2}$), which was used in \eq{SCTM2} ~(thin solid line)}
	\label{fig:SmalFy}}

\FIGURE[h]
	{\centerline{\epsfig{file=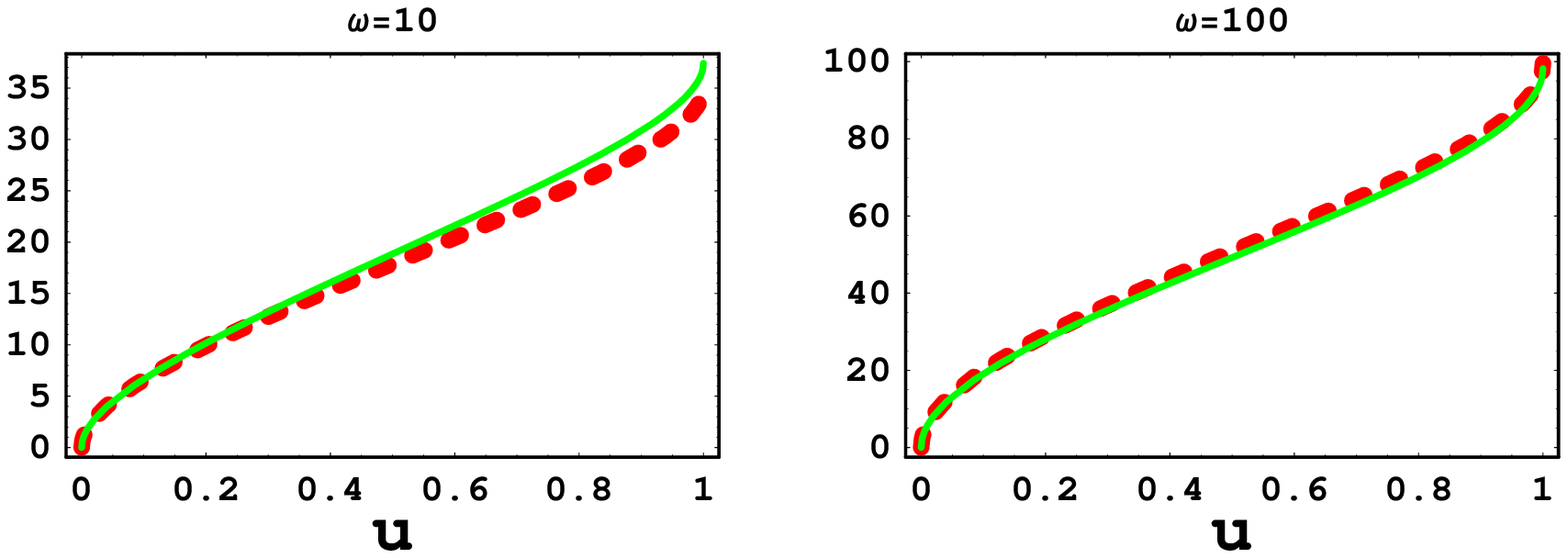,width=15cm, height=5cm}}
		\caption{At this plot we compare positive branch of our exact result (thick doted line) with for $\phi(\omega)$
 with approximation for large values of $\omega$ (i.e. $\omega > \frac{\kappa_1}{\kappa_2}$), which was used in \eq{SCTM5} 
~(thin solid line)}
	\label{fig:LargeFy}}

\end{document}